\documentclass[usenatbib]{mnras}
\usepackage{natbib,amssymb,amsmath,times,graphicx,url,aas_macros}
\usepackage{setspace}      
\usepackage{epstopdf}       
\usepackage{verbatim}       
\usepackage{bm}		  
\usepackage{fleqn}		  

\newcommand{\sm}{$\sim$}
\newcommand{\appx}{$\approx$}

\newcommand\kms{km~s$^{-1}$}

\newcommand\dtscapprox[1]{d$t_\text{sc} \approx {#1}$~yr}
\newcommand\dtsclt[1]{d$t_\text{sc} \le {#1}$~yr}

\newcommand\zetacr{$\zeta_\text{cr}$}
\newcommand\zetam[1]{$\zeta_{#1}$}  
\newcommand\zetams[2]{$\zeta_{#1}^{#2}$}  
\newcommand\zetaeq[1]{$\zeta_\text{cr}=10^{#1}$ s$^{-1}$}  
\newcommand\zetaapprox[1]{$\zeta_\text{cr}\approx10^{#1}$ s$^{-1}$}  
  
\newcommand\zetage[1]{$\zeta_\text{cr} \gtrsim 10^{#1}$ s$^{-1}$}

\newcommand\zetageq[1]{$\zeta_\text{cr} \ge 10^{#1}$ s$^{-1}$}  

\newcommand\rhoeq[1]{$\rho_\text{max} = 10^{#1}$ g~cm$^{-3}$}  
\newcommand\rhoapprox[1]{$\rho_\text{max} \approx 10^{#1}$ g~cm$^{-3}$}  
  
\newcommand\rhole[1]{$\rho_\text{max} \lesssim 10^{#1}$ g~cm$^{-3}$}

\defcitealias{PriceBate2007}{PB07}
\defcitealias{WPB2016}{WPB16}

\title[Collapse to stellar densities using radiative non-ideal MHD]{The collapse of a molecular cloud core to stellar densities using radiation non-ideal magnetohydrodynamics}

\author[Wurster, Bate \& Price]{James Wurster$^{1,2}$\thanks{j.wurster@exeter.ac.uk}, Matthew R. Bate$^{1,2}$\thanks{mbate@astro.ex.ac.uk}, and Daniel J. Price$^{2}$ \\
$^{1}$School of Physics and Astronomy, University of Exeter, Stocker Rd, Exeter EX4 4QL, UK \\
$^{2}$Monash Centre for Astrophysics and School of Physics and Astronomy, Monash University, Vic 3800, Australia \\
}
\date{Submitted: Revised: Accepted: }
\pagerange{\pageref{firstpage}--\pageref{lastpage}} \pubyear{2017}

\begin{document}
\label{firstpage}
\bibliographystyle{mnras}
\maketitle

\begin{abstract}
We present results from radiation non-ideal magnetohydrodynamics (MHD) calculations that follow the collapse of rotating, magnetised, molecular cloud cores to stellar densities.  These are the first such calculations to include all three non-ideal effects: ambipolar diffusion, Ohmic resistivity and the Hall effect.  We employ an ionisation model in which cosmic ray ionisation dominates at low temperatures and thermal ionisation takes over at high temperatures.  We explore the effects of varying the cosmic ray ionisation rate from $\zeta_\text{cr}= 10^{-10}$ to $10^{-16}$~s$^{-1}$.  Models with ionisation rates $\gtrsim 10^{-12}$~s$^{-1}$ produce results that are indistinguishable from ideal MHD. Decreasing the cosmic ray ionisation rate extends the lifetime of the first hydrostatic core up to a factor of two, but the lifetimes are still substantially shorter than those obtained without magnetic fields.  Outflows from the first hydrostatic core phase are launched in all models, but the outflows become broader and slower as the ionisation rate is reduced.  The outflow morphology following stellar core formation is complex and strongly dependent on the cosmic ray ionisation rate.  Calculations with high ionisation rates quickly produce a fast ($\approx 14$~km~s$^{-1}$) bipolar outflow that is distinct from the first core outflow, but with the lowest ionisation rate a slower ($\approx 3-4$~km~s$^{-1}$) conical outflow develops gradually and seamlessly merges into the first core outflow.
\end{abstract}


\section{Introduction}
\label{intro}

Almost fifty years ago, \cite{Larson1969} performed the first numerical calculations to follow a molecular cloud as it collapsed to form a protostar.  \citeauthor{Larson1969} identified several distinct phases during the evolution.  The initial collapse was found to proceed almost isothermally due to the low optical depths at the long wavelength of the radiation.  Once the inner regions began to trap radiation effectively, they began to evolve almost adiabatically producing a pressure-supported object known as the first hydrostatic core. This first core had a typical radius of $\approx 5$~au and an initial mass of a few Jupiter-masses [$M_{\rm J}]$.  The first core grew in mass as it accreted material from the envelope until its central temperature reached $\approx 2000$~K, whereupon molecular hydrogen began to dissociate triggering a second phase of dynamical collapse.  Once the hydrogen had become mostly atomic, a second hydrostatic core, also known as the stellar core, formed with an initial radius $\approx 2$~R$_\odot$ and mass $\approx 1.5~M_{\rm J}$.  The stellar core subsequently accreted the remaining envelope to produce a young star.

This general picture has been confirmed by more recent one-dimensional \citep{MasInu2000,Commerconetal2011b,Vaytetetal2012,Vaytetetal2013} and multi-dimensional calculations.  However, multi-dimensional calculations also allow for the effects of additional physical processes to be studied.  Introducing rotation changes the structure of the first hydrostatic core and also allows the possibility of fragmentation.  Rotating first hydrostatic cores become disc-like in morphology, as demonstrated in two-dimensional calculations \citep{Larson1972, Tscharnuter1987, Tscharnuteretal2009}.  In fact, with sufficient initial rotation, the stellar core forms within a pre-stellar disc \citep{Bate1998, Bate2011, MacInuMat2010}.  If a first core rotates rapidly enough, three-dimensional calculations show that it may become bar-mode unstable and form trailing spiral arms \citep*{Bate1998, SaiTom2006, SaiTomMat2008, MacInuMat2010, Bate2010, Bate2011}.  Gravitational torques from these spiral arms remove angular momentum from the inner regions of the first core.  This expedites the second collapse and helps prevent close binary formation by fragmentation during the second collapse phase \citep{Bate1998}.  With even greater initial rotation, the disc may fragment on scales of tens of au to produce additional first cores \citep[e.g.][]{Bate2011}. 

The introduction of magnetic fields provides another mechanism to transport angular momentum, reducing the rotation rates of first hydrostatic cores.  Magnetic fields can also drive outflows.  Outflows can be launched from the first core with typical speeds of $v \sim 2$~km s$^{-1}$ \citep{Tomisaka2002, Machidaetal2005, BanPud2006, MacInuMat2006, MacInuMat2008, HenFro2008, Commerconetal2010, Burzleetal2011b, PriTriBat2012}.  After the formation of the stellar core, outflows with speeds of $v\approx10-30$~km s$^{-1}$ have been obtained in magnetohydrodynamical (MHD) simulations \citep{BanPud2006, MacInuMat2006, MacInuMat2008}.

Many of the three-dimensional calculations mentioned above used approximate barotropic equations of state to model the thermal evolution of the gas.  The first three-dimensional calculations to follow the collapse of a molecular cloud core to stellar densities while including a realistic equation of state and radiative transfer were those of \cite{WhiBat2006} and \cite{Stamatellosetal2007}.   \cite{Bate2010, Bate2011} showed that the high accretion rates immediately following the formation of the stellar core could produce temperatures sufficient to launch short-lived bipolar outflows even without magnetic fields \cite[see also][]{SchTsc2011}.   However, in reality, magnetic fields are expected to be the primary mechanism for generating outflows from low-mass protostars.

\citeauthor{Tomidaetal2010a} (\citeyear{Tomidaetal2010a}, \citeyear{Tomidaetal2010b}) and \citeauthor{Commerconetal2010} (\citeyear{Commerconetal2010}, \citeyear{Commerconetal2012}) have studied first core formation and the associated magnetically-driven outflows using calculations that include both magnetic fields and radiative transfer.  Recently, \cite{Tomidaetal2013} and \cite*{BatTriPri2014} performed radiation magnetohydrodynamics (RMHD) calculations that followed the collapse to stellar core formation and the launching of both the slow outflow from the first core, and the faster outflow from the vicinity of the stellar core.  While the former of these studies was only able to follow the fast outflow for a fraction of an au, the latter followed the fast outflow until it had escaped the remnant of the first core ($\approx 4$~au).  \citeauthor{BatTriPri2014} used ideal RMHD, while \citeauthor{Tomidaetal2013} performed both ideal RMHD calculations and some that included physical Ohmic resistivity. 

Most recently, attention has turned to the effects resulting from partial ionisation, initially in an attempt to prevent the magnetic braking catastrophe --- the failure to produce rotationally supported Keplerian discs when magnetic field with realistic strengths \citep[e.g.][]{HeiCru2005} are accounted for \citep[e.g.][]{AllLiShu2003,PriBat2007,MelLi2008,HenFro2008,WurPriBat2016}.  In addition to Ohmic resistivity, the magnetic field evolution is affected by ion-neutral (ambipolar) diffusion and the Hall effect.  \cite{TsukamotoEtAl2015a} and \cite{WurPriBat2016} performed non-ideal MHD calculations that followed collapse to the scales of the first hydrostatic core.  They showed that the Hall effect promotes disc formation when the magnetic field is anti-aligned with the rotation axis, while it inhibits disc formation when the field and rotation axes are aligned, confirming earlier analytic studies \citep[e.g.][]{BraWar2012}.  \cite{TsukamotoEtAl2015b} performed non-ideal RMHD calculations that followed the collapse until just before stellar core formation that included both Ohmic resistivity and ambipolar diffusion (but not the Hall effect).  They found that Ohmic resistivity dramatically reduce the magnetic field strength in the first hydrostatic core compared to using ideal RMHD, and also prevented the outflow from the first core.

In this paper, we follow up \cite{BatTriPri2014} with non-ideal RMHD calculations that include all three effects from partial ionisation, namely Ohmic resistivity, ambipolar diffusion and the Hall effect. The calculations were performed using smoothed particle radiation magnetohydrodynamics (SPRMHD), combining the radiation hydrodynamics algorithm from \cite*{WhiBatMon2005} and \citet{WhiBat2006}, and the non-ideal MHD algorithm from \citet{WurPriAyl2014} and \citet{Wurster2016}, an extension of the ideal SPMHD method of \cite{TriPri2012}.

 We focus on the evolution of the magnetic field and the characteristics of the outflows during and after the formation of the first hydrostatic core and the stellar core. We describe our method in Section \ref{sec:numerics}, initial conditions in Section \ref{sec:ic}, results in Section \ref{sec:results}, and conclusions in Section \ref{sec:conclusion}.


\section{Numerical method}
\label{sec:numerics}
\subsection{Radiation non-ideal magnetohydrodynamics}

We solve the equations of self-gravitating, radiation non-ideal magnetohydrodynamics in the form
\begin{eqnarray}
\frac{{\rm d}\rho}{{\rm d}t} & = & -\rho \nabla\cdot \bm{v}, \label{eq:cty} \\
\frac{{\rm d} \bm{v}}{\rm{d} t} & = & -\frac{1}{\rho}\nabla \cdot \left[\left(p+\frac{B^2}{2}\right)I - \bm{B}\bm{B}\right] \notag \\
 &-& \nabla\Phi + \frac{\kappa \mbox{\boldmath$F$}}{c}, \label{eq:mom} \\
\rho \frac{\rm d}{{\rm d}t} \left(\frac{\bm{B} }{\rho} \right) & = & \left( \bm{B} \cdot \nabla \right) \bm{v} + \left.\frac{{\rm d} \bm{B}}{{\rm d} t}\right|_\text{non-ideal} \label{eq:ind}, \\
\rho \frac{\rm d}{{\rm d}t}\left( \frac{E}{\rho}\right) & = & -\nabla\cdot \bm{F} - \mbox{$\nabla \bm{v}${\bf :P}} + 4\pi \kappa \rho B_\text{P} - c \kappa \rho E, \label{eq:radiation} \\
\rho \frac{{\rm d}u}{{\rm d}t} & = & -p \nabla\cdot\bm{v} - 4\pi \kappa \rho B_\text{P} + c \kappa \rho E+ \rho \left.\frac{\text{d} u}{\text{d} t}\right|_\text{non-ideal}, \label{eq:matter} \\
\nabla^{2}\Phi & = & 4\pi G\rho, \label{eq:grav}
\end{eqnarray}
where ${\rm d}/{{\rm d}t} \equiv \partial/\partial t  + \bm{v}\cdot \nabla$ is the Lagrangian derivative,  $\rho$ is the density, ${\bm  v}$ is the velocity, $p$ is the gas pressure, ${\bm B}$ is the magnetic field, $\Phi$ is the gravitational potential, $B_\text{P}$ is the frequency-integrated Plank function, $E$ is the radiation energy density, $\mbox{\boldmath $F$}$ is the radiative flux, {\bf P} is the radiation pressure tensor, $c$ is the speed of light, and $G$ is the gravitational constant, and $I$ is the identity matrix.  Non-ideal MHD contributes to both the induction equation \eqref{eq:ind} and the energy equation \eqref{eq:matter} via \citep{WurPriAyl2014}
\begin{eqnarray}
\label{eq:nimhd}
\left.\frac{{\rm d} \bm{B}}{{\rm d} t}\right|_\text{non-ideal} = &-&\bm{\nabla} \times \left[  \eta_\text{OR}      \left(\bm{\nabla}\times\bm{B}\right)\right] \notag \\
                                                                                            &-& \bm{\nabla} \times \left[  \eta_\text{HE}       \left(\bm{\nabla}\times\bm{B}\right)\times\bm{\hat{B}}\right] \notag \\
                                                                                            &+&  \bm{\nabla} \times \left\{ \eta_\text{AD}\left[\left(\bm{\nabla}\times\bm{B}\right)\times\bm{\hat{B}}\right]\times\bm{\hat{B}}\right\},\end{eqnarray}
and
\begin{eqnarray}
\label{eq:ni:u}
\frac{\text{d} u}{\text{d} t}_\text{non-ideal} &=& \frac{\eta_\text{OR}}{\rho}\left|\bm{\nabla}\times\bm{B}\right|^2 \notag \\
                                                                  &+&\frac{\eta_\text{AD}}{\rho}\left\{ \left|\bm{\nabla}\times\bm{B}\right|^2 - \left[\left(\bm{\nabla}\times\bm{B}\right)\cdot\hat{\bm{B}}\right]^2 \right\}, 
\end{eqnarray}
respectively, where $\eta_\text{OR}$, $\eta_\text{HE}$ and $\eta_\text{AD}$ are the non-ideal MHD coefficients for Ohmic resistivity, the Hall effect and ambipolar diffusion, respectively.  Our previous studies, \citet{WurPriBat2016,WurPriBat2017}, did not include \eqref{eq:ni:u} since we assumed a barotropic equation of state.  We assume units for the magnetic field such that the Alfv{\'e}n speed is $v_{\rm A} = \vert B\vert/\sqrt{\rho}$ (see \citealp{PriMon2004b}).

We use the same flux-limited diffusion method to model radiation transport that we used in \cite{BatTriPri2014}.  Further details of the method can be found in that paper and in \cite{WhiBatMon2005} and \cite{WhiBat2006}. Briefly, we employ an ideal gas equation of state that assumes a 3:1 mix of ortho- and para-hydrogen (see \citealp{Boleyetal2007}) and treats the dissociation of molecular hydrogen and the ionisations of hydrogen and helium.  The mean molecular weight is taken to be $\mu_{\rm g} = 2.38$ at low temperatures, and we use opacity tables from \cite{PolMcKChr1985} and \cite{Alexander1975}.

We use Version 1.2.1 of the {\sc Nicil} library \citep{Wurster2016} to calculate the non-ideal MHD coefficients. The thermal ionisation processes can ionise hydrogen once, and ionise helium, sodium, magnesium and potassium twice; the mass fractions of the five elements are $0.747$, $0.252$, $2.96\times 10^{-5}$, $7.16\times 10^{-4}$ and $3.10\times 10^{-6}$, respectively  (e.g. \citealt{AsplundEtAl2009}; \citealt{KeiWar2014}). Cosmic rays have the ability to remove an electron to create an ion, which may be absorbed by a dust grain. We assume that two species of ions can be created: a heavy ion represented by magnesium \citep{AsplundEtAl2009} and a light ion representing hydrogen and helium compounds whose mass is calculated from the hydrogen and helium mass fractions.  We model a single grain species, $n_\text{g}$, with a radius and bulk density of $a_{\rm g} = 0.1 \mu$m and $\rho_{\rm b} = 3$ g~cm$^{-3}$ \citep{PollackEtAl1994}, respectively; the grain number density is calculated from the local gas density, assuming a dust-to-gas ratio of 0.01.  The grain species has three populations with charges $Z=-1,0,+1$, respectively, where $n_\text{g} = n_\text{g}^- + n_\text{g}^0 + n_\text{g}^+$ to conserve grain density.

\subsection{Smoothed particle radiation non-ideal magnetohydrodynamics}

Our numerical method is almost identical to that used by \cite{BatTriPri2014}, but includes non-ideal MHD effects.  We use {\sc sphNG}, a three-dimensional smoothed particle hydrodynamics (SPH) code that originated from \citet{Benz1990}, but has been substantially extended to include individual particle timesteps, variable smoothing lengths, radiation and magnetohydrodynamics as described below.

The density of each SPH particle is computed by summation over nearest neighbouring particles.  The smoothing length of each particle is variable in time and space, iteratively solving $h = 1.2 (m/\rho)^{1/3}$ where $m$ and $\rho$ are the SPH particle's mass and density, respectively \citep{PriMon2004b, PriMon2007}. Gravitational forces are calculated using a binary tree.  The gravitational potential is softened using the SPH kernel such that the softening length is equal to the smoothing length  \cite[see][for further details]{PriMon2007}.

We solve the MHD equations using a standard smoothed particle magnetohydrodynamics (SPMHD) scheme, evolving $\mbox{\boldmath $B$}/\rho$ as the magnetic field variable (Eq.~\ref{eq:ind}), using the \cite{BorOmaTru2001} source-term approach for stability.  We use the constrained hyperbolic divergence cleaning method of \cite{TriPriBat2016} to maintain the solenoidal constraint on the magnetic field.  This is an adaptation of a similar method developed for grid-based codes \citep{Dedneretal2002}.  
Artificial viscosity and resistivity terms are added to capture shocks and magnetic discontinuities, respectively \citep{PriMon2005,Price2012}.  The artificial viscosity and resistivity parameters are spatially varying and time dependent as described in \cite{Price2012}, using the \cite{MorMon1997} viscosity switch and the \citet{TriPri2013} resistivity switch whereby the resistivity parameter is set as $\alpha_{\rm B}=h \vert \nabla \mbox{\boldmath $B$} \vert / \vert \mbox{\boldmath $B$} \vert$. We use values of $\alpha_{\rm AV} \in [0.1,1]$ and $\alpha_{\rm B} \in [0,1]$.  

The resistive timestep for each non-ideal MHD term is 
\begin{equation}
\label{num:dtnimhd}
\text{d}t_\text{non-ideal} = C_\text{non-ideal}\frac{h^2}{\left|\eta\right|},
\end{equation}
where $C_\text{non-ideal}=1/2\pi$ is a constant equivalent to the Courant number.  Given the $h^2$ dependence, evolving on this explicit timestep is very slow when $\left| \eta \right|$ is large \citep[e.g.][]{MacNorKonWar1995}. Rather than use super-timestepping \citep{AleAmiGre1996} as in our previous studies \citep{WurPriBat2016,WurPriBat2017}, we implemented an implicit solver for the evolution of Ohmic resistivity since this term has the most restrictive timestep during the first hydrostatic core phase. Ambipolar diffusion and the Hall effect are evolved explicitly.  Our implicit solver is described in Appendix~\ref{app:implicit}, where we also compare the implicit and explicit solvers.  The models presented in Section~\ref{res:Halldirection} were calculated using implicit Ohmic resistivity, while the remainder simulations in this paper were calculated using explicit timestepping for all terms.  

The matter and radiation energy equations (\ref{eq:radiation},\ref{eq:matter}) are solved using the method of \cite{WhiBatMon2005} and \cite{WhiBat2006}, except that the standard explicit SPH contributions to the gas energy equation due to the work and artificial viscosity are used when solving the (semi-)implicit energy equations to provide better energy conservation. 

 We employ a second-order Runge-Kutta-Fehlberg integrator \citep{Fehlberg1969} with individual time steps for each particle \citep{BatBonPri1995}.

\section{Initial conditions}
\label{sec:ic} 
Our initial conditions are similar to those used in our previous studies (e.g. \citealt{PriBat2007}; \citealt{PriTriBat2012}; \citealt{BatTriPri2014}; \citealt{WurPriBat2016}).    We set up a dense, cold, spherical, uniform density, slowly rotating molecular cloud core of mass $M=1$~M$_{\odot}$ and radius $R=4\times~10^{16}$~cm = 0.013~pc. The initial density is $\rho_0=7.43\times~10^{-18}$~g~cm$^{-3}$, giving a gravitational free-fall time of $t_{\rm ff} = 2.4 \times 10^{4}$ yr. We use an initial (isothermal) sound speed $c_{\rm s} = \sqrt{p/\rho}= 2.19\times 10^{4}$~cm~s$^{-1}$, corresponding to a gas temperature of $T_{\rm g}=14$~K and $u=4.8 \times 10^8$ erg~g$^{-1}$. The spherical core is placed in pressure equilibrium inside a larger, cubic domain with a side length of $l=8 \times 10^{16}$ cm and a $30:1$ density ratio between the core and the warm ($T_{\rm g}=323$~K) ambient medium. The initial radiation energy density in both the dense core and the ambient medium is set such that it is in thermal equilibrium with the gas in the dense core.  Neither the gas or radiation temperatures of the particles modelling the ambient medium evolve --- their internal energies and radiation temperatures are fixed.  For simplicity we use periodic but non-self-gravitating boundary conditions on the global domain; the large density ratio ensures that the ambient medium does not contribute significantly to the self-gravity of the cloud.  The core is set in solid body rotation about the $z$-axis with $\Omega = 1.77 \times 10^{-13}$~rad s$^{-1}$,  corresponding to a ratio of rotational to gravitational energy $\beta_{\rm r} \simeq 0.005$ and $\Omega t_{\rm ff} = 0.14$. 

\begin{figure}
\centering
\includegraphics[width=\columnwidth]{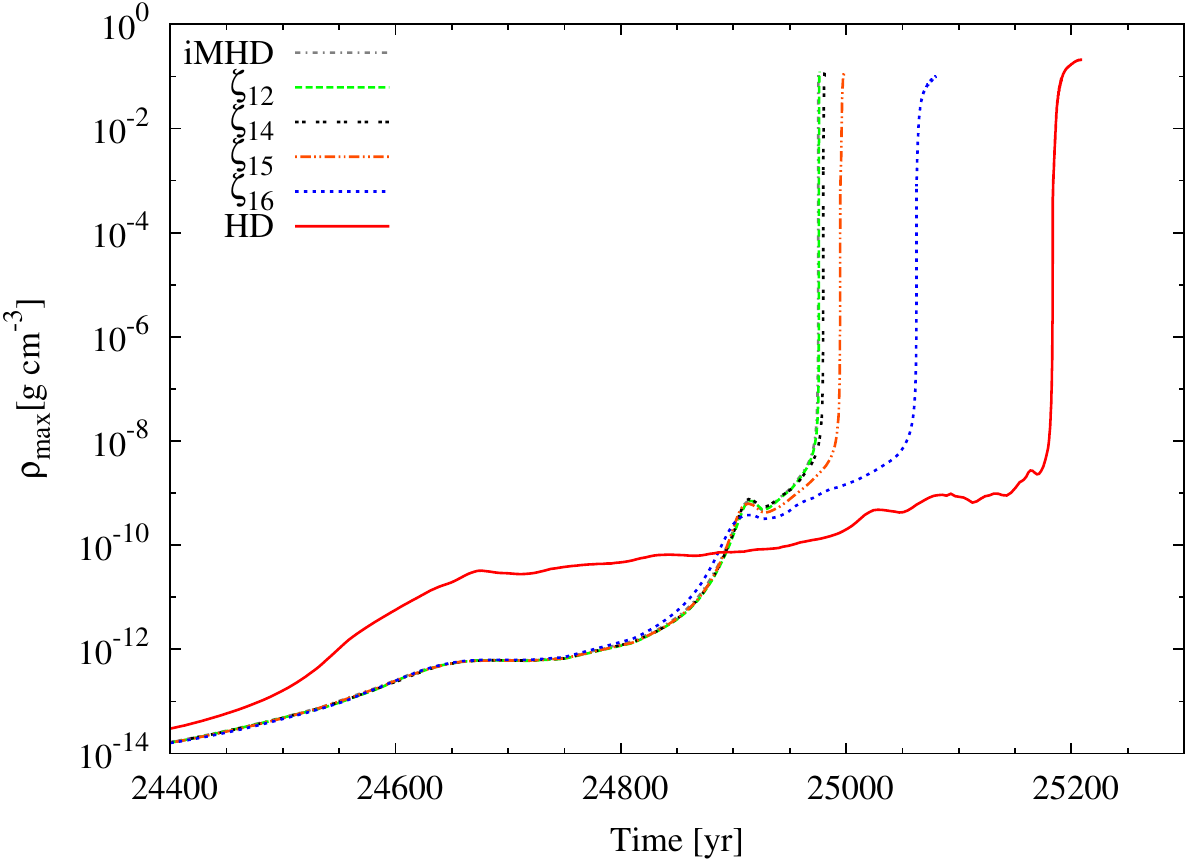}
\caption{Collapse to stellar densities: Maximum density as a function of time.  The hydrodynamical model, HD, is from \citet{BatTriPri2014}.  Decreasing the cosmic ray ionisation rate yields longer-lived first cores.  The non-ideal MHD model with \zetaeq{-12} (named \zetam{12}) is indistinguishable from ideal MHD model (named iMHD).}
\label{fig:rhoVStime}
\end{figure} 

\begin{figure}
\centering
\includegraphics[width=\columnwidth]{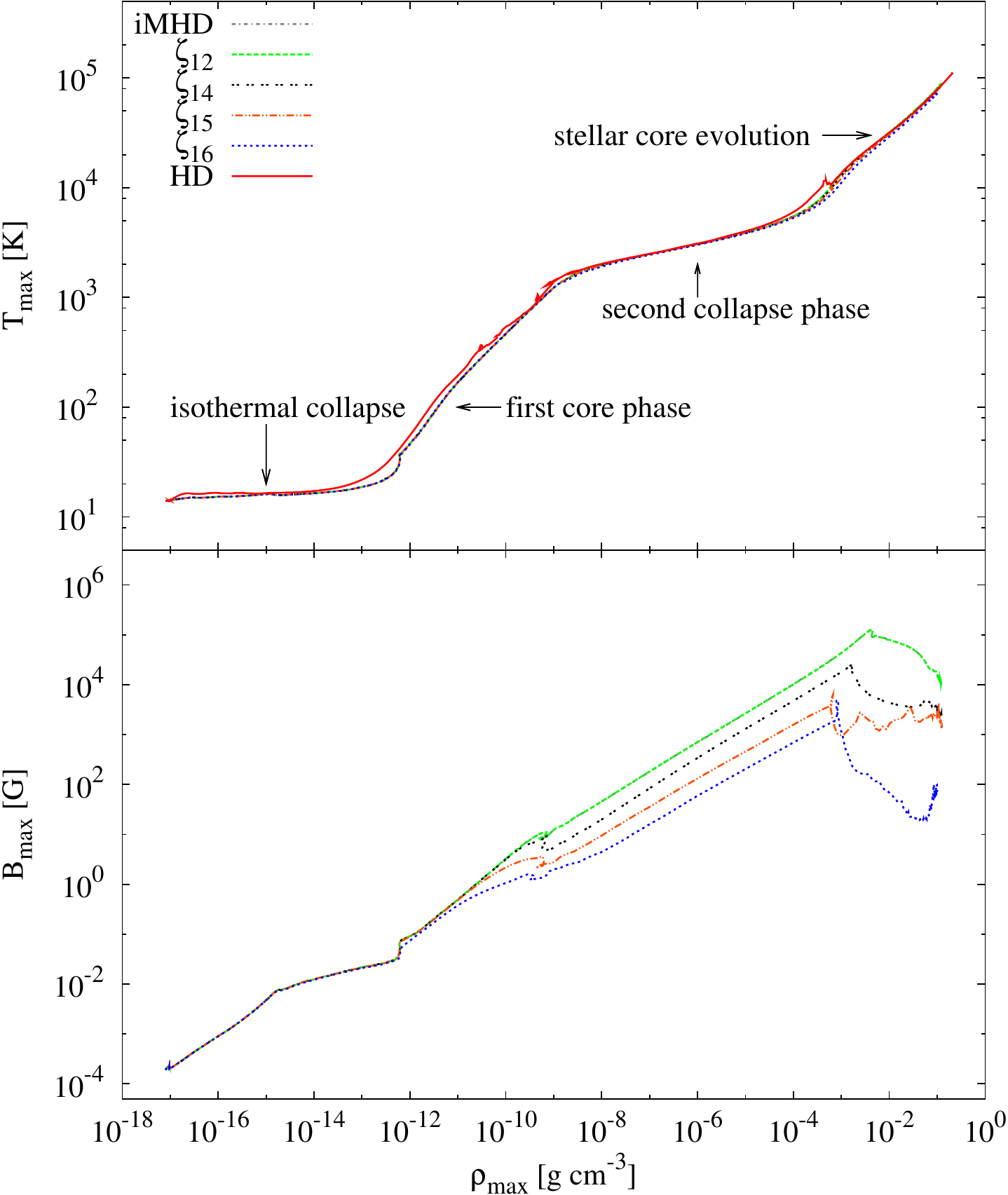}
\caption{Maximum temperature (top panel) and maximum magnetic field strength (bottom panel) as a function of maximum density (a proxy for time) in the collapsing molecular cloud cores.  The magnetised models have an initial magnetic field strength of $1.63 \times 10^{-4}$ G, which is 5 times the critical mass-to-flux ratio.  The different phases of evolution are labelled on the temperature plot.  Varying the ionisation rate affects the evolution of the maximum temperature by less than 20 per cent across all magnetised models (top panel). The maximum magnetic field strength differs between models only after the first core begins to form at \rhoapprox{-12} (bottom panel).  Throughout the evolution, the differences between iMHD and \zetam{12} are less than 10 per cent.  After the formation of the stellar core at $\rho_\text{max} \approx~10^{-3}$~g~cm$^{-3}$, the maximum magnetic field strength in the non-ideal MHD models is spatially offset from the density maximum by 2--20 R$_\odot$.}
\label{fig:VSrho}
\end{figure}

By default, we thread the entire domain with a uniform magnetic field that is anti-aligned with the axis of rotation of the spherical core, i.e. $B_\text{0,x} = B_\text{0,y} = 0$, $B_\text{0,z} = -B_0\hat{\bm{z}}$; this orientation promotes disc formation in the presence of the Hall effect assuming a low ionisation rate and given our initial direction of rotation (e.g. \citealt{BraWar2012}; \citealt{TsukamotoEtAl2015b}; \citealt{WurPriBat2016}).  We also perform one calculation with a low cosmic ray ionisation rate in which the field direction is aligned with the axis of rotation to investigate how the different manifestation of the Hall effect alters the results.  We choose an initial magnetic field strength of $B_0 = 1.63 \times 10^{-4}$ G, which corresponds to a normalised mass-to-magnetic flux ratio of $\mu_0 = 5$, expressed in units of the critical value for a uniform spherical cloud \citep{Mestel1999,MacKle2004},
\begin{equation}
\mu_0 \equiv \left(\frac{M}{\Phi_\text{B}}\right)_0 / \left(\frac{M}{\Phi_\text{B}}\right)_\text{crit},
\end{equation}
where
\begin{equation}
\left(\frac{M}{\Phi_\text{B}}\right)_0 \equiv \frac{M}{\pi R^{2} B_{0}}; \hspace{5mm} \left(\frac{M}{\Phi_\text{B}}\right)_\text{crit} = \frac{c_{1}}{3} \sqrt{\frac{5}{G}},
\label{eq:mphicrit}
\end{equation}
where $\Phi_\text{B}$ is the magnetic flux threading the surface of the (spherical) cloud, $c_{1}\simeq 0.53$ is a parameter determined numerically by \cite{MouSpi1976}, and  $\left(M/\Phi_\text{B}\right)_\text{crit}$ is written assuming cgs units.  

The non-ideal MHD algorithm uses the default values of the \textsc{Nicil} library \citep{Wurster2016}, except we that test several different cosmic ray ionisation rates, $\zeta_\text{cr}$.  Cosmic ray ionisation is the dominant ionisation source for $T \lesssim 1000$~K, above which thermal ionisation is the dominant source, independent of $\zeta_\text{cr}$.  Throughout this paper, unless explicitly stated, when we refer to ionisation rate, we are referring to the initial cosmic ray ionisation rate, $\zeta_\text{cr}$.  The \textsc{Nicil} library calculates the ionisation fractions and non-ideal MHD coefficients on the fly using a limited chemical network.  A more complex network would offer slightly different ionisation fractions with the same $\zeta_\text{cr}$, thus leading to different non-ideal MHD coefficients.  However, at low temperatures, previous tests of the \textsc{Nicil} library showed that modifying $\zeta_\text{cr}$ had a larger effect on the coefficients than modifying the chemical network.  Thus, although different chemical networks may yield different coefficients \citep[e.g.][]{TsukamotoEtAl2015a,MarchandEtAl2016}, the general trends we find in the study below should be independent of which network is ultimately chosen.

We use $3 \times 10^{6}$ equal-mass SPH particles in the core, and $1.46 \times 10^{6}$ particles in the external medium, both initialised on cubic lattices.  This resolution was found to be adequate to capture the evolution in the ideal MHD calculations \citep{BatTriPri2014}.  Resolving the Jeans length requires $\gtrsim 10^5$ particles per solar mass \citep{BatBur1997}, so the Jeans mass is well resolved at all times.


\begin{figure*}
\centering
\includegraphics[width=1.0\textwidth]{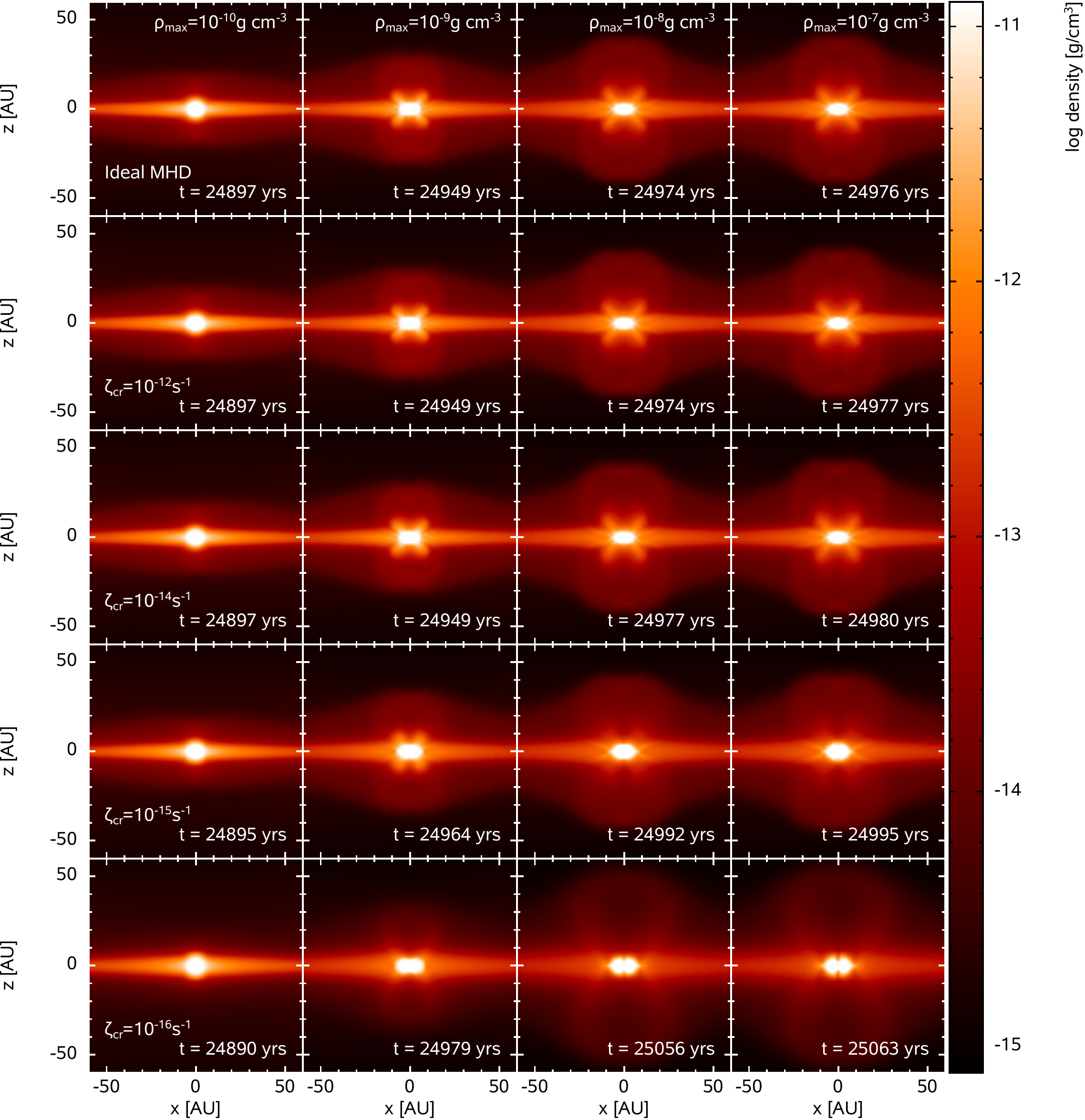}
\caption{The evolution of the first hydrostatic core and its outflow: Gas density cross sections taken through the centre of the collapsing core parallel to the rotation axis.  The rows show the models with decreasing ionisation rates (top to bottom), as functions of increasing maximum density (left to right).  The times at which each maximum density is reached differ for each model, with the corresponding times given in each frame.  Models iMHD and \zetam{12} are essentially identical, with \zetam{14} following a similar, but slightly delayed, evolution.  A broader outflow is launched in \zetam{15} and \zetam{16}.}
\label{fig:fhc:rho:map}
\end{figure*} 

\begin{figure}
\centering
\includegraphics[width=0.393\columnwidth]{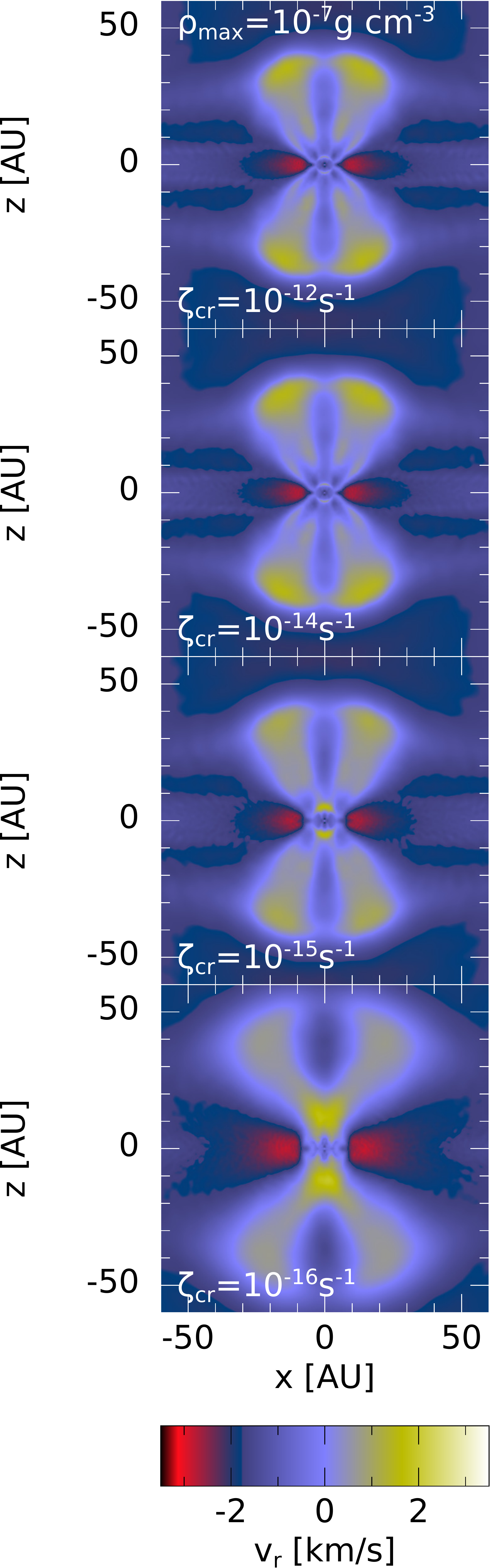}
\includegraphics[width=0.28\columnwidth,trim={6cm 0 0 0},clip]{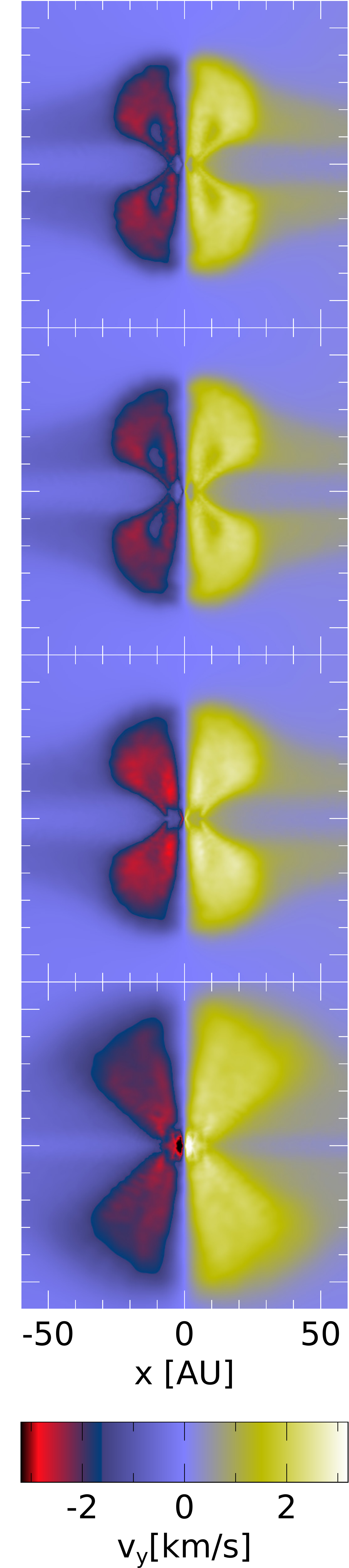}
\includegraphics[width=0.295\columnwidth,trim={6cm 0 0 0},clip]{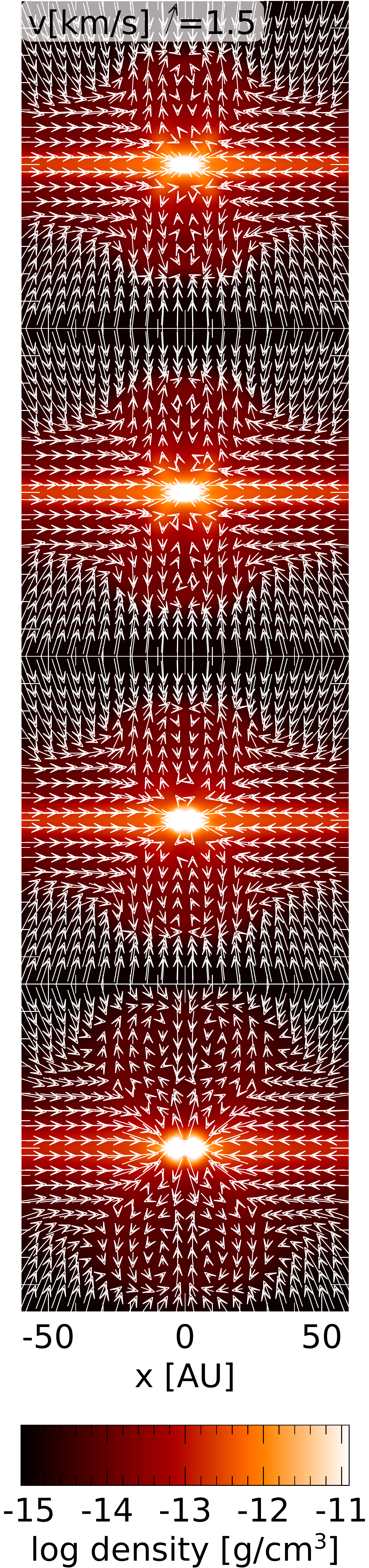}
\caption{Velocity structure of the first core outflows: Gas velocity cross sections taken through the centre of the collapsing core parallel to the rotation axis at the end of the first hydrostatic core phase at \rhoapprox{-7}.  Model iMHD is almost identical to \zetam{12} and is therefore not shown.  From left to right are the radial velocity $v_\text{r}$ (where $v_\text{r} < 0$ represents infall and $v_\text{r} > 0$ represent outflow), rotational velocity $v_\text{y}$ about the axis of rotation, and gas density over-plotted with velocity vectors.  In general, the rotational velocity is faster than the infall/outflow velocities.  The outflows have a conical structure, with the opening angles of the outflow and the infall down the rotation axis both increasing as the cosmic ray ionisation rate $\zeta_\text{cr}$ is reduced.  The large-scale outflow is slower for lower ionisation rates, however a faster small-scale outflow from the poles of the first core develops at the lowest ionisation rates.}
\label{fig:fhc:velocities}
\end{figure}  

\begin{figure}
\centering
\includegraphics[width=0.392\columnwidth]{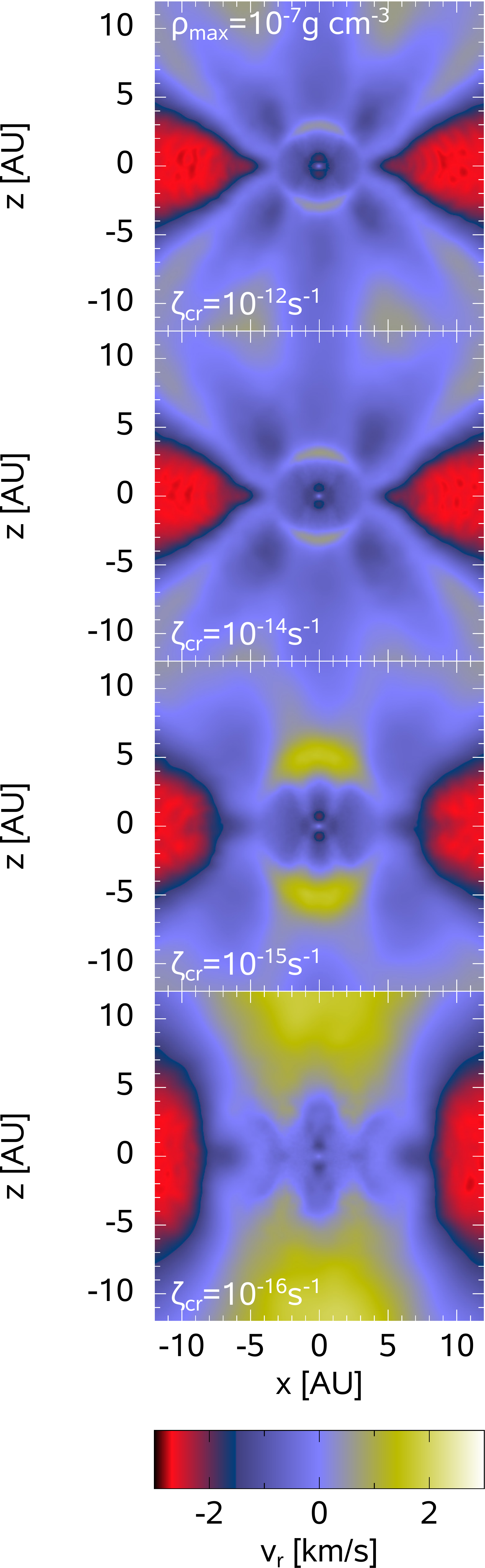}
\includegraphics[width=0.28\columnwidth,trim={6cm 0 0 0},clip]{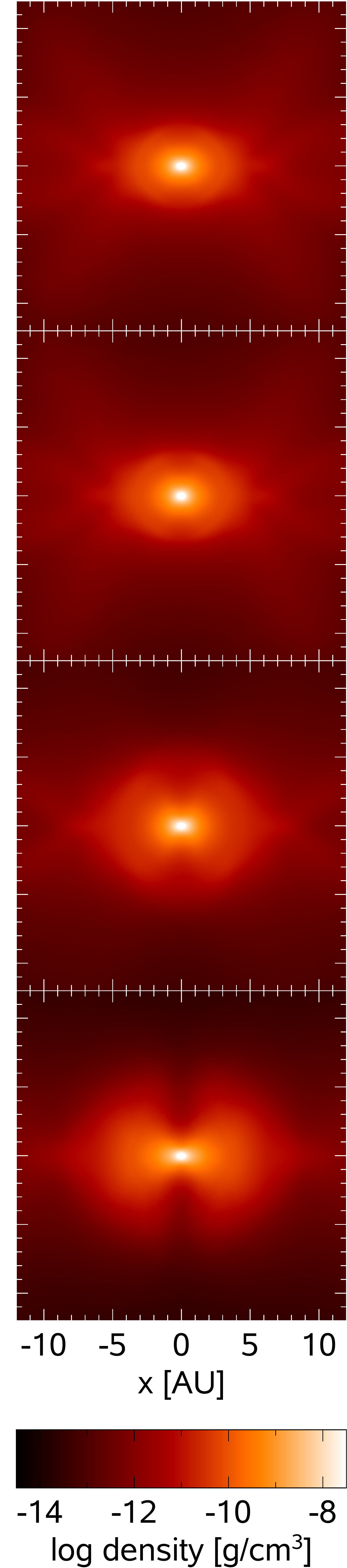}
\includegraphics[width=0.28\columnwidth,trim={6cm 0 0 0},clip]{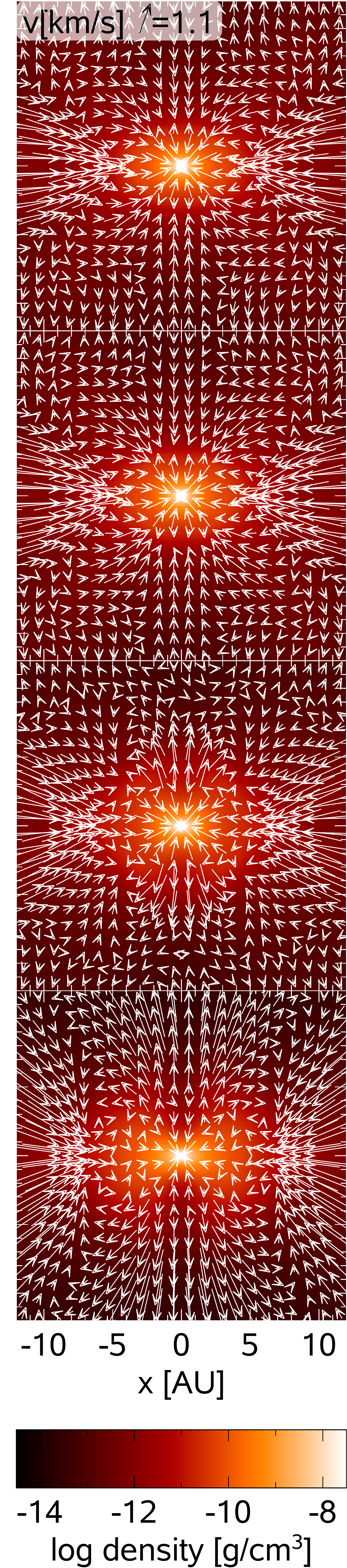}
\caption{The launching region at the base of the first core outflows:  From left to right is the radial velocity $v_\text{r}$, gas density, and velocity vectors plotted over gas density.  Each frame is zoomed in compared to Fig.~\ref{fig:fhc:velocities} to show the gas motion around the first hydrostatic core.  All MHD models display the large conical outflows, but only the models with low ionisation rates launch outflows from the poles of the first core ($r \approx 2.5$ au from the centre of the core).  These outflows are faster with lower ionisation rates ($\approx 1$~km~s$^{-1}$ for \zetam{15} and $\approx 2.3$~km~s$^{-1}$ for \zetam{16}).  Because of the outflows from the poles, gas is only accreted onto the core through the midplane in the low ionisation rate models.}
\label{fig:fhc:velocities:small}
\end{figure}  
\begin{figure*}
\centering
\includegraphics[width=0.24\textwidth]{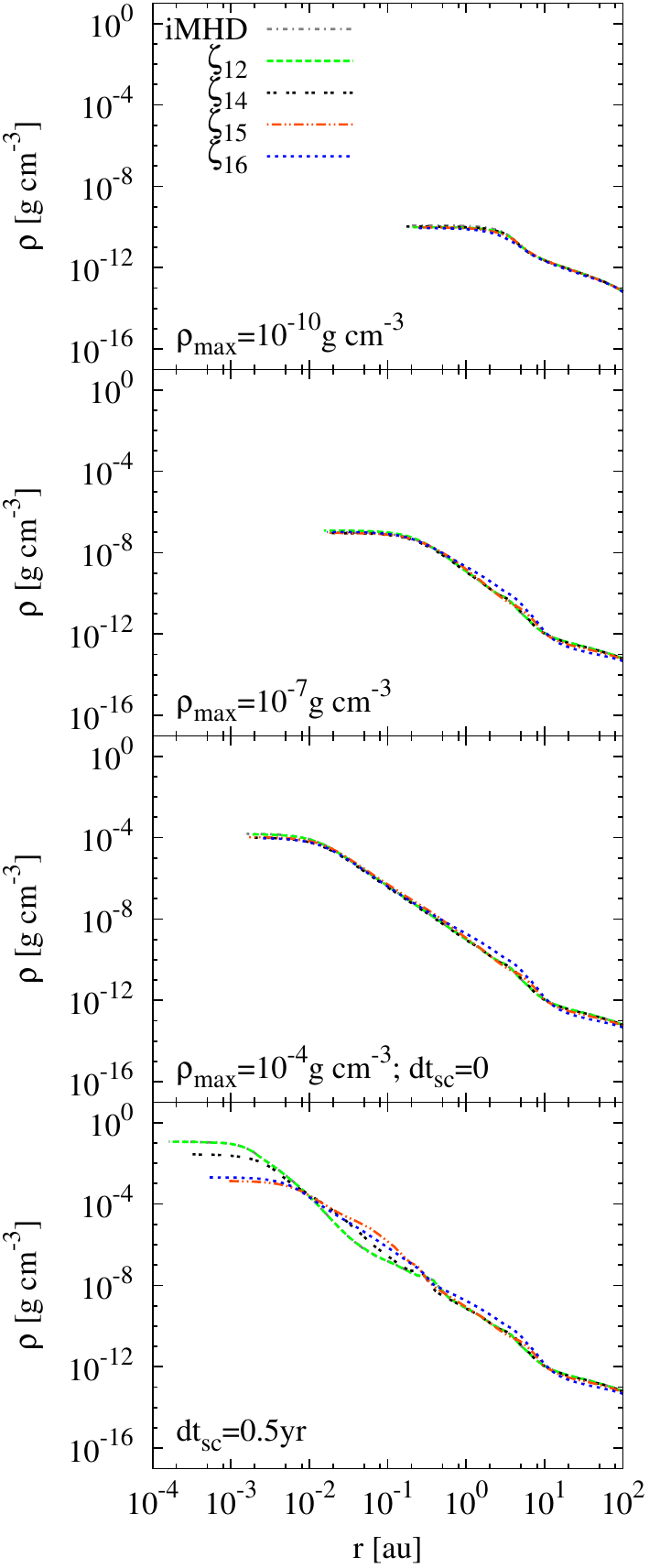}
\includegraphics[width=0.24\textwidth]{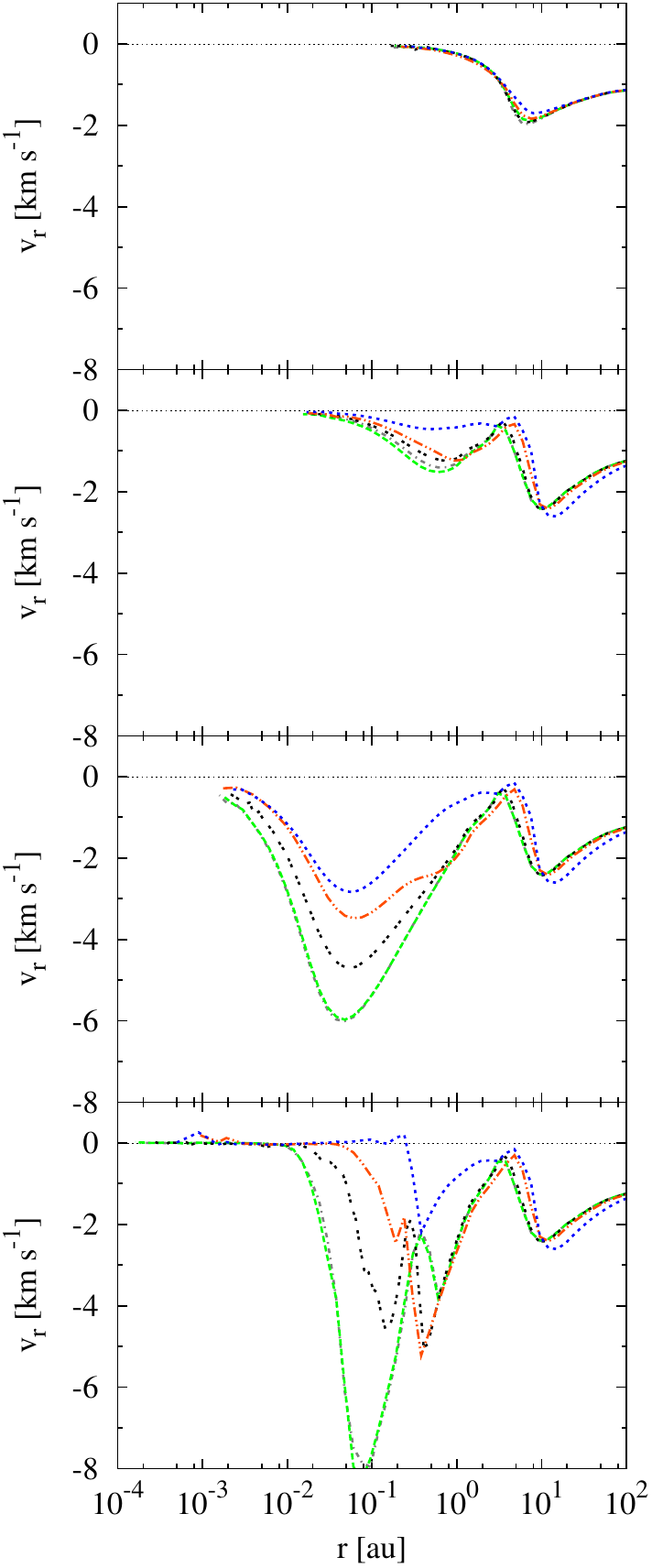}
\includegraphics[width=0.24\textwidth]{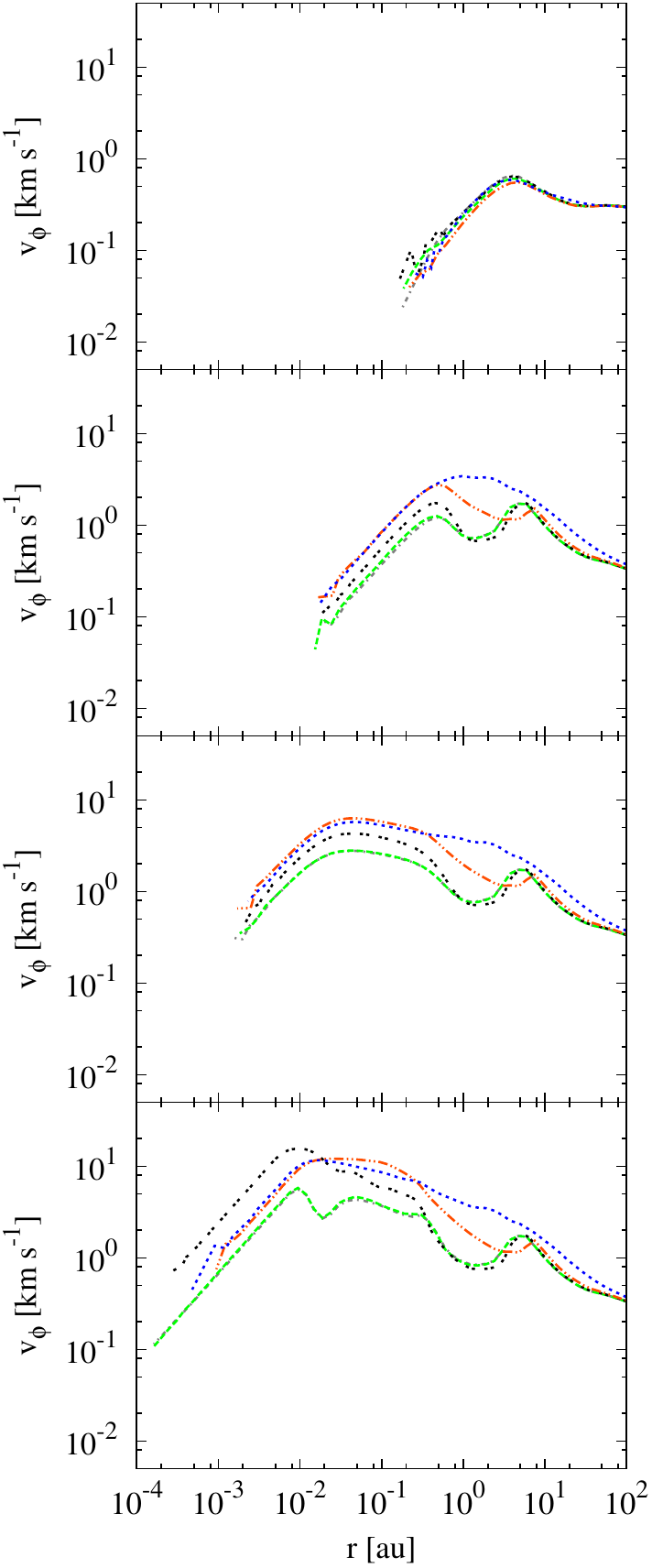}
\includegraphics[width=0.239\textwidth]{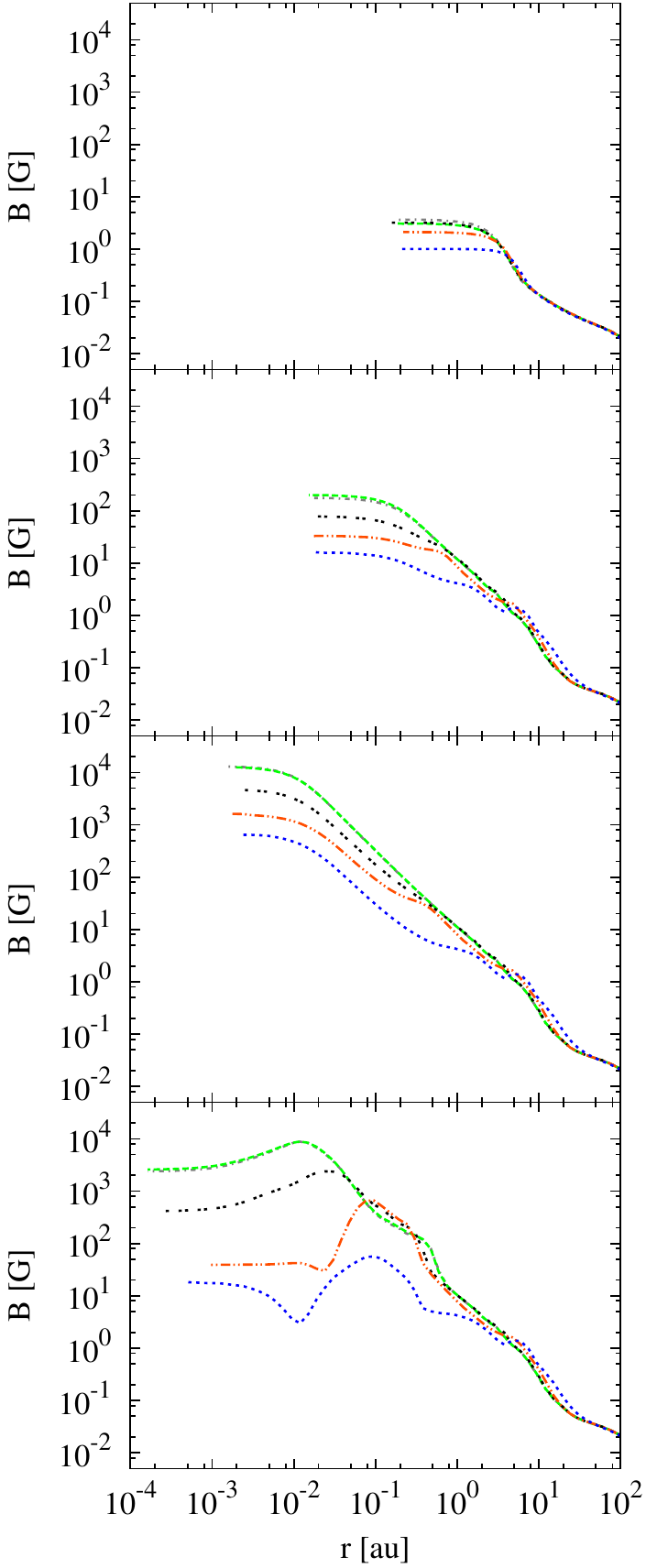}
\caption{Azimuthally averaged gas density, radial and azimuthal velocities, and magnetic field strength for the gas within 20$^\circ$ of the midplane.  From top to bottom, the plots are at $\rho_\text{max} \approx 10^{-10}$, $10^{-7}$, $10^{-4}$  g cm$^{-3}$ and \dtscapprox{0.5} after the formation of the stellar core.  The top two rows are during the first core phase and the second two rows are just before and just after stellar core formation.  The gas in the less ionised models generally has a weaker magnetic field strength.  As a consequence, reduced magnetic braking leads to higher rotation speeds and reduced infall speeds in the less ionised models.}
\label{fig:radialprofiles}
\end{figure*}  

\section{Results}
\label{sec:results}

Our primary suite of models includes an ideal MHD model (named iMHD), and four non-ideal MHD models with $\zeta_\text{cr} = \left\{10^{-12}, 10^{-14}, 10^{-15}, 10^{-16}\right\}$ s$^{-1}$, which we name \zetam{12}, \zetam{14}, \zetam{15} and \zetam{16}, respectively.  We perform an additional non-ideal MHD calculation with \zetaeq{-10}, which we do not discuss because the results are identical to iMHD.  We briefly compare our results to the hydrodynamical model (named HD) from \cite{BatTriPri2014}.

Our lowest cosmic ray ionisation rate is still higher than the typically accepted local rate of \zetaeq{-17} $\exp\left(-\Sigma/\Sigma_\text{cr}\right)$, where $\Sigma$ is the gas surface density and $\Sigma_\text{cr}$ is the characteristic attenuation depth for cosmic rays \citep{SpiTom1968,UmeNak1981}. In particular, many previous studies used a fixed rate of \zetaeq{-17} \citep[e.g.][]{LiKraSha2011,TsukamotoEtAl2015a,WurPriBat2016,WurPriBat2017,TsukamotoEtAl2017}. Our restriction to \zetageq{-16} is purely due to computational limitations --- as the ionisation rate is decreased, the increasing non-ideal MHD coefficients result in shorter timesteps that continue to decrease during the first collapse phase as the density increases (see Eqn.~\ref{num:dtnimhd}).  At our chosen spatial resolution, we have not yet been able to follow a \zetaeq{-17} model with anti-aligned magnetic field past the first core phase, even employing implicit resistivity.

Fig.~\ref{fig:rhoVStime} shows the evolution of the maximum density as a function of time for each calculation.  The magnetised models all reach the first hydrostatic core phase within \sm10 yr of one another, and the less ionised models remain in this phase longer.  This is consistent with \citet{BatTriPri2014} who found that cores with weaker magnetic fields collapsed more slowly. The hydrodynamic model, for reference, collapsed to the first core phase faster than the magnetised models, but remained in this phase longer.  Although increasing the ionisation rate increases the length of time the model exists in the first core phase, the first core lifetime remains shorter than in the absence of magnetic fields.

Fig.~\ref{fig:VSrho} shows the evolution of the maximum temperature and magnetic field strength as a function of maximum density (we use maximum density as a proxy for time because it is a better representation of the evolutionary state of the protostar).  The various phases of protostellar collapse are visible in the temperature plot (top panel of Fig.~\ref{fig:VSrho}): the almost isothermal collapse at $\rho_\text{max}/(\text{g cm}^{-3}) \lesssim 10^{-13}$, the first core phase from $10^{-12} \lesssim \rho_\text{max}/(\text{g cm}^{-3}) \lesssim 10^{-8}$, the second collapse phase from $10^{-8} \lesssim \rho_\text{max}/(\text{g cm}^{-3}) \lesssim 10^{-3}$, and the formation of the second (stellar) core at $\rho_\text{max}/(\text{g cm}^{-3}) \gtrsim 10^{-3}$ (e.g. \citealp{Larson1969,MasInu1999}).  The temperature evolution is only weakly dependent on the ionisation rate, with the maximum temperatures between iMHD and \zetam{16} differing by less than 20 per cent. The maximum temperature and density occur in the centre of the core at all times. 

The magnetic field strength begins to diverge between models once the first hydrostatic core forms at \rhoapprox{-12}.  The magnetic field strength grows throughout the first core phase, but it grows more rapidly with higher ionisation rates.  By the end of the first core phase, the maximum field strength is approximately an order of magnitude stronger in iMHD or \zetam{12} ($B_\text{max}\approx 50$~G) compared to \zetam{16} ($B_\text{max}\approx 5$~G). \cite{TsukamotoEtAl2015b} showed that this difference in the magnetic field growth during the first core phase compared to that seen in ideal MHD calculations is primarily due to Ohmic resistivity rather than ambipolar diffusion.  During the second collapse phase, the maximum magnetic field strength grows by 3 orders of magnitude as the field is dragged in by the collapsing gas ($B_\text{max} \propto \rho^{0.6}$).  The maximum field strength that is attained is a factor of \sm2000 greater in iMHD and \zetam{12} compared to \zetam{16}.  Once the stellar core has formed, the magnetic field decreases by 1--2 orders of magnitude within a few years. This decrease occurs in all models, with a faster decrease in the centre of the core than compared to the surrounding gas.  As a result, after the stellar core has formed, the maximum magnetic field strengths are spatially offset by a distance of between $0.01$ and $0.1$ au (2--20 R$_\odot$) from the centre of the stellar core.  This decrease of magnetic field within the stellar core is largely due to numerical resistivity \citep{BatTriPri2014}.

Both iMHD and HD remain in the second collapse phase for \appx3 yr.  The less ionised models spend slightly longer in the second collapse phase, with \zetam{16} remaining there for \appx7 yr.  
 
\subsection{The first hydrostatic core}
\label{sec:fhc}

First hydrostatic cores produced by unmagnetised or weakly magnetised (e.g. $\mu_0 = \infty, 100$) rotating molecular cloud cores are oblate, disc-like objects and with rapid enough rotation may undergo bar instability \citep[e.g.][]{Bate1998,Bate2011,BatTriPri2014}.  As the initial magnetic field strength is increased, the angular momentum transport provided by magnetic braking decreases radius of the disc \citep{BatTriPri2014}.  With significant magnetic fields ($\mu \lesssim 20$), bipolar outflows are magnetically launched above and below the first hydrostatic core.  These have speed of $\approx 1-2$~km~s$^{-1}$ and tend to be broader with weaker initial magnetic field strengths.

Fig.~\ref{fig:fhc:rho:map} shows the evolution of the gas density in cross sections parallel to the rotation axis during the first core phase. Rather than compare models at the same time, we use maximum density as a proxy for time, showing results at $\rho_\text{max} \approx 10^{-10}, 10^{-9},10^{-8} $ and $10^{-7}$ g cm$^{-3}$.  The latter maximum density is reached just as the second collapse phase begins.

Until the start of the first core phase, the evolution is approximately independent of the cosmic ray ionisation rate, $\zeta_\text{cr}$, with all models having similar characteristics at \rhoapprox{-10} (first column of Fig.~\ref{fig:fhc:rho:map}).   The only slight difference is that the shock above and below the pseudo-disc that surrounds the first hydrostatic core perpendicular to the initial magnetic field and rotation axis is weaker in the models with lower ionisation (see the density cross sections in the first column of Fig.~\ref{fig:fhc:rho:map}, and the velocity vectors in the third column of Fig.~\ref{fig:fhc:velocities}).  

The evolution of iMHD and \zetam{12} throughout the first core phase are essentially identical.  By the end of the first collapse phase, \zetam{14} has similar outflows at similar maximum densities, although it takes $\approx4$ yr longer to collapse to the given density.  The vertical evolution is different for \zetam{15} and \zetam{16}, which fail to form the strong conical density enhancements that are associated with the opening angle of the gas outflow (this conical density enhancement appears as an `X'-shaped pattern in the cross sections of Fig.~\ref{fig:fhc:rho:map}).

\begin{figure*}
\centering
\includegraphics[width=0.255\textwidth]{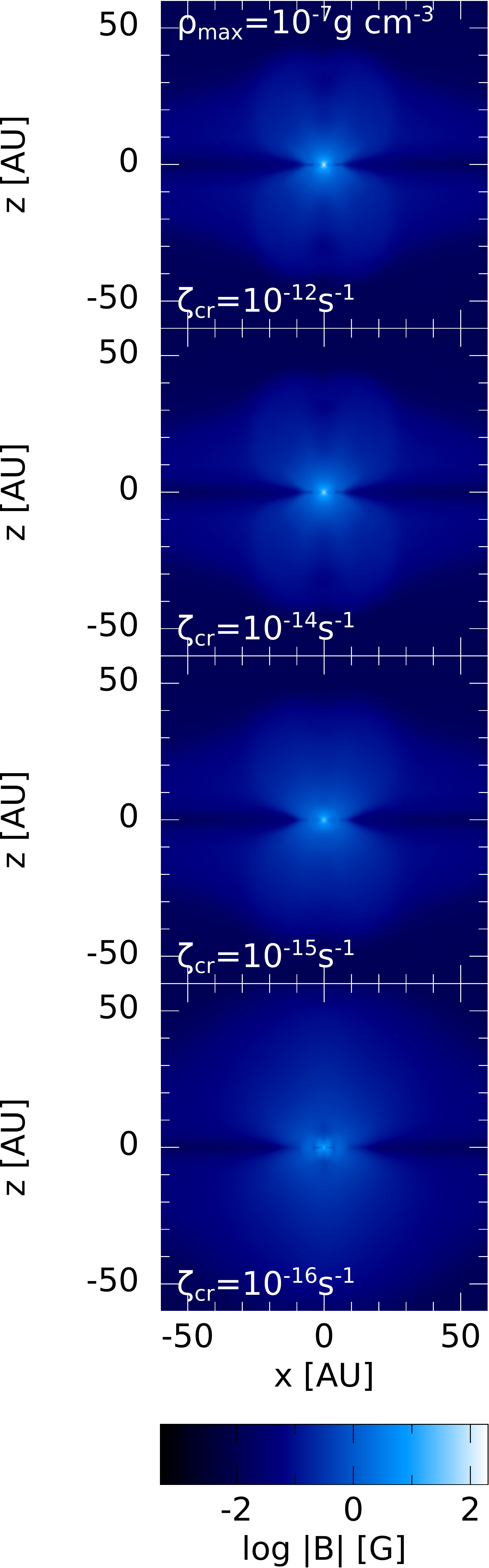}
\includegraphics[width=0.17\textwidth,trim={7cm 0 0 0},clip]{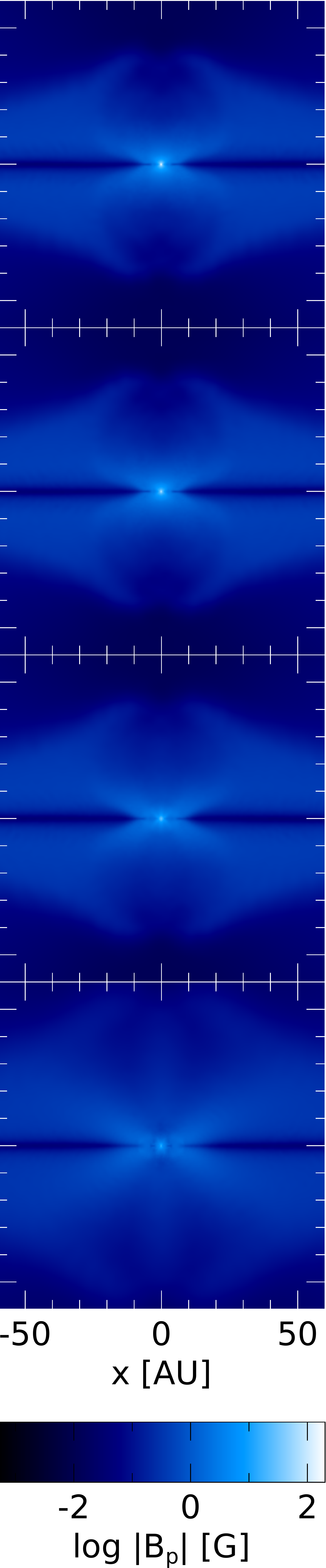}
\includegraphics[width=0.17\textwidth,trim={7cm 0 0 0},clip]{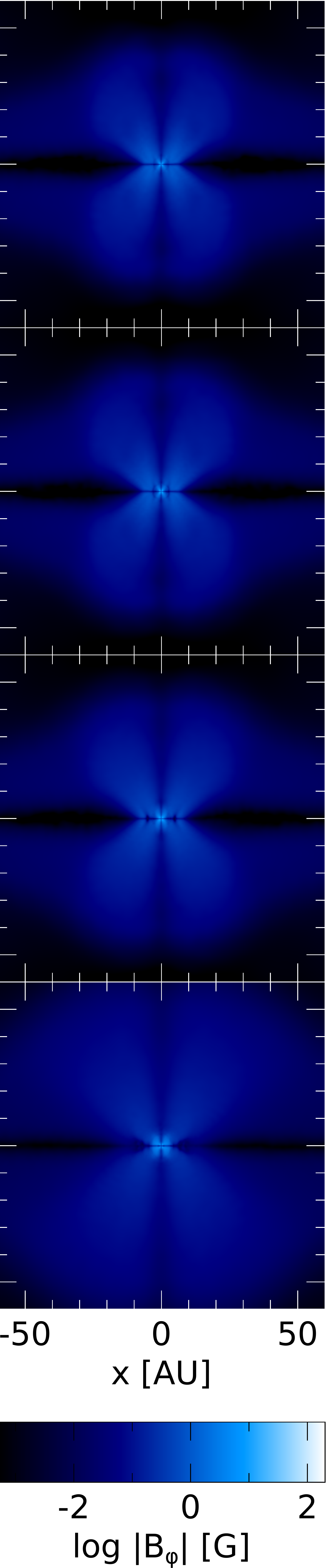}
\includegraphics[width=0.17\textwidth,trim={7cm 0 0 0},clip]{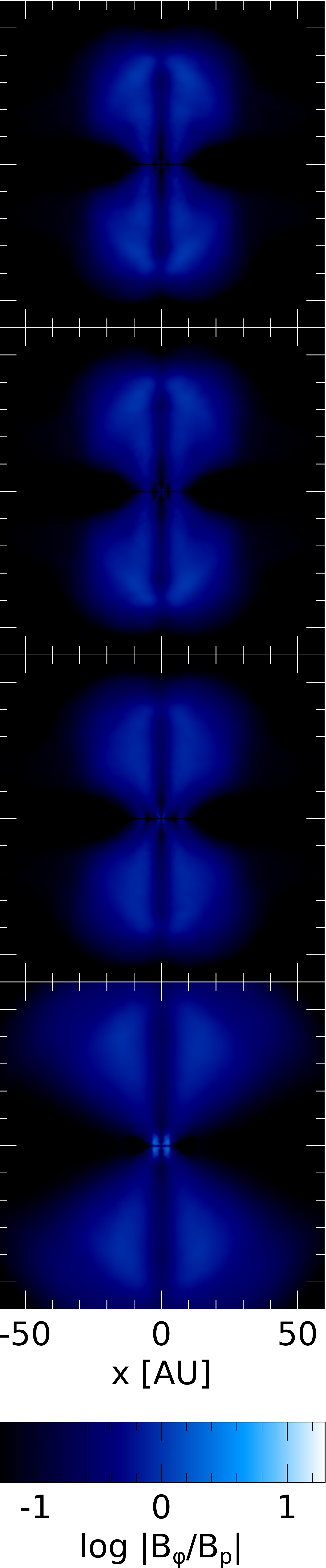}
\includegraphics[width=0.17\textwidth,trim={7cm 0 0 0},clip]{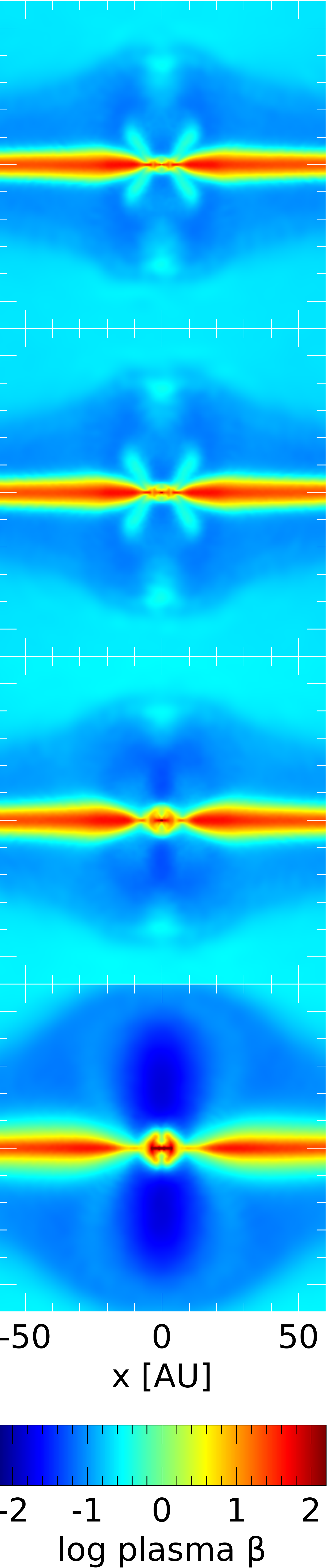}
\caption{Magnetic structure of the first core outflows:  From left to right are the cross sections of the total magnetic field strength, magnitude of the poloidal field $|B_\text{p}| = \sqrt{B_\text{r}^2 + B_\text{z}^2}$, magnitude of the toroidal/azimuthal field $|B_\phi |$, the ratio $|B_\phi/B_\text{p}|$, and plasma $\beta$ in the outflows from the first core for the partially ionised models.  The images are taken at \rhoapprox{-7}. The magnetic field is weaker in the first core and its immediate surroundings with lower ionisation rates, with the decrease mostly occurring in $|B_\text{p}|$.  Except for the range $0.7 \lesssim r/\text{au} \lesssim 4$  in \zetam{16}, $|B_\phi/B_\text{p}| \lesssim 1$.  Magnetic towers exists in all models, which become more dominated by magnetic pressure as \zetacr \ is decreased. }
\label{fig:fhc:B:crosssection}
\end{figure*} 

\begin{figure}
\centering
\includegraphics[width=0.494\columnwidth]{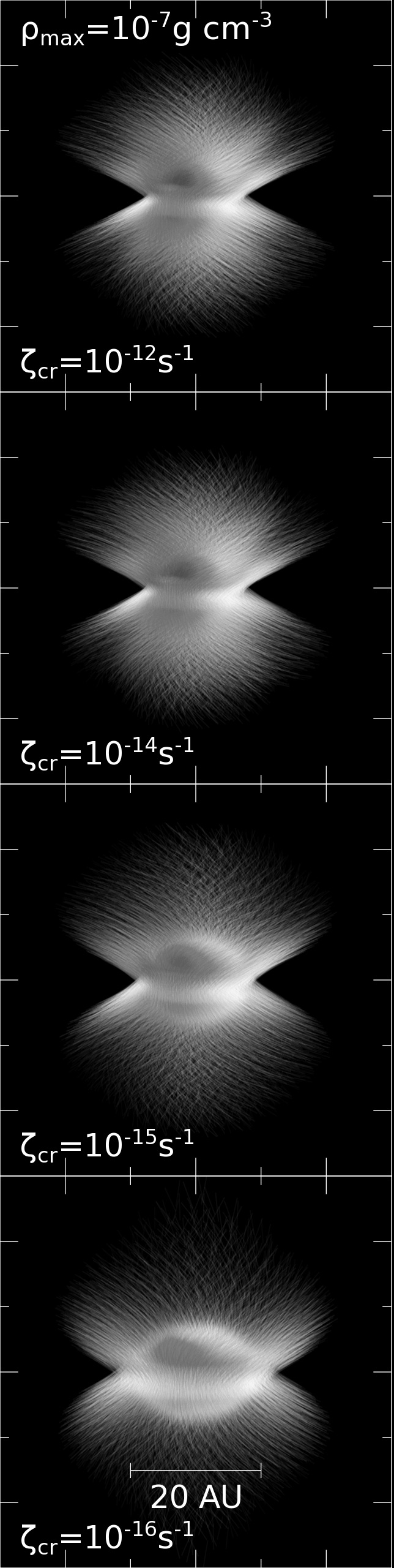}
\includegraphics[width=0.492\columnwidth]{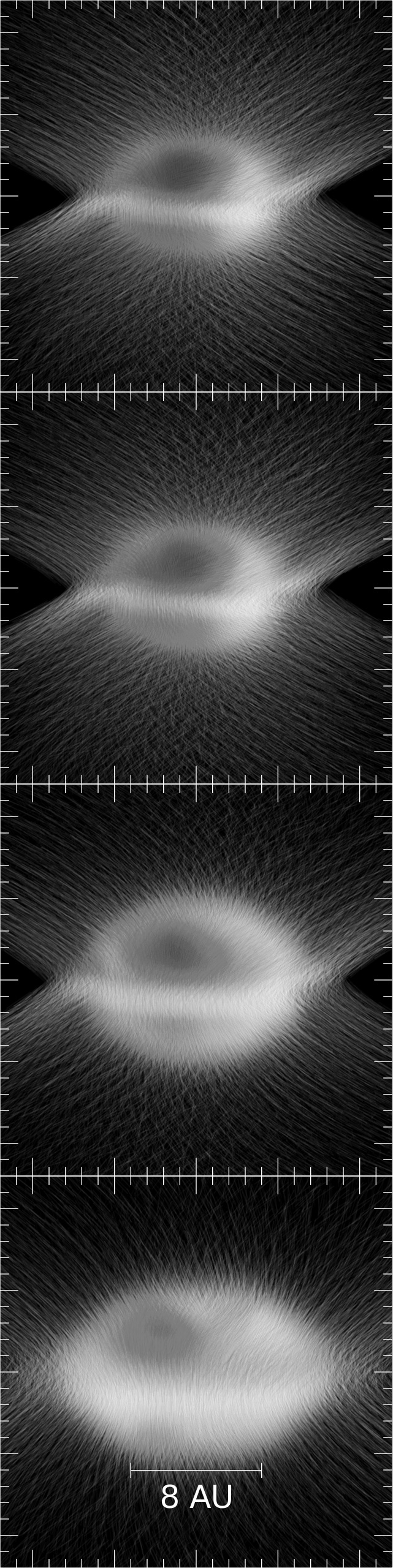}
\caption{Visualisations of the magnetic field geometry in the first core outflows, for $0.2 < |B|/\text{G} < 200$.  The images are inclined by 10$^\circ$ out of the page, and the panels in the left- (right-) hand column have a horizontal dimension of 60 au (24 au).  As the ionisation rate is decreased, the magnetic field becomes less `pinched' and the enhancement extends further above and below the midplane.}
\label{fig:fhc:B:ironfilings}
\end{figure} 

\subsubsection{Gas velocities}

Fig.~\ref{fig:fhc:velocities} shows the velocities in cross sections through the centres of the cores at \rhoapprox{-7} (at the end of the first collapse phase) for the ionised models.  Models iMHD and \zetam{12} have the same outflows, so for clarity we excluded the iMHD results from the figure.  The left-hand column renders the radial velocity $v_\text{r}$, where $v_\text{r} < 0$ represents infall, the middle column gives the rotational velocity about the rotation axis, $v_\text{y}$, and the right-hand column plots velocity vectors over density maps.

All of the magnetised models have `large-scale' bipolar outflows extending to $r \gtrsim 40$ au by the end of the first core phase.  The outflow velocities are slower in the lower ionisation models, reaching $v_\text{r} \approx 0.9$ km s$^{-1}$ in \zetam{16} compared to $v_\text{r} \approx 1.7$ km s$^{-1}$ in \zetam{12}.  The outflow has progressed further in \zetam{16}, but this is a result of the additional $\approx 86$ yr of evolution in the first core phase compared to \zetam{12}.

As the outflows form and expand, the ambient gas continues to collapse.  Once the gas enters the pseudo-disc surrounding the first core, it spirals onto the core since it is rotating at sub-Keplerian speeds.

Fig.~\ref{fig:fhc:velocities} also shows that some gas continues to fall inwards towards the first hydrostatic core near to the axis of rotation.  This gives the outflows their conical geometry and produces the `X' shape in the density cross sections in Fig.~\ref{fig:fhc:rho:map} with high ionisation rates; with the lowest ionisation rates, this morphology is much weaker.  The spread of the inflowing gas along the axis of rotation is larger at lower ionisation rates, and outflows in these models also have slightly larger opening angles.  In the ideal MHD models of \cite{BatTriPri2014}, weaker initial magnetic fields and the associated reduced magnetic braking produced more rapidly-rotating first cores with larger radii and broader outflows.  A similar weaker effect is at work here --- reduced ionisation results in less magnetic braking, more rapid rotation and slightly broader outflows.

There is, however, a significant difference at the base of the first core outflows between the high ionisation and low ionisation models.  Fig.~\ref{fig:fhc:velocities:small} shows a zoom in of the radial velocity, gas density and velocity vectors in a cross section through the centre of the core at \rhoapprox{-7} for the ionised models.  In \zetam{12} and \zetam{14} (and with iMHD) the gas infalling along the rotation axis reaches the poles of the first hydrostatic core.  However, in the lower ionisation rate models \zetam{15} and \zetam{16}, the outflow begins above and below the first core (including at the poles) and the gas infalling along the axis of rotation collides with the outflow and its collapse is arrested.  This difference in the morphology of the outflow on small scales can also be clearly seen in the radial velocity plots in Fig.~\ref{fig:fhc:velocities} (first column).  Unlike the outflows on large scales, the small-scale regions of the outflows are faster with reduced ionisation rates, reaching up to $v_\text{r} \approx 2.3$ km s$^{-1}$ in \zetam{16}.

The outflows around the core are rotating in the same sense as the initial rotation of the cloud, and the rotational speeds are similar to (but slightly faster than) the outflow speeds.  In general, the rotation speed of the outflows increases as the ionisation rate decreases.  

When the Hall effect is included and the initial magnetic field and axis of rotation are anti-aligned, counter-rotating envelopes have formed in previous numerical studies \citep[e.g.][]{KraLiSha011,LiKraSha2011,TsukamotoEtAl2015a,WurPriBat2016,TsukamotoEtAl2017} at $r \sim 100$ au.  These studies used the lower ionisation rate of \zetaapprox{-17}, and the counter-rotating envelope formed to conserve angular momentum as a result of the Hall effect spinning up the disc.  Our models do not form counter rotating envelopes.  Given that our minimum ionisation rate is \zetaeq{-16}, it is likely that the Hall effect is simply not strong enough in our models to require the counter-rotating envelope to conserve angular momentum.

The azimuthally averaged radial and azimuthal velocities of the gas within 20$^\circ$ of the midplane are shown in the middle two columns of Fig.~\ref{fig:radialprofiles}, with the top two rows showing the profiles near the beginning and end of the first collapse phase.  The velocity profiles of the midplane are similar for all models at \rhoapprox{-10}.  At \rhoapprox{-7}, the gas has a similar radial velocity along the midplane for all the models, but slightly decreasing for decreasing ionisation rates.  The lower ionisation models also have greater rotational velocities due to reduced magnetic braking because of ambipolar diffusion and Ohmic resistivity and probably also the action of the Hall effect which acts to promote rotation when the initial magnetic field is anti-aligned with the rotation axis.

\begin{figure*}
\centering
\includegraphics[width=0.3\textwidth]{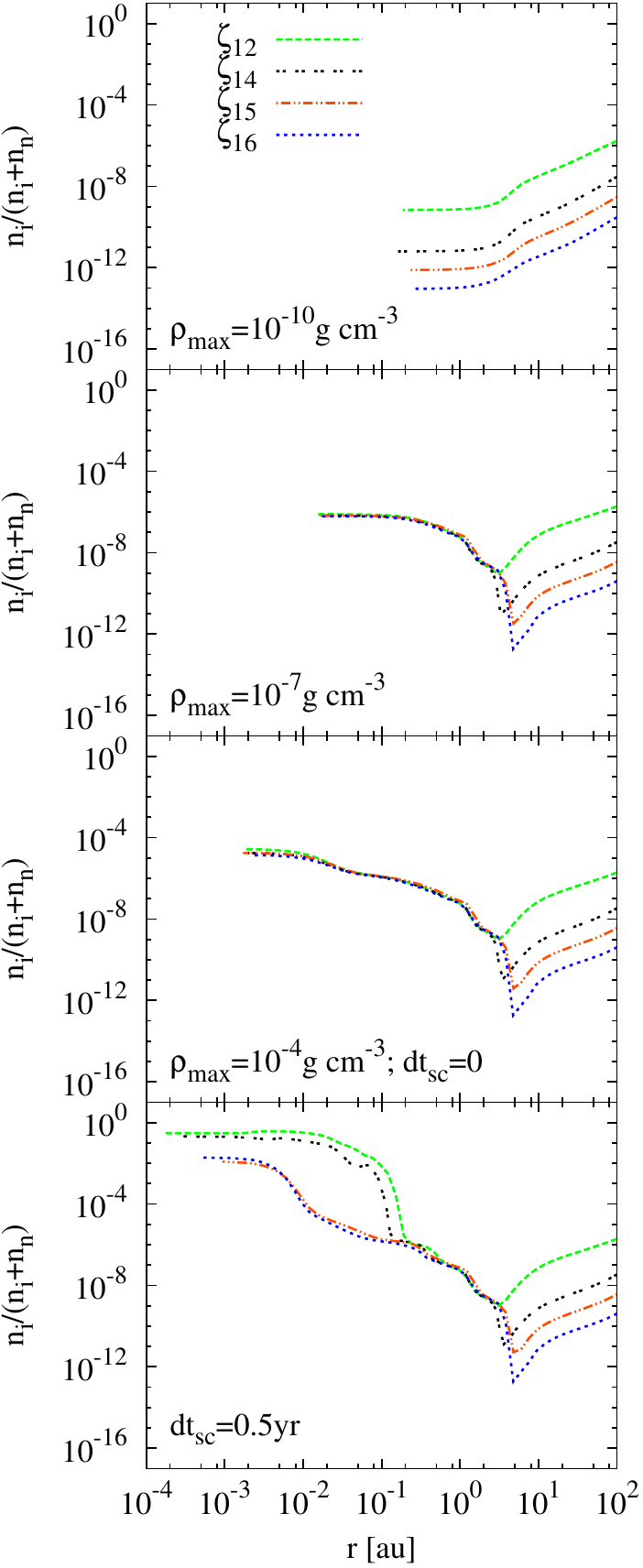}
\includegraphics[width=0.3\textwidth]{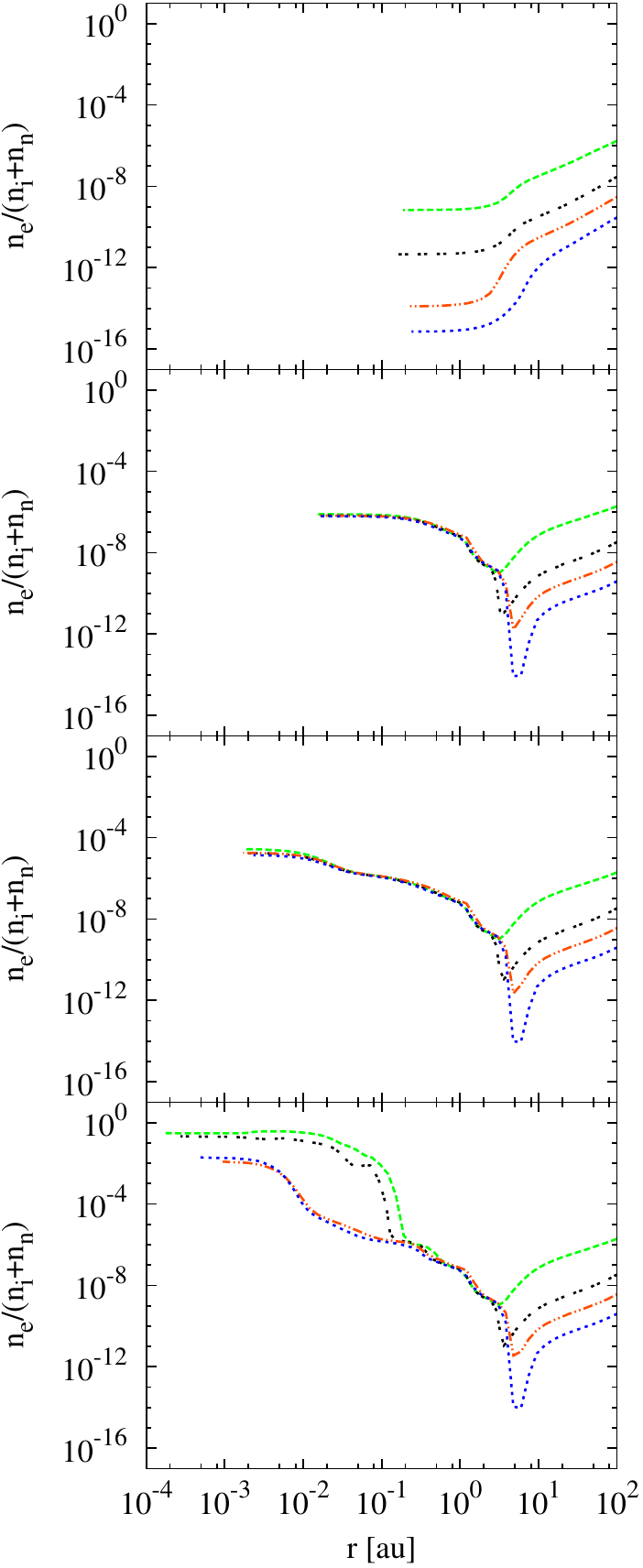}
\includegraphics[width=0.3\textwidth]{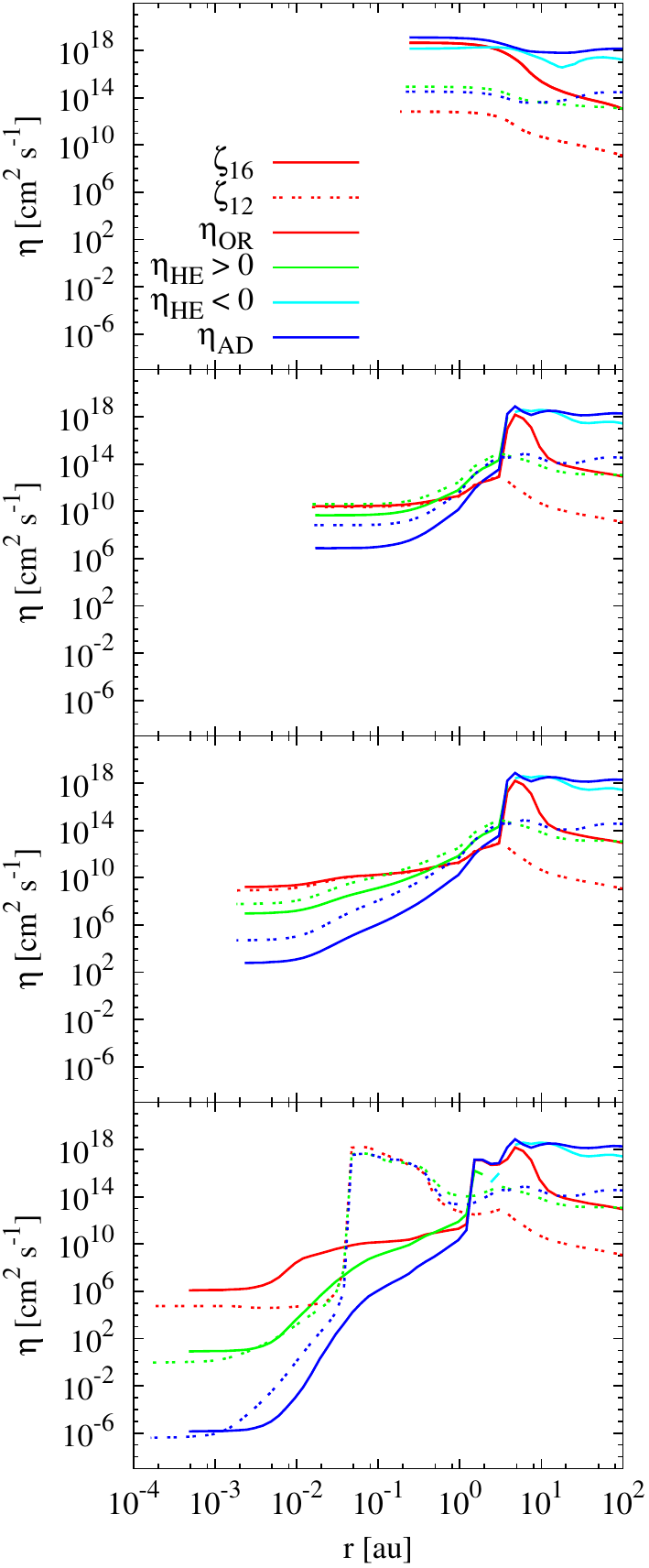}
\caption{Azimuthally-averaged ion and electron fractions for our suite of partially ionised models (first two columns), and the non-ideal MHD coefficients for \zetam{16} (solid lines) and \zetam{12} (dashed lines; third column) for the gas within 20$^\circ$ of the midplane; the rows are as in Fig.~\ref{fig:radialprofiles}.  The ion and electron number density are not necessarily equal since grains can absorb electrons, and ions can be doubly ionised at high temperatures.  The number densities converge when the temperature is $T \gtrsim 1000$~K where thermal ionisation becomes the dominant ionisation process.  At $T \lesssim 1000$ K, reducing the ionisation rate decreases the coefficients, but the decrease is not linear with \zetacr.}
\label{fig:ionisation}
\end{figure*}  

In summary, decreasing the initial ionisation rate reduces the magnetic braking.  This results in more rapid rotation of the first core, and during the second collapse the rotational velocity is higher and the radial velocity is lower.  The outflow from the first core is also broader.

\subsubsection{Magnetic fields}
\label{res:fhc:magnetic}

The right-hand column of Fig.~\ref{fig:radialprofiles} shows the azimuthally-averaged magnetic field strength of the gas within 20$^\circ$ of the midplane.  The non-ideal MHD effects diffuse the magnetic field out of the centre of the cloud so that, at any particular point during the collapse, the maximum field strength decreases with decreasing ionisation rate.

Fig.~\ref{fig:fhc:B:crosssection} shows the magnetic field strengths in cross sections through the centres of the cores at \rhoapprox{-7} for the ionised models; from left to right is the total magnetic field strength, the magnitude of the poloidal field $|B_\text{p}| = \sqrt{B_\text{r}^2 + B_\text{z}^2}$, the magnitude of the toroidal/azimuthal field $|B_\phi |$, the ratio $|B_\phi/B_\text{p}|$, and plasma $\beta$.  Recall that the initial conditions are $\bm{B}_\phi  = 0$ and $\bm{B}_\text{p} = -(1.63 \times 10^{-4}$ G$)\hat{\bm{z}}$.  Fig.~\ref{fig:fhc:B:ironfilings} shows visualisations of the magnetic field geometry at two different frame sizes.

Given the initial magnetic field geometry, the majority of the magnetic field strength is from the poloidal component, which decreases in strength with decreasing ionisation rates.  The initially rotating cloud winds the magnetic field to convert the poloidal component into the toroidal component, as seen in Fig.~\ref{fig:fhc:B:ironfilings}.

As the ionisation rate is decreased, the central magnetic fields becomes less `pinched,' and the region of enhanced magnetic field strength extends further above and below the midplane.  In the large-scale, `X'-shaped outflows in all four models, the poloidal and toroidal components are similar in strength, but the poloidal component is generally stronger (i.e. $|B_\phi/B_\text{p}| \lesssim 1$).  This is also true for the inner, small scale ($r \sim 1$~au) outflows in \zetam{12}, \zetam{14} and \zetam{15}.  Despite the weaker toroidal field in \zetam{16} compared to the higher ionisation rate models, the toroidal component is slightly stronger than the poloidal component in the small scale outflows at $0.7 \lesssim r/\text{au} \lesssim 4$.  \citet{BatTriPri2014} found that by decreasing the initial magnetic field strength, the outflows were more likely to exhibit $|B_\phi/B_\text{p}| > 1$, have the magnetic field enhancement extend further above and below the midplane, and yield slower and broader outflows.  Thus, we find a similar, albeit weaker, effect by decreasing the ionisation rate.

Throughout the first core phase, plasma $\beta > 1$ in the midplane, thus the gas is always supported by gas pressure rather than magnetic pressure.  The disc is more dependent on gas pressure than magnetic pressure for the lower ionisation rate models, which is expected since both ambipolar diffusion and Ohmic resistivity act to reduce the strength of the magnetic field.

By \rhoapprox{-7}, a magnetic tower \citep{Lyndenbell2003,KatMinShi2004} has formed in which plasma $\beta < 1$; this corresponds to the region of low-velocity infall and low rotational velocities (see Fig.~\ref{fig:fhc:velocities}).  For all our models, the magnetic tower contains a toroidal magnetic field, which is still weaker than the poloidal component.  For decreasing ionisation rates, the tower becomes broader and more magnetically dominated, and is no longer confined by the magnetically dominated conical (`X'-shaped) winds seen in \zetam{12} and \zetam{14}.  This suggests that these outflows are powered by magnetic pressure.  Similar large-scale outflows have been previously seen in simulations \citep[e.g.][]{Tomisaka1998,AllLiShu2003,BanPud2006,BatTriPri2014}.

The launching region of the outflows (see Fig.~\ref{fig:fhc:velocities:small}) is at the interface where plasma $\beta \sim 1$, with plasma $\beta > 1$ closer to the core.  In \zetam{15} and \zetam{16}, the interface is sharper and the toroidal magnetic field is piling up near the core (where plasma $\beta > 1$).  This prevents further infall from distant, magnetically supported gas (plasma $\beta < 1$), and results in the launching of the polar outflows from scales of a few au.

In summary, by the end of the first core phase, the magnetic field remains mostly poloidal, and is less pinched for models with lower ionisation rates.  All models have formed magnetic tower outflows, launched from the surface of the first hydrostatic core.  The outflows have lower plasma $\beta$ (i.e. they are more magnetically dominated) for lower ionisation rates.

\subsubsection{Non-ideal MHD effects}
\label{res:fhc:magnetic:ni}

The first two columns of Fig.~\ref{fig:ionisation} show the azimuthally-averaged ion and electron fractions for the gas within 20$^\circ$ of the midplane; the fractions are $f_s\equiv n_s/(n_\text{i} + n_\text{n})$ for $s \in\{\text{i},\text{e}\}$, thus $n_s/(n_\text{i} + n_\text{n}) \approx 1$ is the totally ionised case representing ideal MHD.  The two ratios, $f_\text{i}$ and $f_\text{e}$, are not necessarily equal since, at cooler temperatures, electrons can be absorbed by grains, while at higher temperatures, elements may be doubly ionised.  The third column in Fig.~\ref{fig:ionisation} shows the azimuthally-averaged non-ideal MHD coefficients for \zetam{16} (solid lines) and \zetam{12} (dashed lines).

Decreasing the ionisation rate decreases the number density of ions and electrons, making the gas more neutral; this in turn increases the effect of the non-ideal MHD coefficients.  The effect is non-linear with \zetacr, and at \rhoapprox{-10} the magnetic field strength in the first hydrostatic core is \appx3.5 times stronger in \zetam{12} than in \zetam{16}.  At this maximum density, the magnetic field strengths are similar for all models at $r \gtrsim 7$ au, where the non-ideal MHD effects are weak enough to only trivially affected the magnetic field.

At \rhoapprox{-10}, $\eta_\text{HE} < 0$ for \zetam{16} but  $\eta_\text{HE} > 0$ for \zetam{12} throughout the midplane.  Thus, in \zetam{16}, the Hall effect is decreasing the toroidal component of the magnetic field, $|B_\phi |$, in the inner $r \lesssim 7$~au where the effect is strong, which reduces the magnetic braking.  Although the Hall effect should increase $|B_\phi |$ and enhance magnetic braking in \zetam{12}, the effect is too weak to make any significant deviation from iMHD.  

At \rhoapprox{-7}, thermal ionisation is the dominant ionisation process in the core ($r \lesssim 7$~au and $T \gtrsim 1000$~K), thus the ion and electron number densities have converged for all models.  By this density, the magnetic field strength in the core is \sm10 times higher for \zetam{12} compared to \zetam{16}.  The non-ideal MHD coefficients are dependent on the magnetic field strengths such that $\eta_\text{OR} \propto B^0$, $\eta_\text{HE} \propto B^1$ and $\eta_\text{AD} \propto B^2$.  Since the higher ionisation rate models have stronger magnetic fields in the core at this density, they also have larger coefficients of $\eta_\text{HE}$ and $\eta_\text{AD}$, while both models have similar values of $\eta_\text{OR}$.  Thus, the models with higher ionisation rates are now more strongly affected by the non-ideal MHD effects in the core than the lower ionisation rate models.

At this density, all models have $\eta_\text{HE} > 0$ in the inner $r \lesssim 7$ au, since $\eta_\text{HE}$ is being calculated based upon the high ionisation fraction from thermal ionisation.  For $r \gtrsim 7$~au, where cosmic ray ionisation remains the dominant ionisation source, the sign of $\eta_\text{HE}$ remains unchanged from the previous snapshot for all models.  Despite the Hall effect contributing to magnetic braking in the core, its contribution is too weak to have any significant effect on the evolution of the magnetic field.  In the surrounding gas, however, the non-ideal effects remain important and the contribution is similar to the previous snapshot. Thus, in \zetam{16} at \rhoapprox{-7}, there exists a sharp transition region between negative and positive $\eta_\text{HE}$, where the Hall effect transitions from increasing to decreasing the toroidal magnetic field strength.  The size and sharpness of the transitions regions varies with both time and \zetacr.

In summary, the evolution through the first core phase is strongly dependent on the external ionisation rate.  Partial ionisation leads to less magnetic braking, slower outflows from the first core, and a different morphology of the outflow, particularly at the base of the outflow in the immediate vicinity of the first hydrostatic core.  The ionisation fraction in the core is dependent on thermal ionisation while the fraction in the surrounding medium is dependent on the cosmic ray ionisation rate. In the core, $\eta_\text{HE} > 0$, while its sign in the surrounding medium is dependent on the cosmic ray ionisation rate, as is the location and sharpness of the turn-over if the two signs are different.  Although the non-ideal MHD coefficients may be higher in the core for the higher ionisation rate models, the coefficients are too small to significantly contributed to the evolution of the magnetic field within the first core.

\subsubsection{Magnetic braking}
\label{res:fhc:magbrake}

Magnetic braking occurs when angular momentum is transported away from the central region by magnetic torques caused by the winding and pinching of the magnetic field lines \citep[e.g.][]{BasMou1994}. The amount of braking depends on both the magnetic field strength and its coupling to the charged particles.  Thus, in ideal MHD where there is perfect coupling between the gas and magnetic field, strong braking is expected, whereas less braking should occur in non-ideal MHD once the drift of the charged and neutral particles is taken into account.  The reduction in angular momentum caused by magnetic braking can prevent a rotationally supported disc from forming, and can cause central objects to rotate slower.

By the end of the first core phase, the azimuthal velocity in the core decreases for increasing cosmic ray ionisation rates (see third column of Fig.~\ref{fig:radialprofiles}); the core in \zetam{16} is spinning \appx2.3 times faster than in iMHD.  At this density, the ionisation fractions in the core are similar for all initial cosmic ray ionisation rates, thus this difference in spin is a result of the initial collapse and the lower angular momentum of the accreting gas.  The rotational velocities in the outer regions are approximately independent of the cosmic ray ionisation rate, indicating that the effect of magnetic braking increases closer to the central object where the magnetic field strength and the rotation rate both increase.  To conserve angular momentum, the reduced azimuthal velocities in the high ionisation rate models require faster outflows;  the large-scale outflows are \appx2.2 times faster for iMHD than for \zetam{16}.

\subsection{The stellar core}
\label{sec:shc}
 
When molecular hydrogen begins to dissociate at $T \approx 2000$~K and \rhoapprox{-8}, the second phase of the collapse begins.  This collapse continues until the stellar core is formed at  $T \approx 5000$~K and \rhoapprox{-3} \citep{Larson1969}.  As shown in Fig.~\ref{fig:rhoVStime}, the onset of the stellar core phase is delayed in the less ionised models due to a longer first core phase, but all models are in the second collapse phase for only 3-7 yr, with the less ionised models collapsing more slowly.  

By definition, all of our cores have the same density and temperature at the start and end of the second collapse phase; as a result, they all have the same ionisation fractions in the core where thermal ionisation is the dominant process.  However, the remainder of the characteristics are dependent on the cosmic ray ionisation rate.  The lower ionisation rate models have weaker magnetic field and slower infall velocities but higher rotational velocities; see third column of Fig.~\ref{fig:radialprofiles}.

\begin{figure}
\centering
\includegraphics[width=0.8\columnwidth]{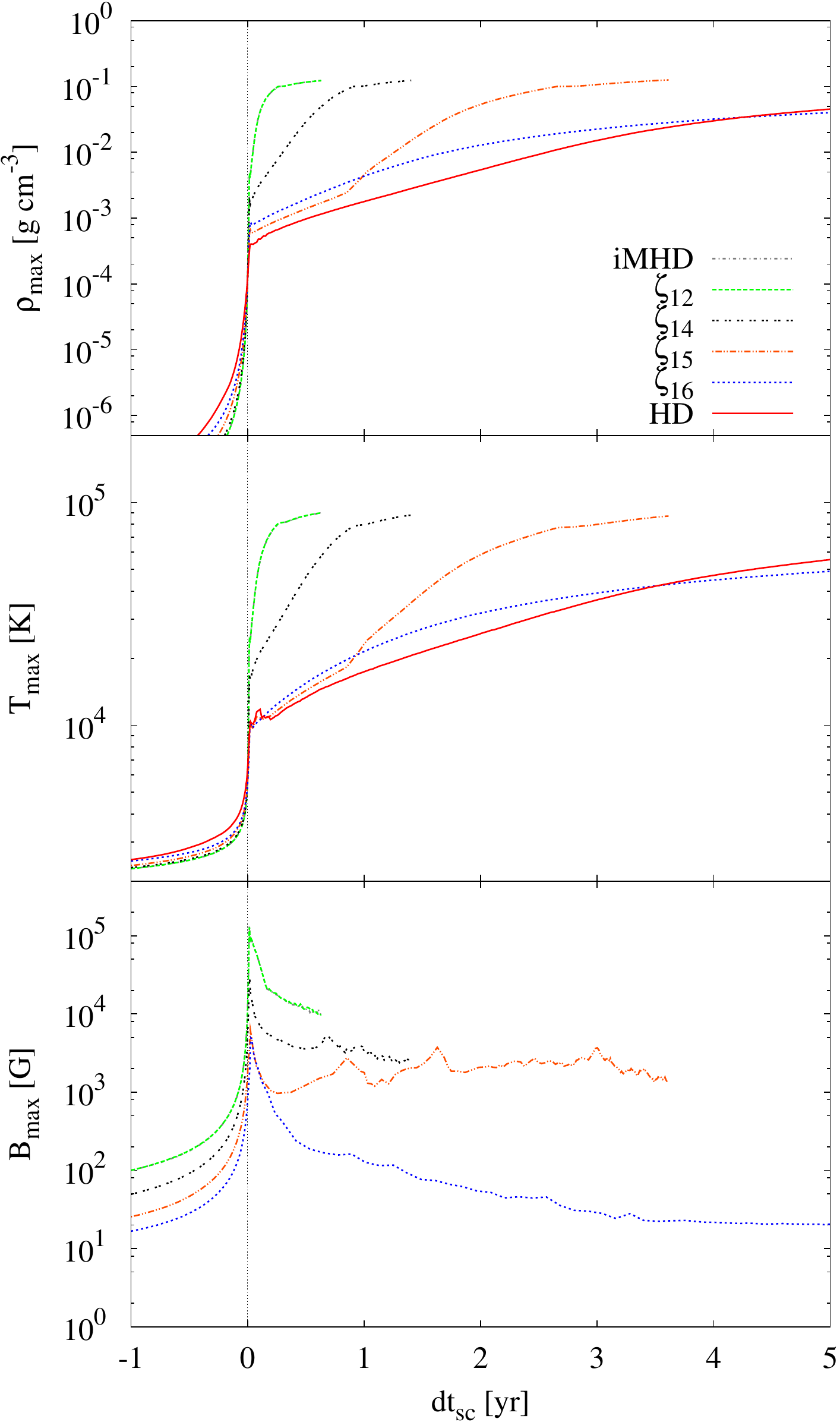}
\caption{The formation and evolution of the stellar core: The time evolution of the maximum density (top), maximum gas temperature (middle), and maximum magnetic field strength (bottom) during the formation of the stellar core (we take the time of formation to be when \rhoeq{-4}).  The growth rate in the maximum density and temperature decreases with decreasing initial cosmic ray ionisation rate.  The magnetic field strength decays more rapidly after stellar core formation with decreasing initial cosmic ray ionisation rate.  The maximum gas density and temperature are always in the centre of the core, whereas the maximum magnetic field strength becomes spatially offset from the density maximum by 2--20 R$_\odot$ after stellar core formation. }
\label{fig:shc:Vtime}
\end{figure}

The evolution of all characteristics of the stellar core and its surroundings begin to diverge after its formation.  Fig.~\ref{fig:shc:Vtime} shows the evolution of the maximum density, gas temperature and magnetic field strength as the stellar core forms and begins growing in mass.  We define the time of stellar core formation as the time at which the maximum density reaches \rhoeq{-4}; although the actual collapse stops at densities ranging from $\rho_\text{max} \approx 4\times 10^{-4}$ to $3\times 10^{-3}$ g~cm$^{-3}$ for the different models, the collapse from \rhoeq{-4} to stellar core formation takes much less than a month, so it is convenient to use the time when \rhoeq{-4} for all models. We denote the time since stellar core formation as d$t_\text{sc}$.

The core in iMHD continues to rapidly accrete, and by d$t_\text{sc} \approx 3.2$ mo has reached $T_\text{max} \approx 80000$~K and \rhoapprox{-1}.  Due to the small timesteps required to evolve such high densities and temperatures, we ended this simulation at d$t_\text{sc} \approx 8$ mo.  For decreasing ionisation rates, the growth rate is slower, with \zetam{16} reaching \rhoapprox{-1} at d$t_\text{sc} \approx 17$ yr; at this density, the temperature is \appx9 per cent cooler than in iMHD.  The slower growth rates with lower ionisation rate are a direct consequence of the higher rotation speeds and lower infall rates that are seen in Fig.~\ref{fig:radialprofiles}.  Given the different mass accretion rates onto the core and that this gas has different characteristics in each model, even if each model ultimately reaches a similar maximum density, the stellar core properties will likely never be identical. This contrasts with the results of \citep{BatTriPri2014} who found no significant variation in the thermodynamic properties of the stellar core from hydrodynamical and ideal MHD calculations with different initial field strengths.

\begin{figure*} 
\centering
\includegraphics[width=1.0\textwidth]{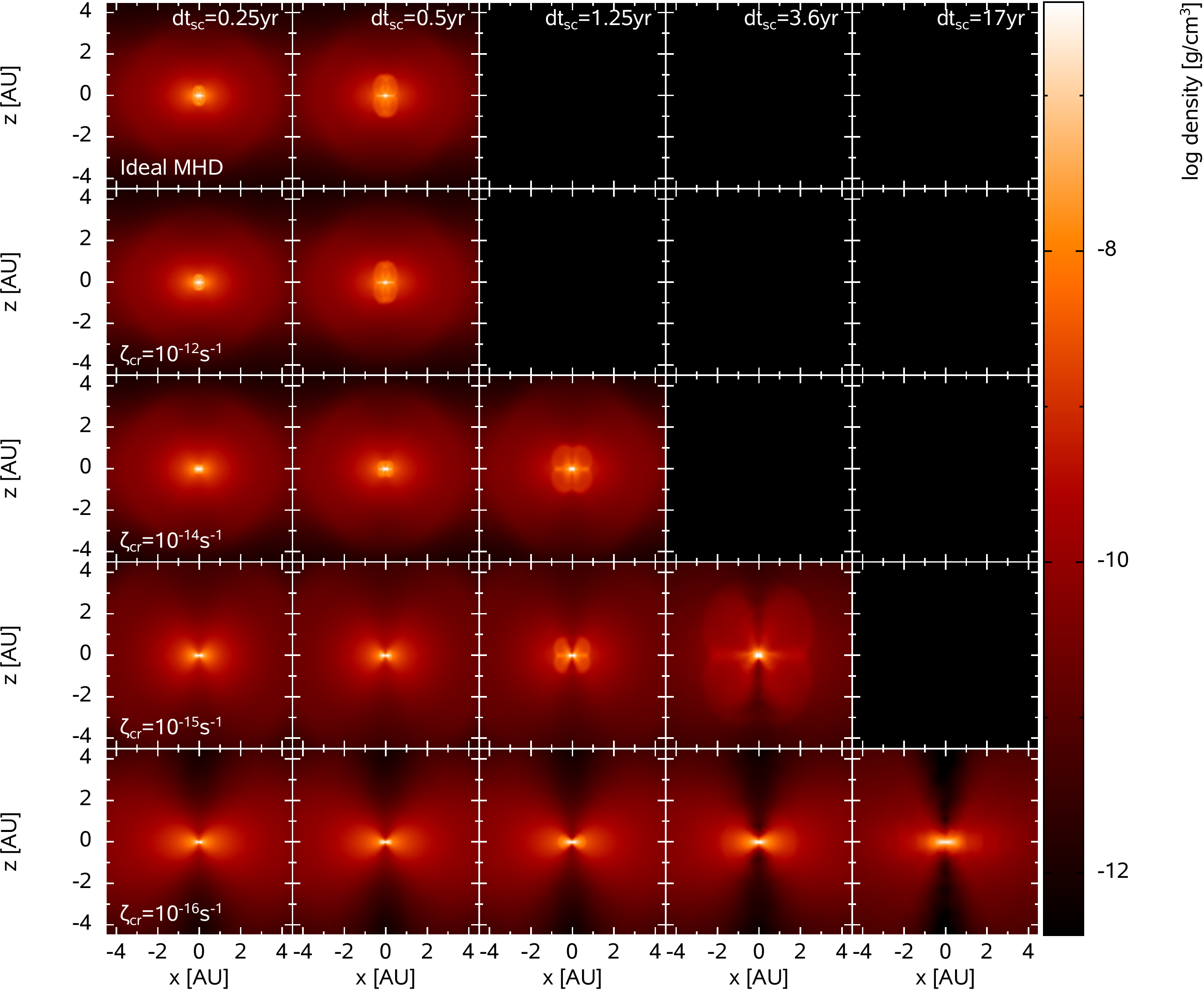}
\caption{The evolution of the stellar core and associated outflows: Gas density cross sections taken through the centre of the stellar core and parallel to the rotation axis at selected times after stellar core formation.  Strong, collimated outflows form in the high-ionisation models whereas \zetam{15} produces a lower-density, slower, conical outflow.  Model \zetam{16} forms a clear circumstellar disc, that slowly develops a broad wind. }
\label{fig:shc:density}
\end{figure*}  

Fig.~\ref{fig:shc:density} shows the evolution of the gas density in cross sections through the centre of the stellar core and parallel to the rotation axis after the formation of the stellar core.  The rows represent the different models in our suite and the columns represent different times since the formation of the stellar core. Over the first d$t_\text{sc} \approx 0.5$ yr, the maximum density grows to be \sm90 times larger in iMHD than in \zetam{16}.  Defining the stellar core to be all the gas with $\rho > 10^{-4}$ g cm$^{-3}$, the stellar core is more massive in iMHD than in \zetam{16} ($M_\text{core} \approx 16$M$_\text{J}$ compared to $3.6$M$_\text{J}$) at this age despite both cores having a similar radius of $r \approx 0.013$ au $\approx 3$~R$_\odot$.  

\subsubsection{Influence of the Hall effect}
\label{sec:shc:Hall}
Until the end of the second collapse, \zetam{12} (or iMHD) and \zetam{16} have produced the extreme values, with a smooth transition between them by varying the cosmic ray ionisation rate; see all previous line graphs at \rhole{-4}.  However, after the formation of the stellar core, the smooth trend between extremes is no longer universal.  For example, for d$t_\text{sc} \lesssim 1$ yr, \zetam{15} has lower $\rho_\text{max}$ and $T_\text{max}$ than \zetam{16} (top and middle panels of Fig.~\ref{fig:shc:Vtime}).

The lack of smooth trends is a result of the Hall effect, whose coefficient can vary in sign  (e.g. \citealp{WarNg1999}).  As discussed in Section~\ref{res:fhc:magnetic}, in our initial models, $\eta_\text{HE} > 0$ for high ionisation rates and  $\eta_\text{HE} < 0$ for low rates.  During the formation of the first hydrostatic core, the ionisation rates in the core increase due to thermal ionisation, thus form a transition region in \zetam{15} and \zetam{16} where $\eta_\text{HE}$ changes sign; after the formation of the stellar core, \zetam{14} develops a `pseudo-transition' region of $\eta_\text{HE} \lesssim 0$ with $\eta_\text{HE} > 0$ on both sides of it.

The Hall coefficient quickly transitions from negative to positive for \zetam{16}, but it is a shallow transition for \zetam{15} with the sign frequently changing over d$r \approx 5$ au.  Thus, in \zetam{15}, the Hall effect is essentially negligible in the large transition region, which would contribute to the growth rate that does not follow the expected pattern for d$t_\text{sc} \lesssim 1$ yr.

\begin{figure}
\centering
\includegraphics[width=0.392\columnwidth]{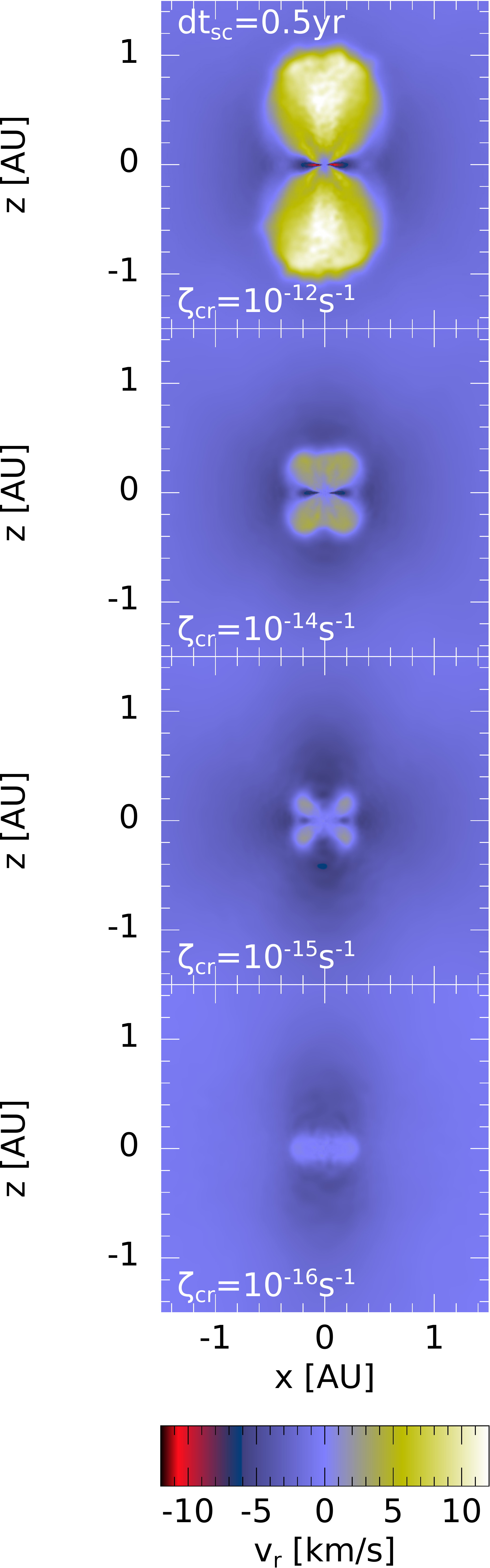}
\includegraphics[width=0.28\columnwidth,trim={6cm 0 0 0},clip]{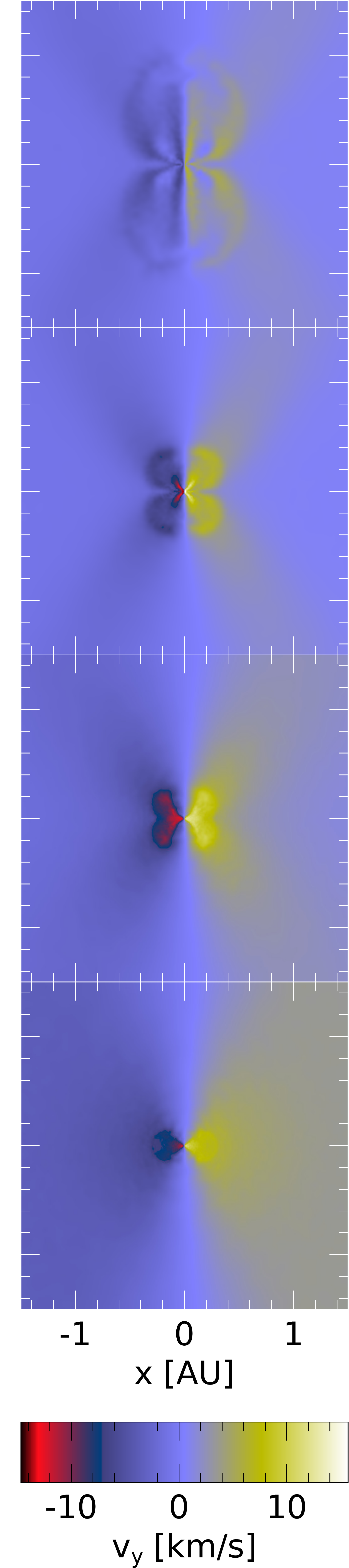}
\includegraphics[width=0.28\columnwidth,trim={6cm 0 0 0},clip]{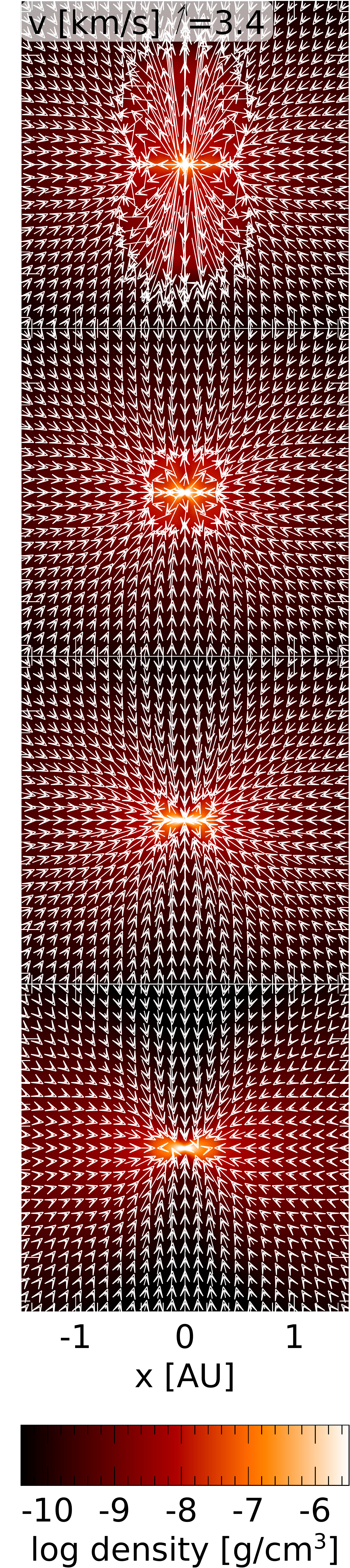}
\caption{Early-time stellar core outflows: Gas velocity cross sections taken through the centre of the stellar core and parallel to the rotation axis at \dtscapprox{0.5} after its formation.  Each frame is smaller than in Fig.~\ref{fig:shc:density} to better show the structure around the core.  From left to right is the radial velocity $v_\text{r}$, rotational velocity $v_\text{y}$, and gas density over-plotted with velocity vectors to trace the flow.  At this early time, there is a fast ($v_\text{r} \approx 14$~km~s$^{-1}$) outflow being launched from the stellar core in \zetam{12}, which is weakly rotating.  Rotational speeds increase and outflow velocities decrease with decreasing initial cosmic ray ionisation rate, such that there is a small circumstellar disc with no outflow in \zetam{16}.}
\label{fig:shc:velocities}
\end{figure} 

\subsubsection{Gas velocities}

Fig.~\ref{fig:shc:velocities} shows the radial and azimuthal velocities and velocity vectors in cross sections through the centres of the stellar cores at \dtscapprox{0.5}; as with the first hydrostatic core, iMHD and \zetam{12} have the same velocities, thus iMHD has been excluded for clarity.  The radial profiles of the gas within 20$^\circ$ of the midplane in shown in the bottom row of Fig.~\ref{fig:radialprofiles}.

The structure in the vicinity of the stellar cores at this time is strongly dependent on the cosmic ray ionisation rate, with iMHD and \zetam{12} producing collimated stellar core outflows \citep[in agreement with][]{Tomidaetal2013,BatTriPri2014}, but with \zetam{14} and \zetam{15} instead launching broader outflows.  The maximum outflow velocities range from $v_\text{r} \approx 14$~km s$^{-1}$ for \zetam{12} to $v_\text{r} \approx 3.2$~km s$^{-1}$ for \zetam{15}.  There is no outflow on sub-au scales in the \zetam{16} model at this time.

The strong, collimated outflow in \zetam{12} redirects the infalling gas along the surface of the outflow towards the midplane.  Despite its weaker outflow, the gas in \zetam{14} has a similar flow pattern; the outflow is broader than in \zetam{12}, but still has a strong vertical component to prevent gas from falling in along the rotation axis.  The stellar outflow in \zetam{15} has a similar morphology to the outflow from the first hydrostatic cores.  The outflow is weaker than in the higher ionisation cases and is predominantly along the diagonals in the cross sections (i.e. a conical outflow), such that the infalling gas is both redirected around the surface of the outflow to the midplane and is funnelled along the rotation axis to the core.  

In contrast to the more ionised cases, in \zetam{16} there is no outflow from the vicinity of the stellar core at \dtscapprox{0.5} (Fig.~\ref{fig:shc:velocities}).  Instead, a stable circumstellar disc is formed around the stellar core.  That this is a circumstellar disc can be clearly seen not only in Fig.~\ref{fig:shc:velocities}, but also in the radial and azimuthal velocity plots in the bottom row of Fig.~\ref{fig:radialprofiles}.  At \dtscapprox{0.5}, this disc has a radius of $r \approx 0.3$~au.  This result is similar to the results of the very first three-dimensional calculations of hydrodynamical collapse to stellar densities of \cite{Bate1998}.  Those calculations also showed the formation of a small circumstellar disc ($r \approx 0.1$~au) around the stellar core inside the remnant of the first core, although they were performed neither with radiative transfer nor magnetic fields. 

After the formation of the stellar core, a large Hall pseudo-transition region forms in \zetam{14}.  In this region, $\eta_\text{HE} < 0$, thus gas that enters it gets spun up by the Hall effect, increasing $v_\phi$.  As the gas migrates though the transition region, it continues to increase its rotational velocity.  Once the gas reaches the core where $\eta_\text{HE} > 0$, its large $v_\phi$ cannot be dissipated, thus, at \dtscapprox{0.5}, the largest rotational velocity is in \zetam{14} (see Fig.~\ref{fig:radialprofiles}).

\begin{figure*}
\centering
\includegraphics[width=0.3\columnwidth]{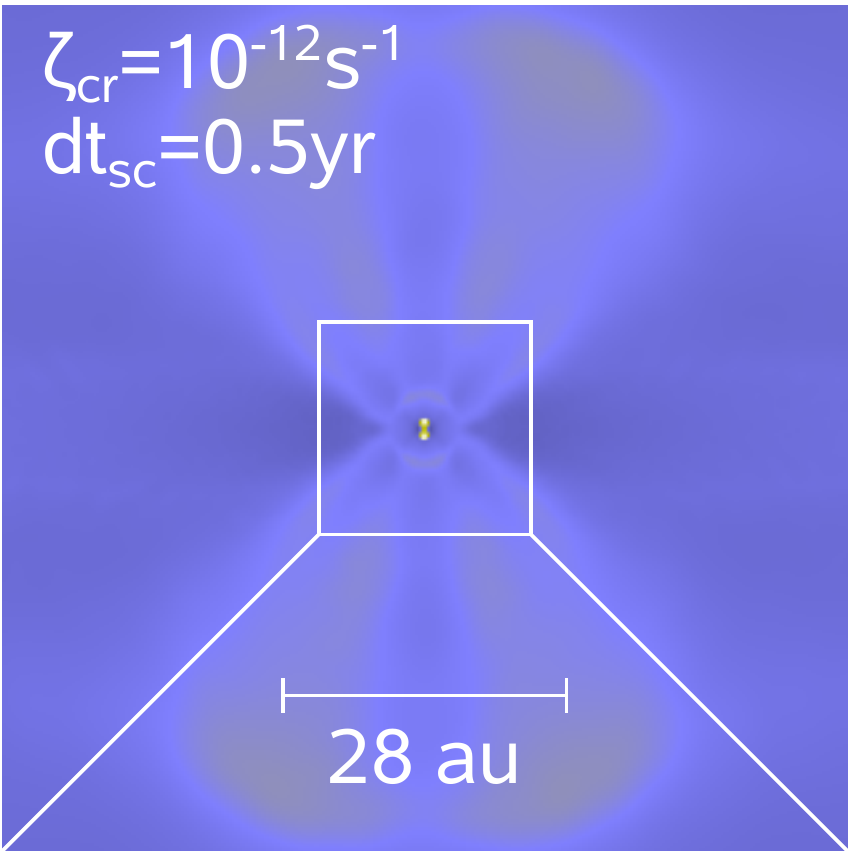}
\includegraphics[width=0.3\columnwidth]{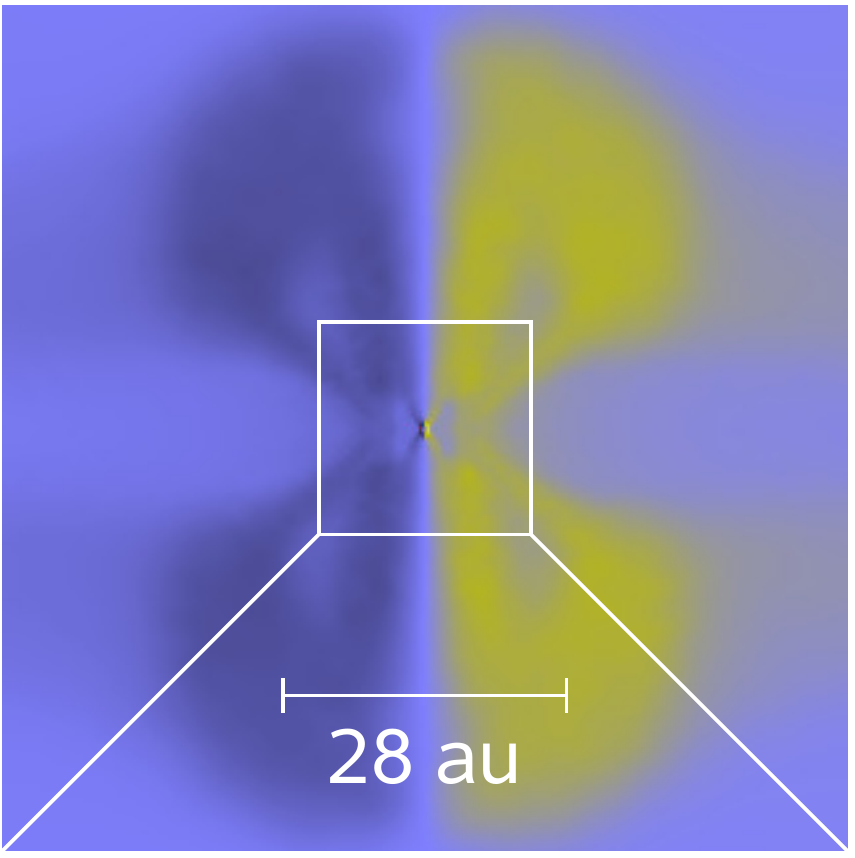}
\includegraphics[width=0.3\columnwidth]{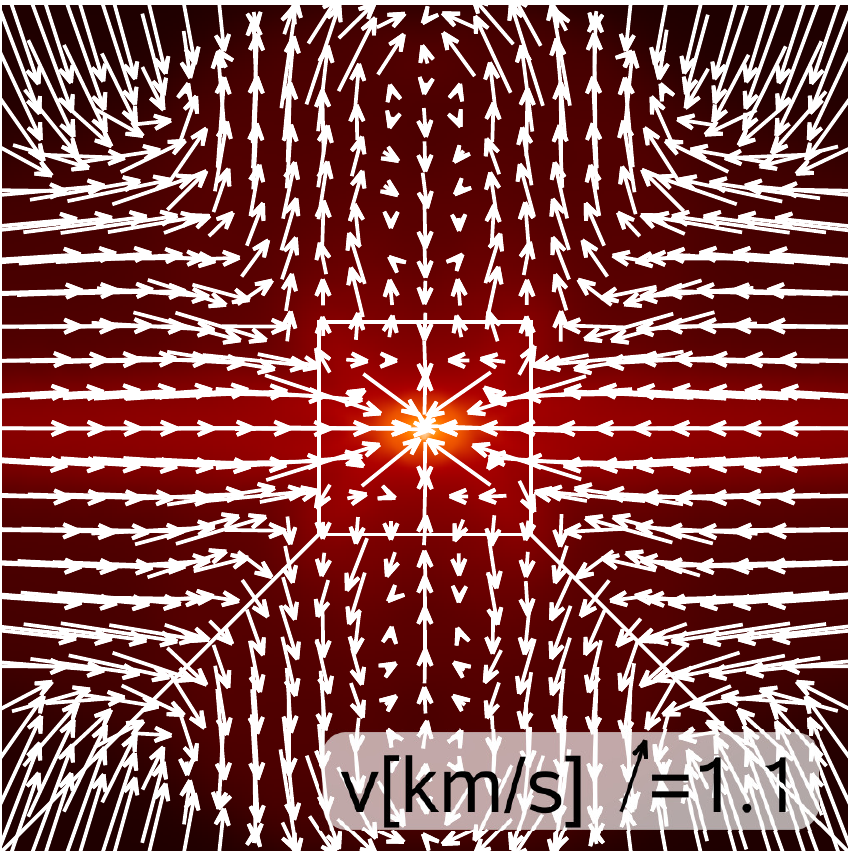}
\hspace{2cm}
\includegraphics[width=0.3\columnwidth]{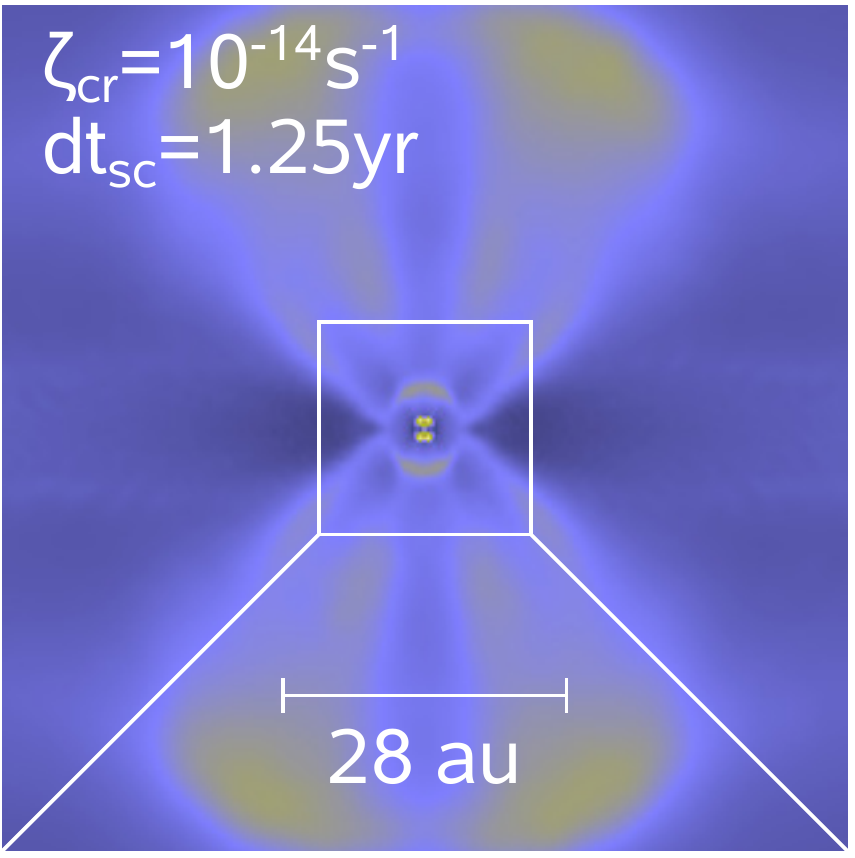}
\includegraphics[width=0.3\columnwidth]{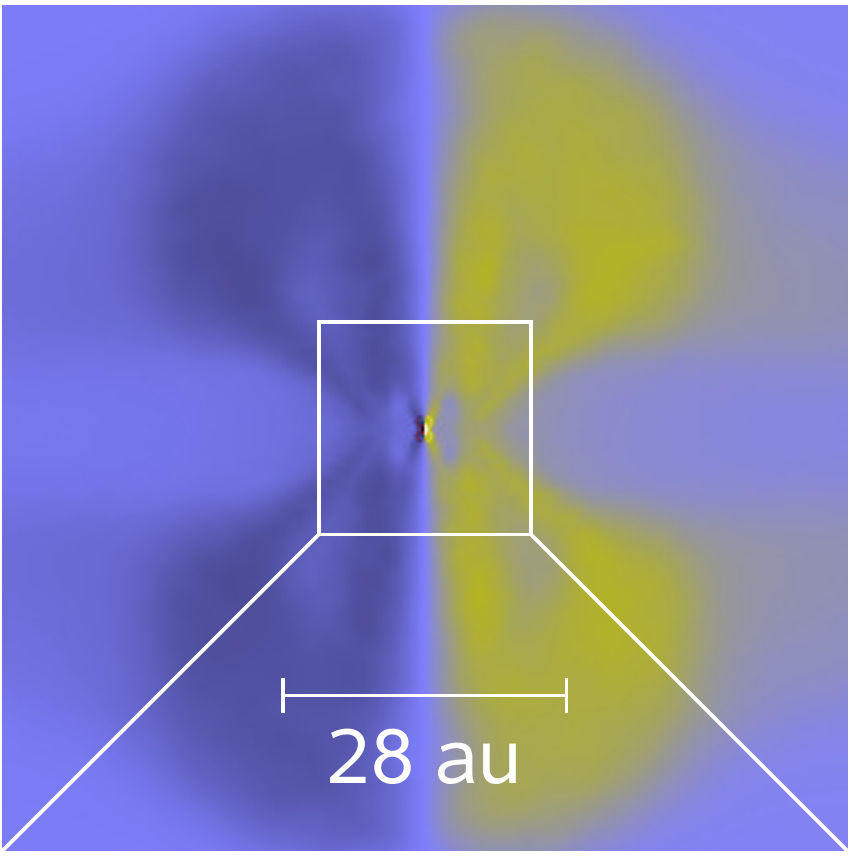}
\includegraphics[width=0.3\columnwidth]{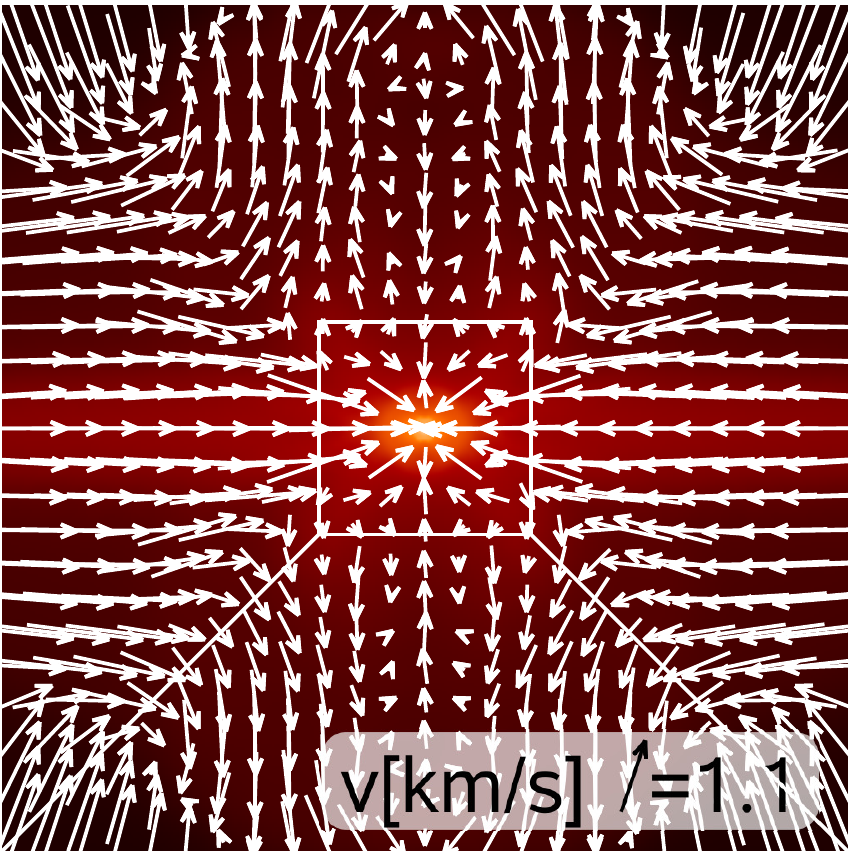}
\includegraphics[width=0.3\columnwidth]{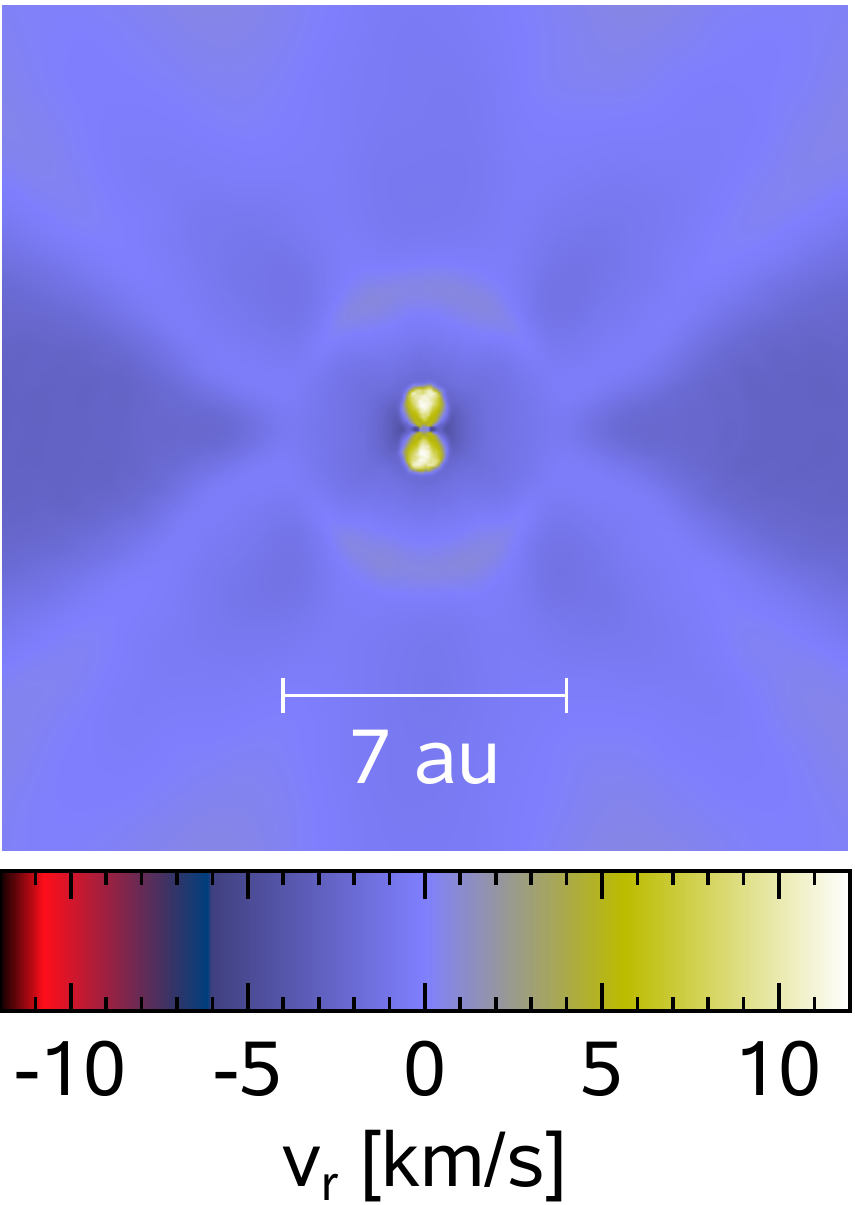}
\includegraphics[width=0.3\columnwidth]{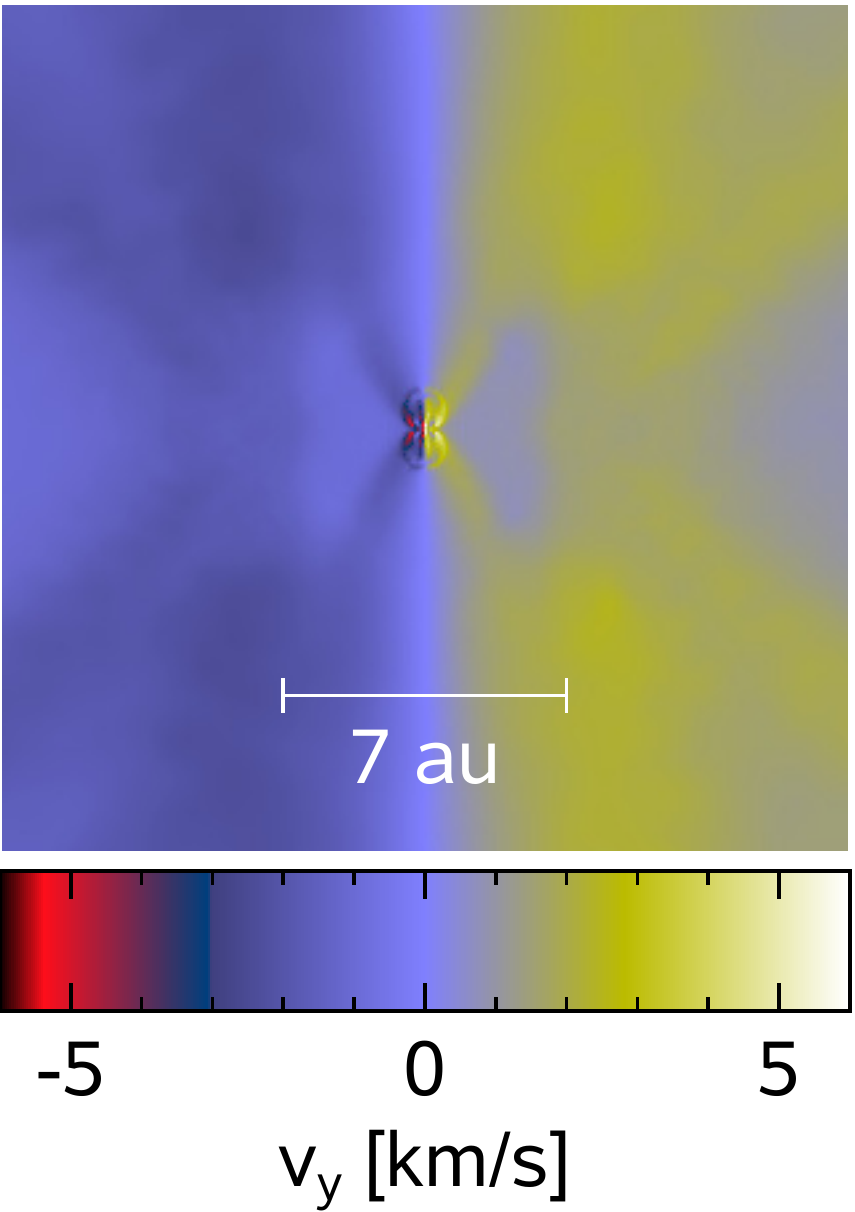}
\includegraphics[width=0.3\columnwidth]{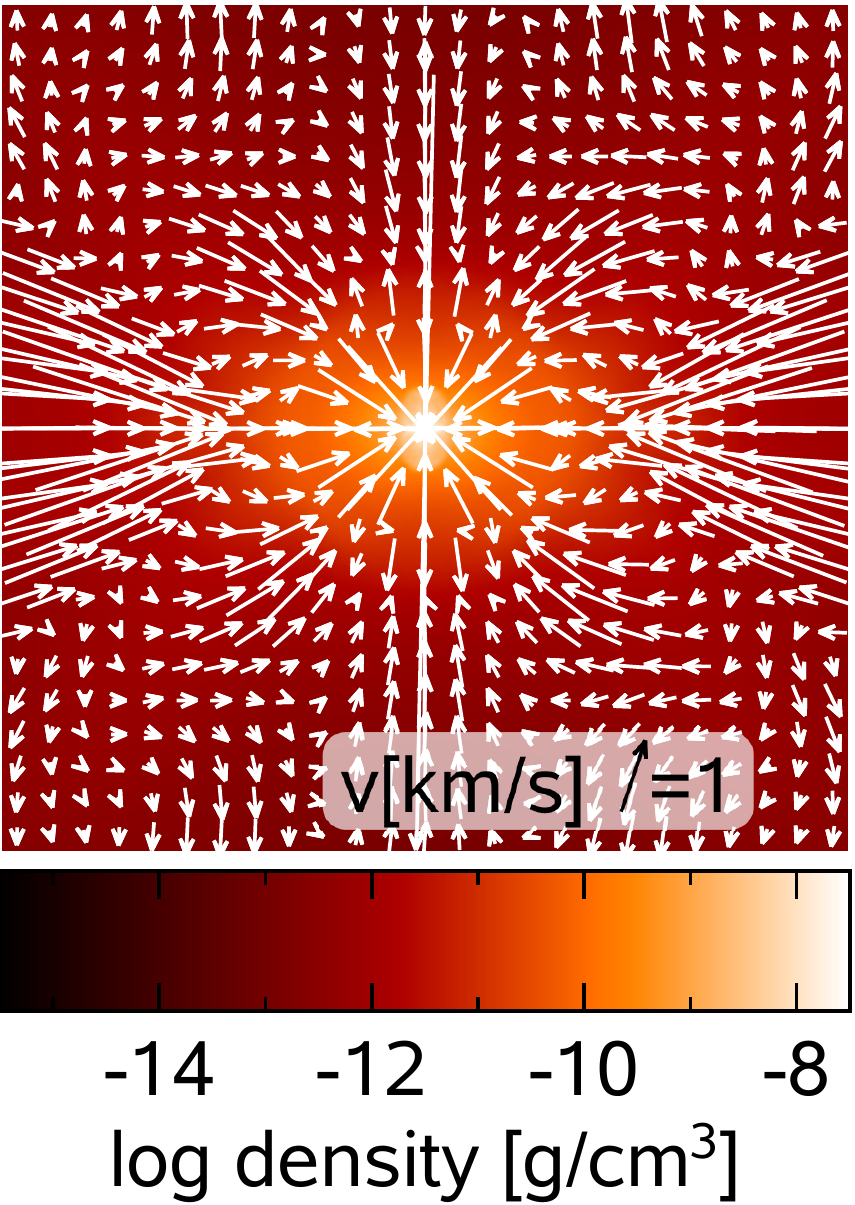}
\hspace{2cm}
\includegraphics[width=0.3\columnwidth]{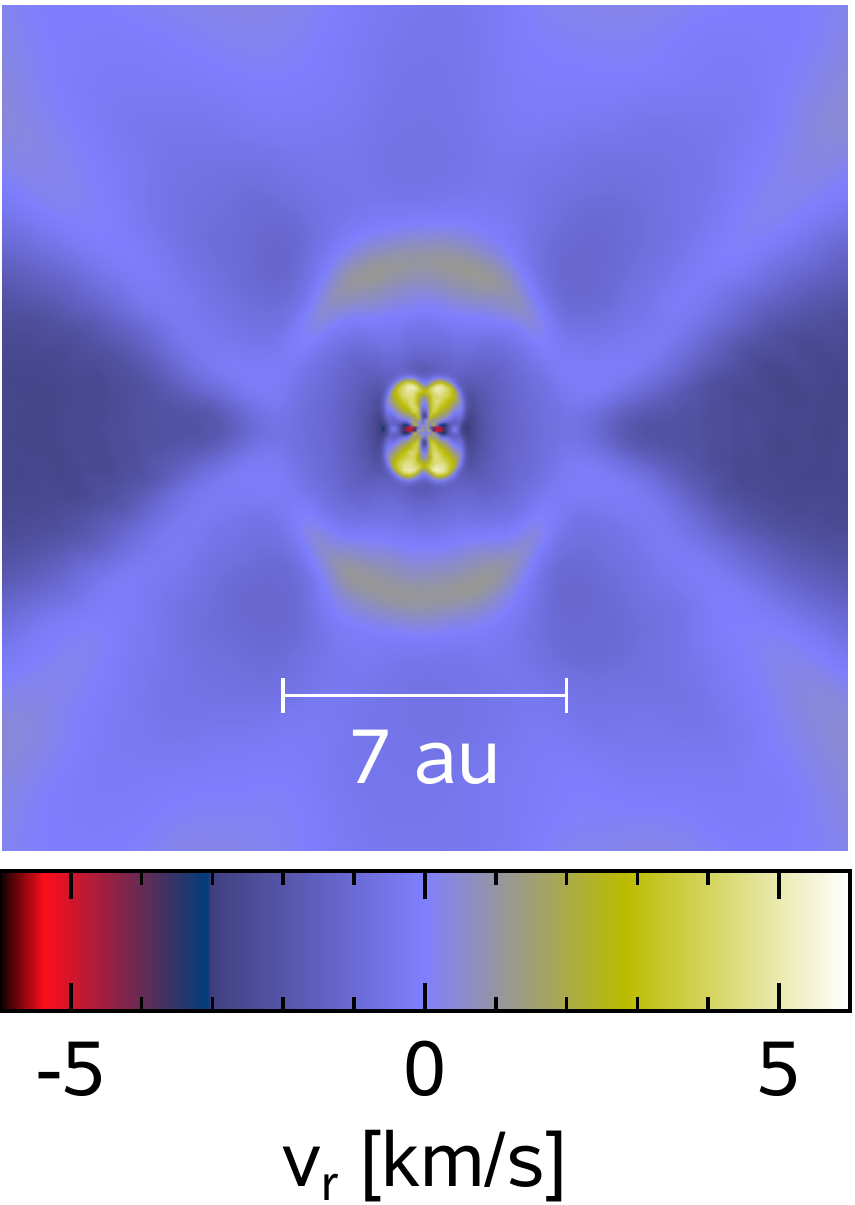}
\includegraphics[width=0.3\columnwidth]{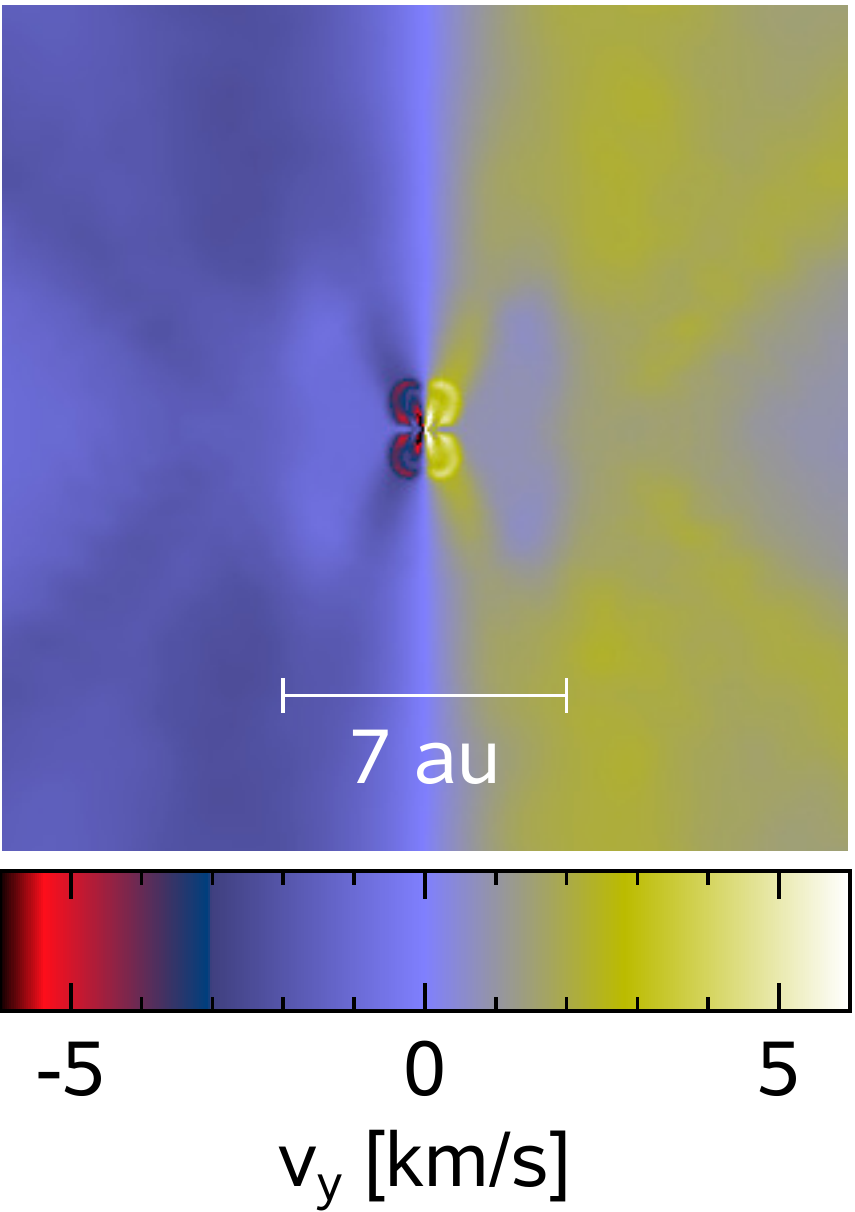}
\vspace{0.5cm}
\includegraphics[width=0.3\columnwidth]{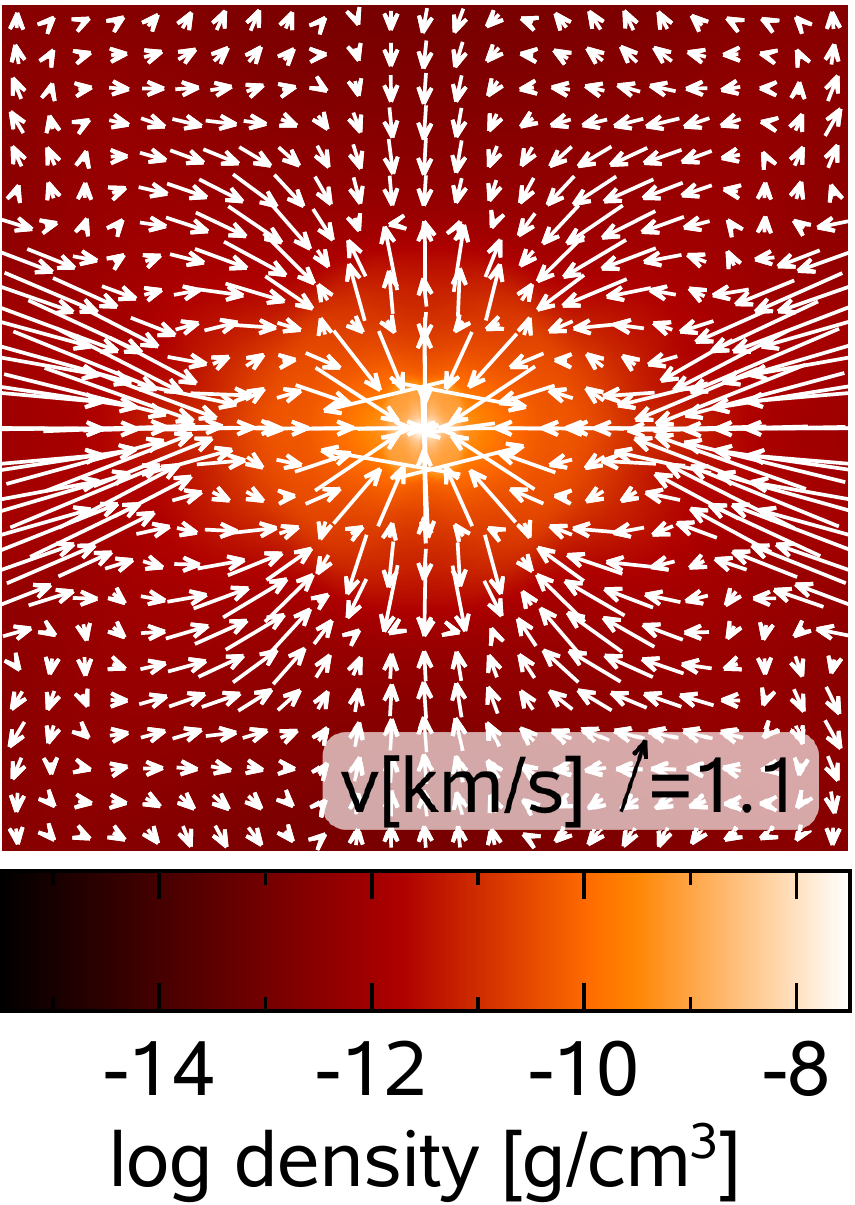}
\includegraphics[width=0.3\columnwidth]{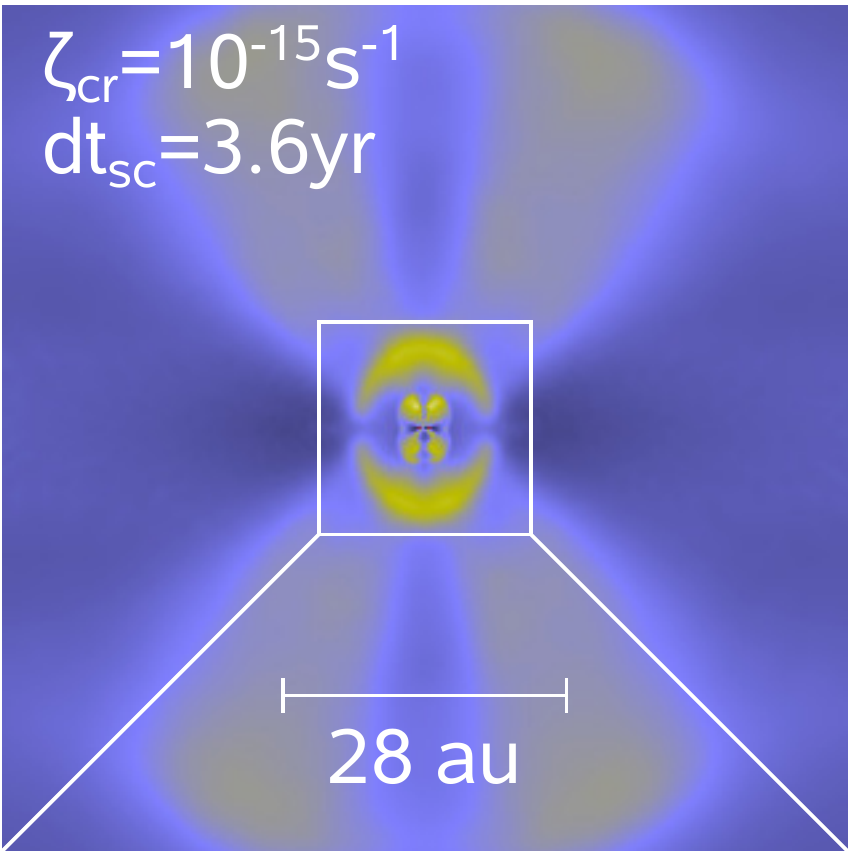}
\includegraphics[width=0.3\columnwidth]{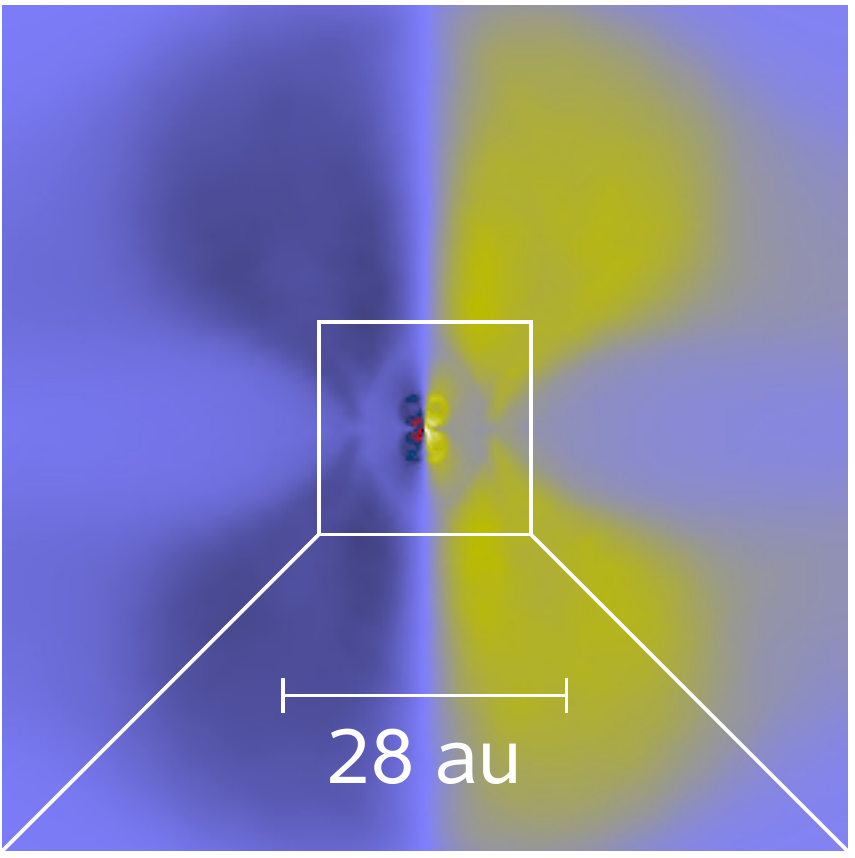}
\includegraphics[width=0.3\columnwidth]{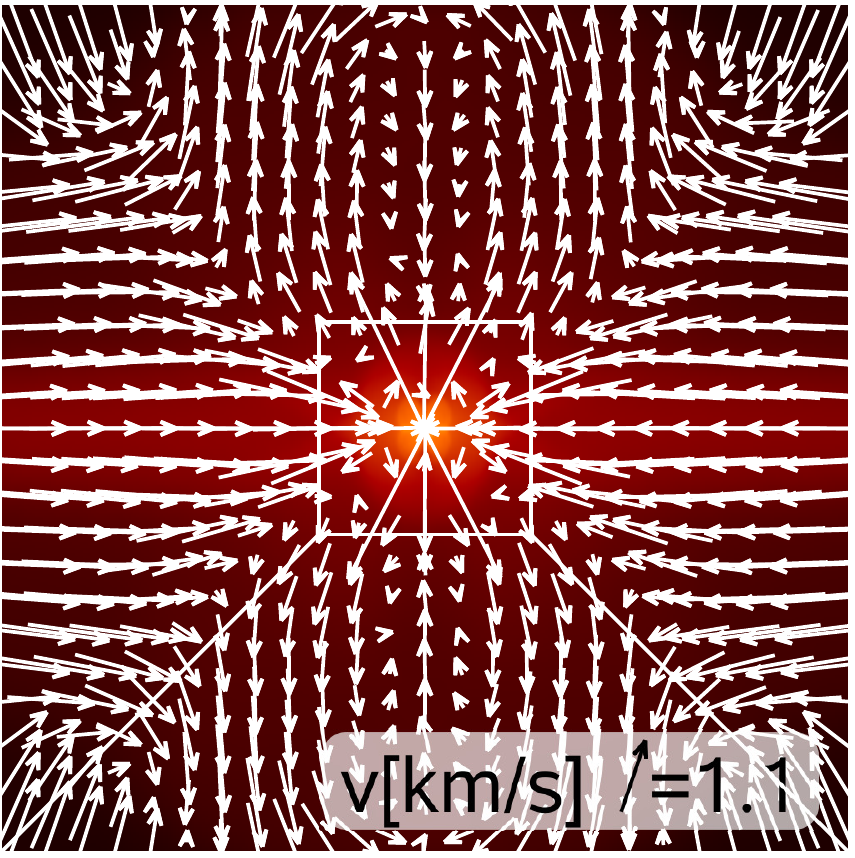}
\hspace{2cm}
\includegraphics[width=0.3\columnwidth]{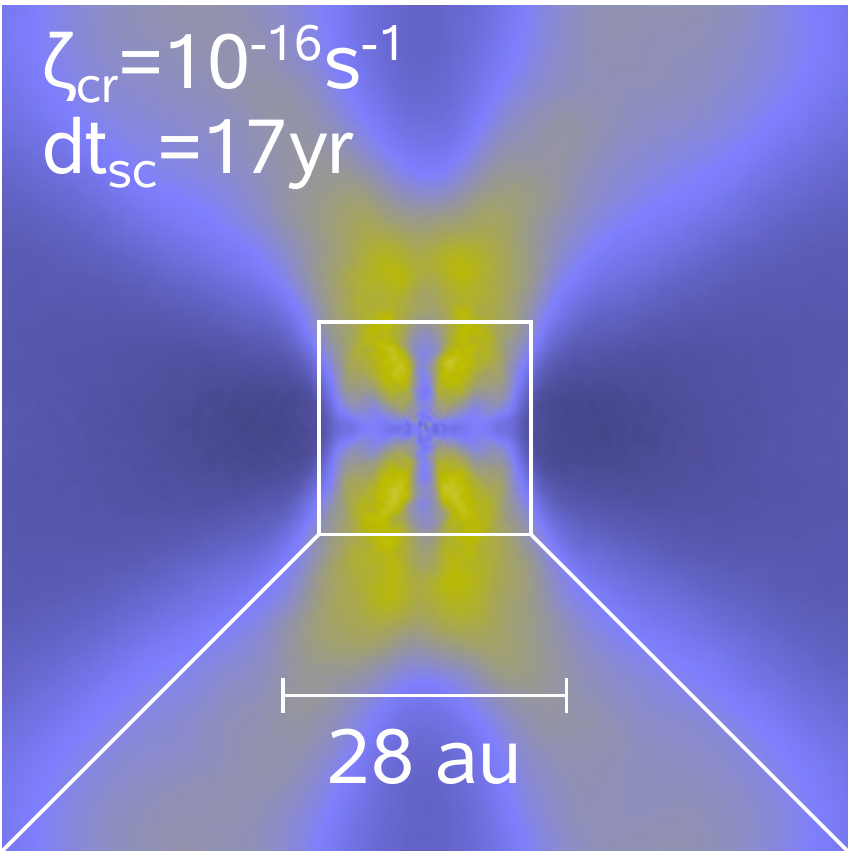}
\includegraphics[width=0.3\columnwidth]{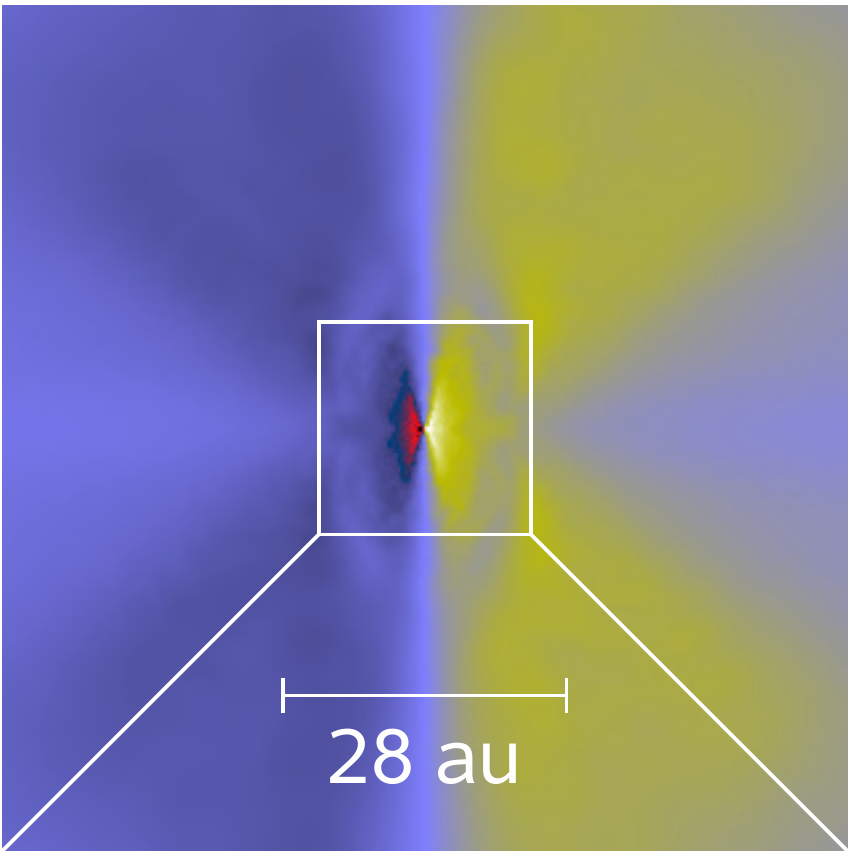}
\includegraphics[width=0.3\columnwidth]{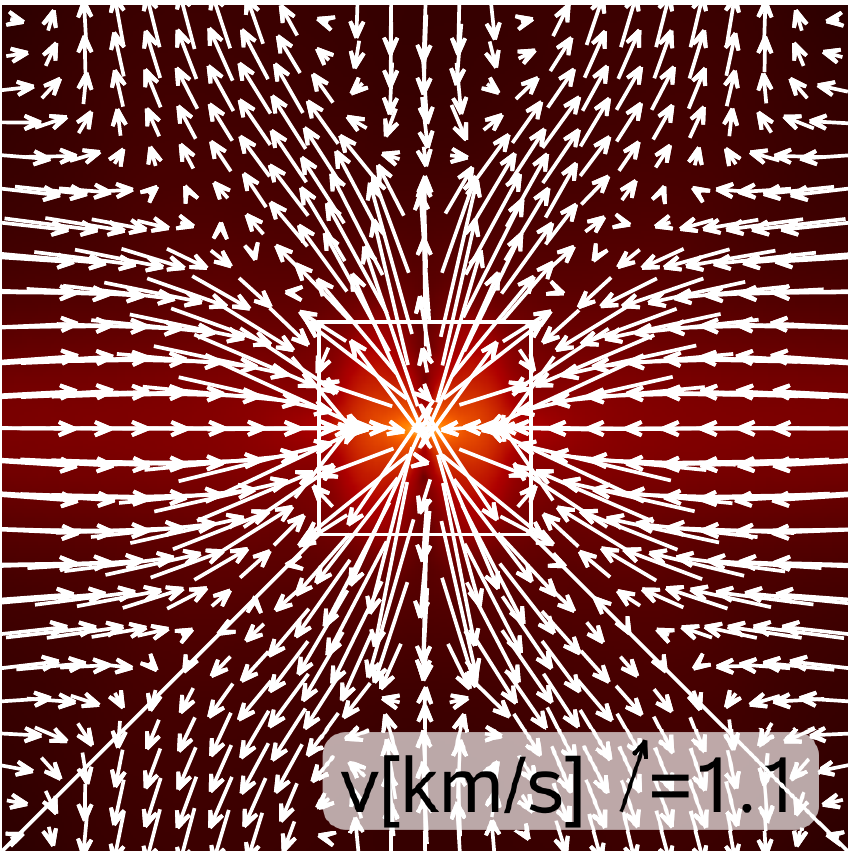}
\includegraphics[width=0.3\columnwidth]{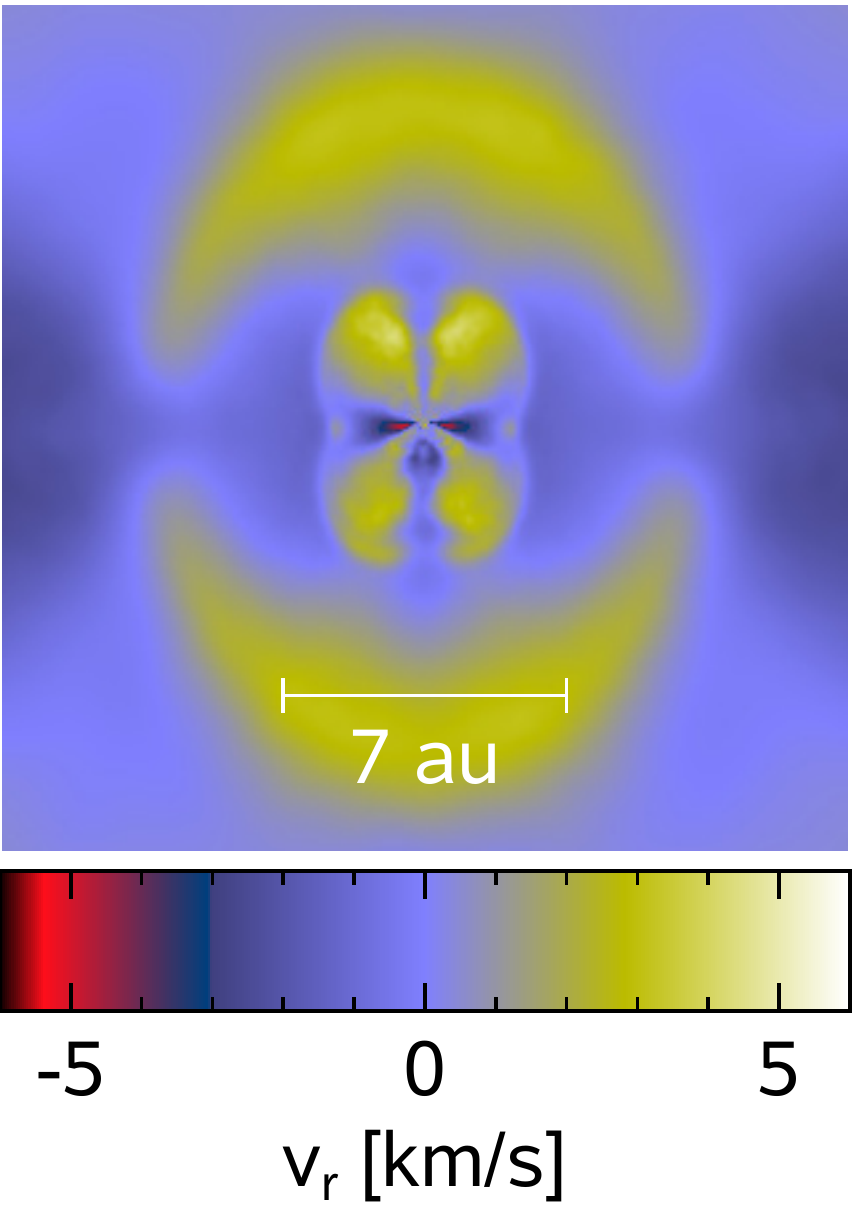}
\includegraphics[width=0.3\columnwidth]{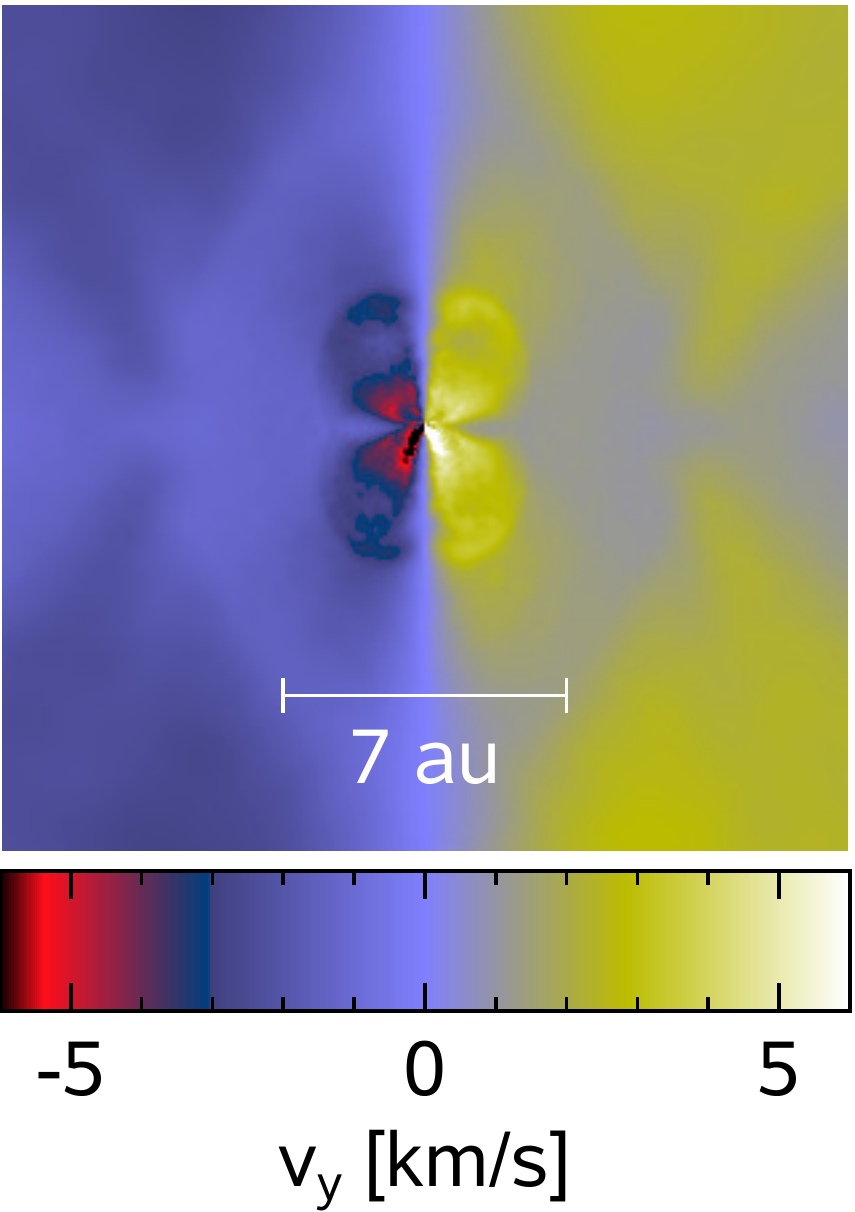}
\includegraphics[width=0.3\columnwidth]{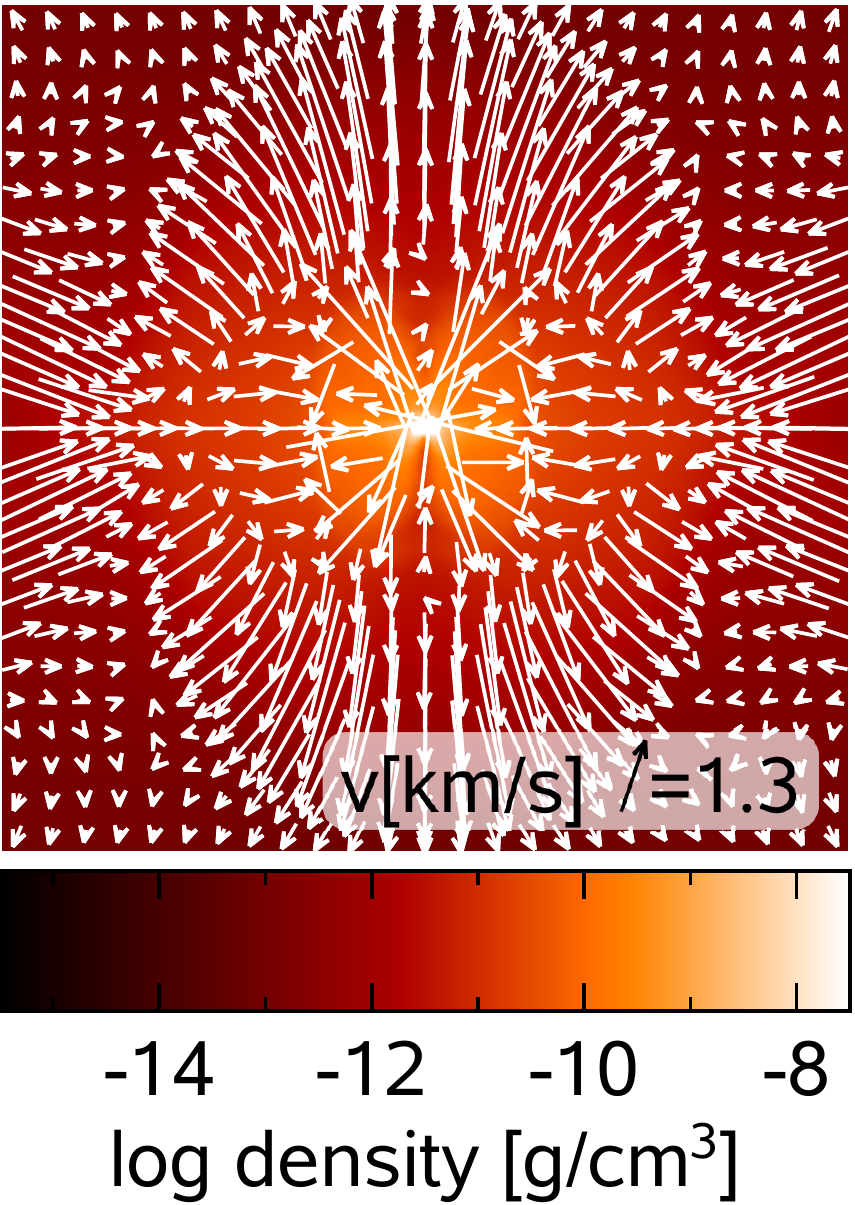}
\hspace{2cm}
\includegraphics[width=0.3\columnwidth]{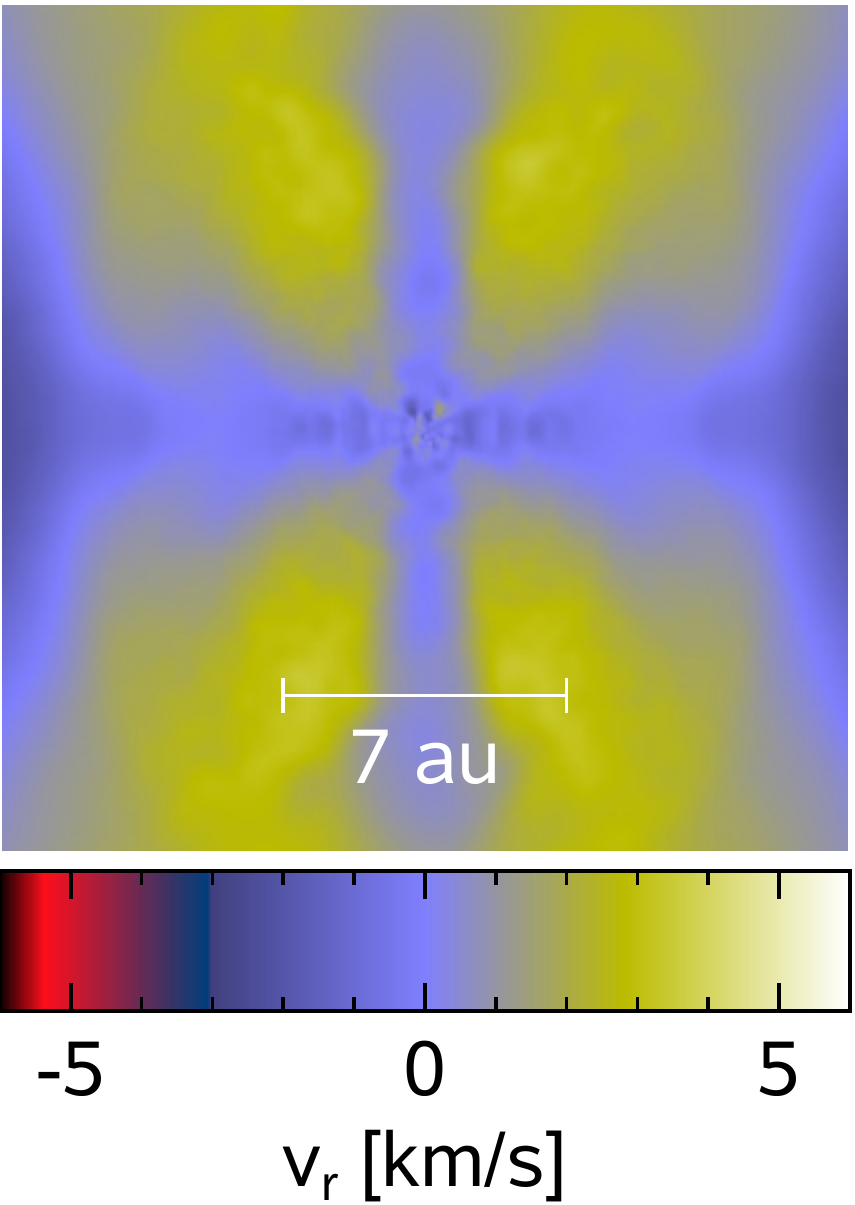}
\includegraphics[width=0.3\columnwidth]{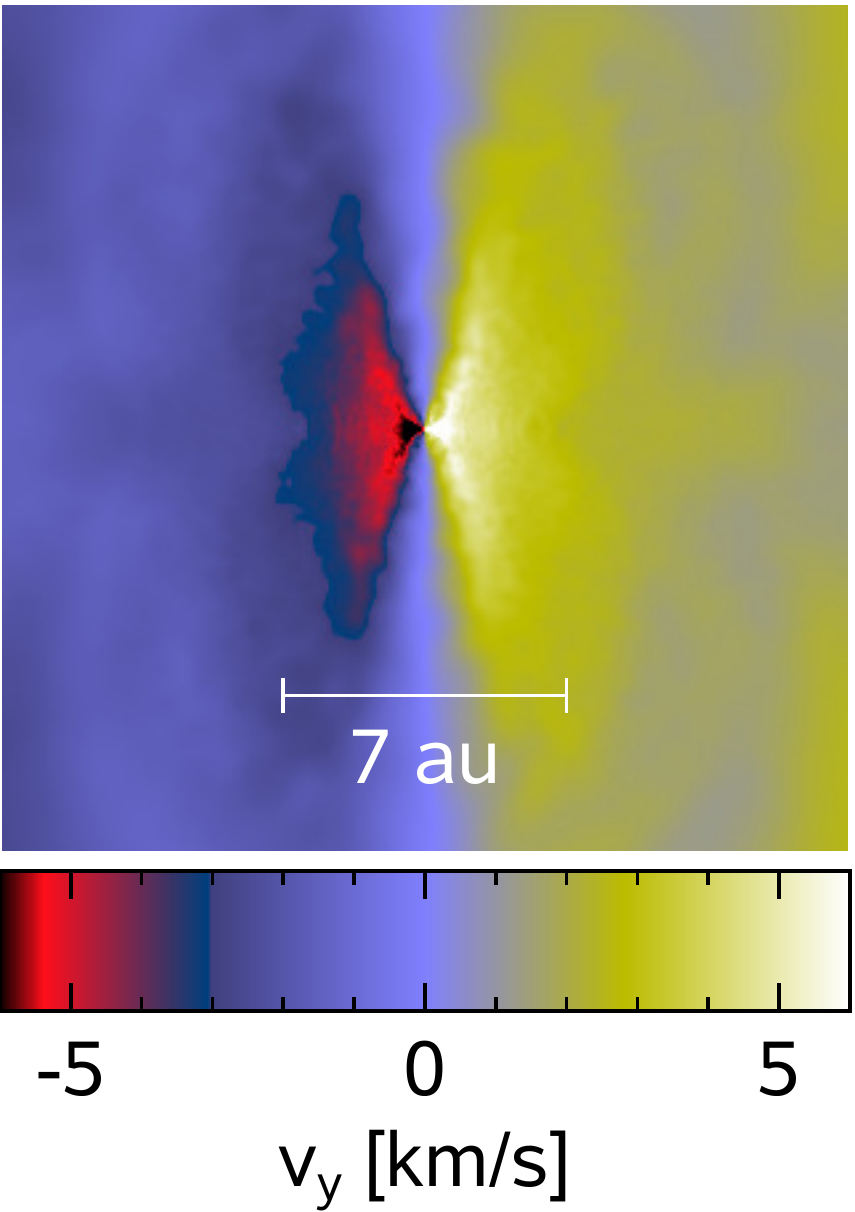}
\includegraphics[width=0.3\columnwidth]{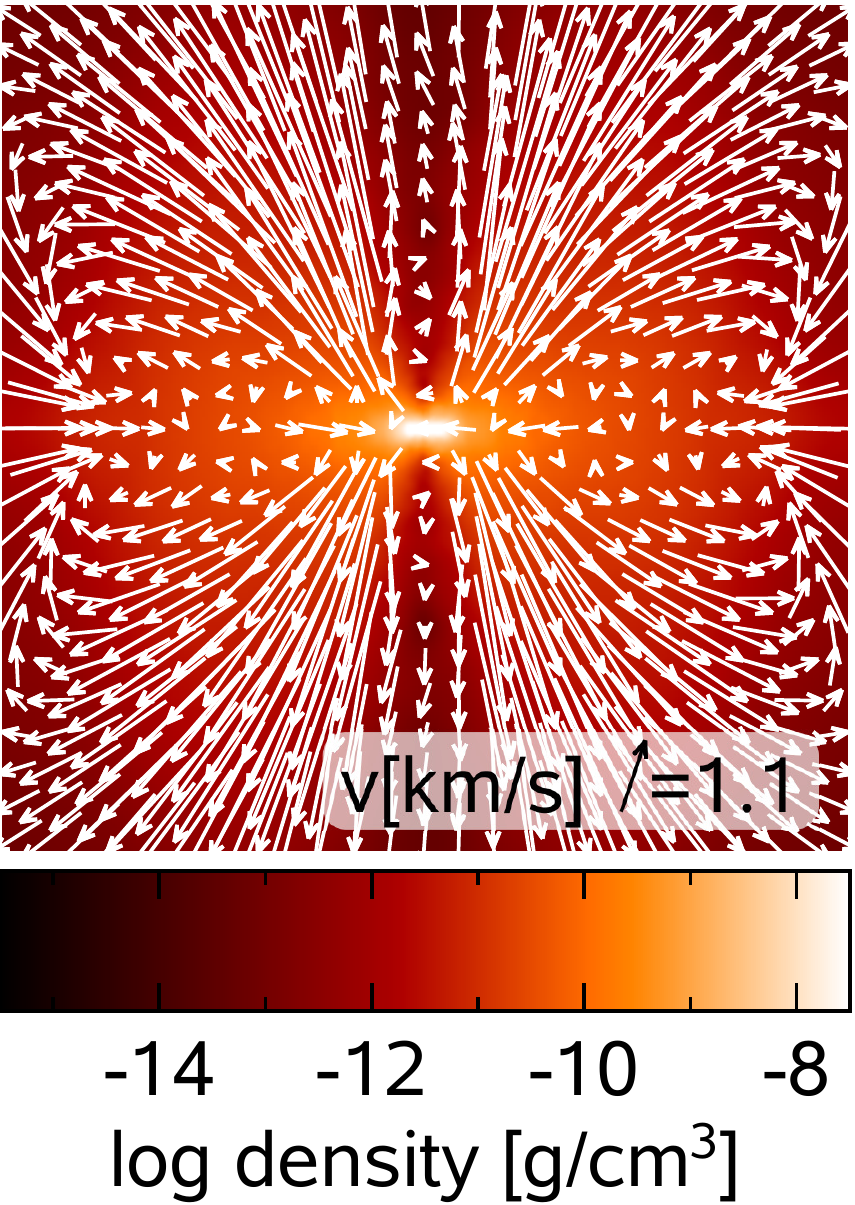}
\caption{Overall outflow morphologies at the end of the calculations:  The four groups of panels depict the end states of \zetam{12} at \dtscapprox{0.5}, \zetam{14} at \dtscapprox{1.25}, \zetam{15} at \dtscapprox{3.6} and \zetam{16} at \dtscapprox{17}.  The colour ranges are consistent across groups, except for \zetam{12}, which extend to $v_\text{r,max} = \pm 12$~\kms \ to show the detail of the outflow from the stellar core.  The velocity vectors are different in each plot to emphasise the gas motion.  The top row in each group has frame size (84 au)$^2$ and the bottom row has frame size (21 au)$^2$; the white box in each upper panel gives the extent of the region shown in the corresponding lower panel.  From left to right in each group is the radial velocity $v_\text{r}$, rotational velocity about the axis of rotation $v_\text{y}$, and gas density over-plotted with velocity vectors to trace the flow.   Model \zetam{12} is presented at the same time as in Fig.~\ref{fig:shc:velocities}, but at different frame sizes. The large-scale conical first core outflows and small-scale stellar core outflows are clearly visible in the models with \zetageq{15}.  The stellar core outflow is strong and collimated at the highest ionisation rate, but becomes slower and broader at lower ionisation rates.  An outflow from the surface of the first core is also present in the \zetam{14} and \zetam{15} models.  With the lowest ionisation rate (\zetaeq{16}) there is no distinct small-scale outflow.  Instead a circumstellar disc drives a $v_\text{r} \approx 4$~km~s$^{-1}$ broad conical outflow.}
\label{fig:shc:late}
\end{figure*}

Each model evolves at a different rate after the formation of the stellar core (see Fig.~\ref{fig:shc:Vtime}), thus, we have evolved the lower ionisation rate models further since they generally have lower temperatures and densities shortly after their formation and, thus, larger timesteps. Fig.~\ref{fig:shc:late} shows the radial and rotational velocities and velocity vectors in a cross section through these cores at the (arbitrary) end of the calculations.  The velocities are presented at two different panel sizes to show both the large and small scale structure of the outflows.  

By the end of the calculation, all models are launching outflows, and the outflows get faster as they evolve.  Only the outflow from the vicinity of the stellar core in \zetam{12} is well collimated, and this outflow reaches $v_\text{r,max} \approx 14$~km s$^{-1}$ and $z \approx 1.1$ au.

In \zetam{14} and \zetam{15}, there are also two distinct outflows: the large-scale first core outflow, and the smaller stellar core outflow.  By \dtscapprox{3.6}, the velocity of the stellar core outflow in \zetam{15} has increased to $v_\text{r,max} \approx 6 $~\kms.  The bow shock near the base of the outflow from the first core that was visible in Fig.~\ref{fig:fhc:velocities:small} at \rhoapprox{-7} has strengthened; its velocity has increased to $v_\text{r,max} \approx 3.6$~\kms.

A rotationally-supported disc has formed in \zetam{16} by \dtscapprox{17}.  Broad winds with $v_\text{r,max} \approx 4$~\kms \ are launched from the disc.

In summary, stellar core outflows are launched at later times, with lower velocities and with less collimation as the ionisation rate is decreased.  Even by \dtscapprox{17}, there is no stellar core outflow in \zetam{16}.

\subsubsection{Magnetic fields}

During the formation of the first and second cores, the maximum magnetic field strength occurs at the highest density.  However, after the formation of the stellar core, the magnetic field strength decreases, but more rapidly in the core than in the surrounding gas.  Thus, at d$t_\text{sc}\approx0.5$ yr, the strongest magnetic field strength is at $0.01 \lesssim r/\text{au} \lesssim 0.1$, and is \sm$3 - 40$ times higher than in the core, depending on the model.

\cite{BatTriPri2014} showed that the magnetic field evolution within the stellar core in the iMHD calculation is resolution-dependent. They found that increasing the resolution from one to three million particles in the initial sphere increases the value of $\bm{B}_\text{max}$ by a factor of \sm20, while increasing the resolution further from three to ten million particles only increases $\bm{B}_\text{max}$ by a factor of two (see their appendix).  Thus, the maximum magnetic field strengths presented here probably converged to within a factor of a few.  However, the subsequent decay of the field is likely dominated by numerical resistivity.  This occurs because physical resistivity becomes negligible in the second core due to thermal ionisation.

\begin{figure*}
\centering
\includegraphics[width=0.255\textwidth]{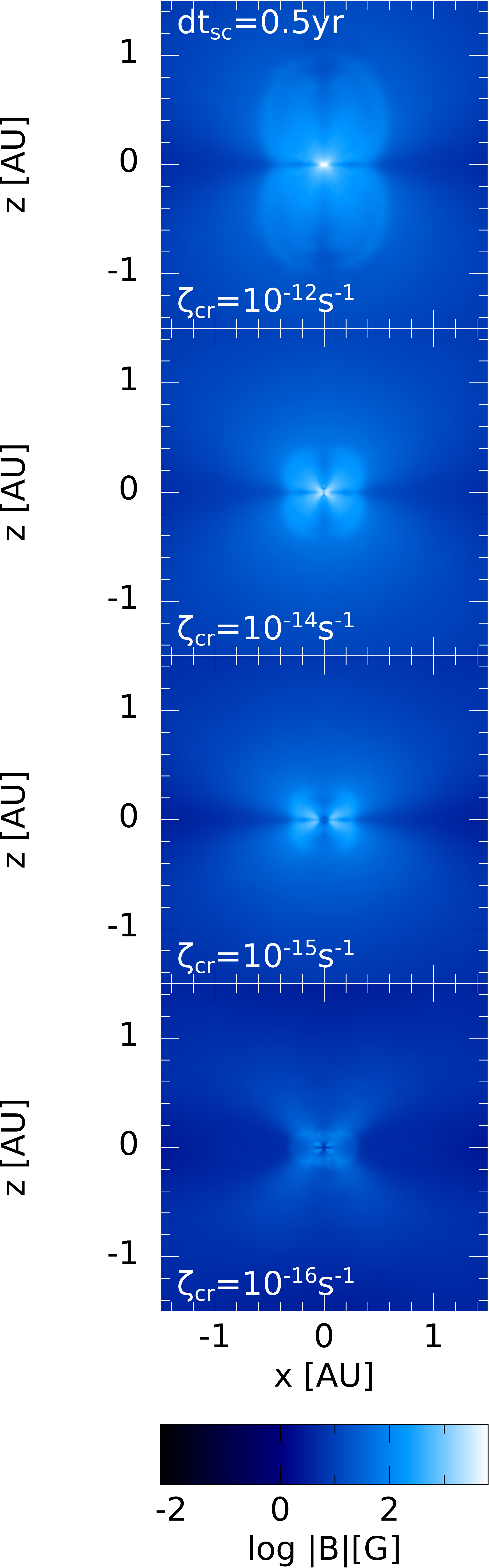}
\includegraphics[width=0.17\textwidth,trim={7cm 0 0 0},clip]{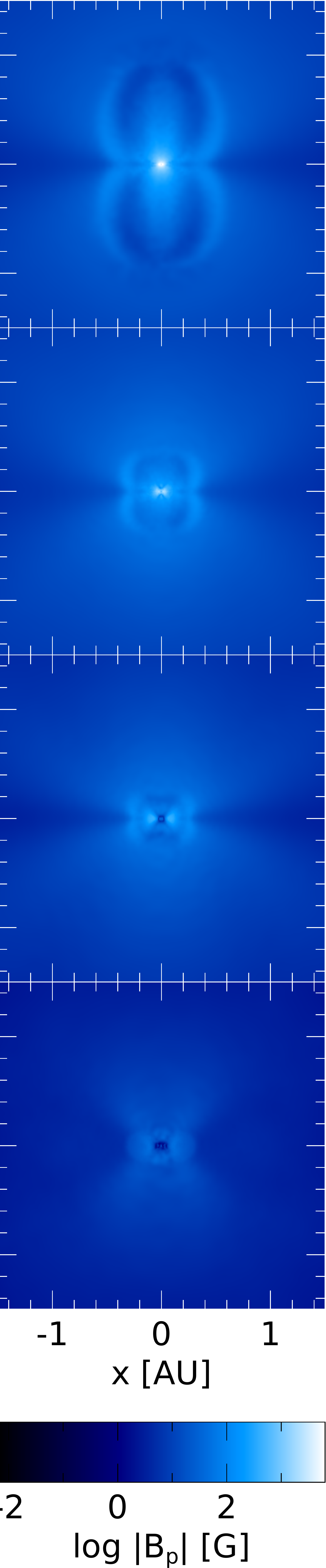}
\includegraphics[width=0.17\textwidth,trim={7cm 0 0 0},clip]{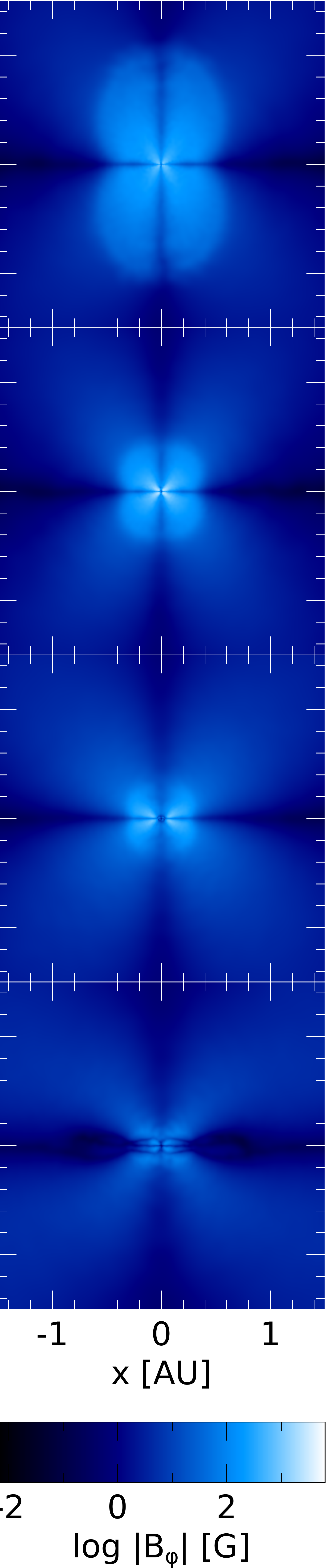}
\includegraphics[width=0.17\textwidth,trim={7cm 0 0 0},clip]{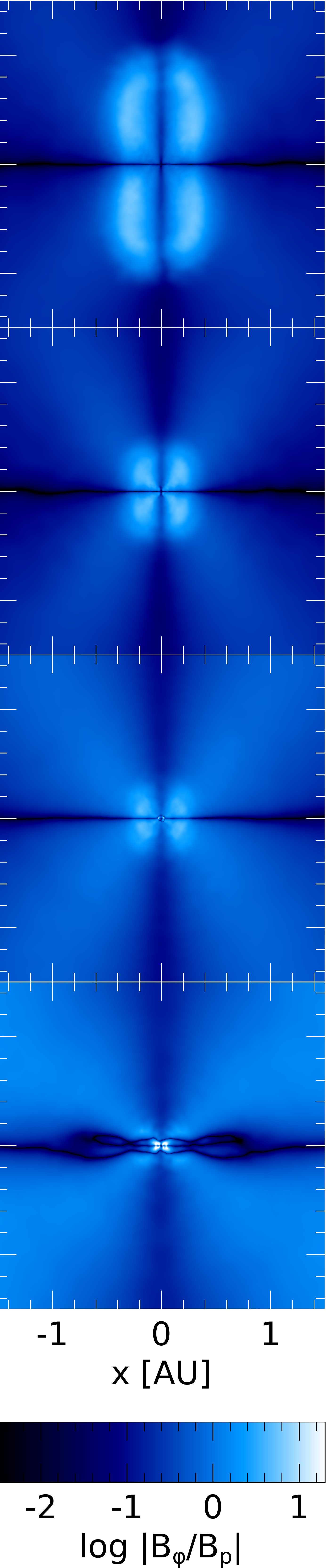}
\includegraphics[width=0.17\textwidth,trim={7cm 0 0 0},clip]{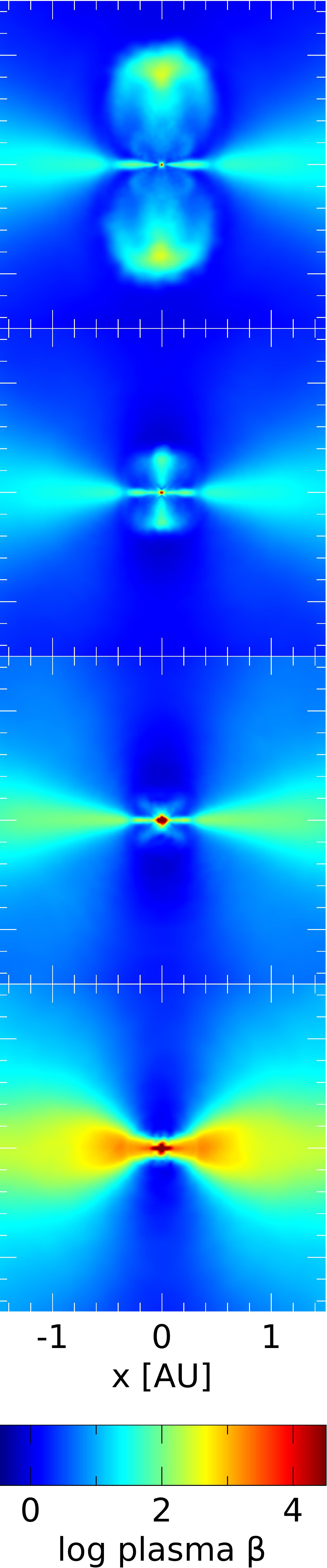}
\caption{Magnetic structure of the outflows from the vicinity of the stellar core: From left to right are the cross sections of the total magnetic field strength, magnitude of the poloidal field $|B_\text{p} |$, magnitude of the toroidal/azimuthal field $|B_\phi |$, the ratio $|B_\phi/B_\text{p}|$, and plasma $\beta$ in the outflows at \dtscapprox{0.5}. The magnetic field strength decreases with decreasing initial ionisation rate.  Unlike the first core outflows, these small-scale outflows are all dominated by the toroidal component, $|B_\phi |$.  Like the first core outflows, these are magnetic tower flows, but there is also significant thermal pressure.} 
\label{fig:shc:B:crosssection}
\end{figure*} 

\begin{figure}
\centering
\includegraphics[width=0.492\columnwidth]{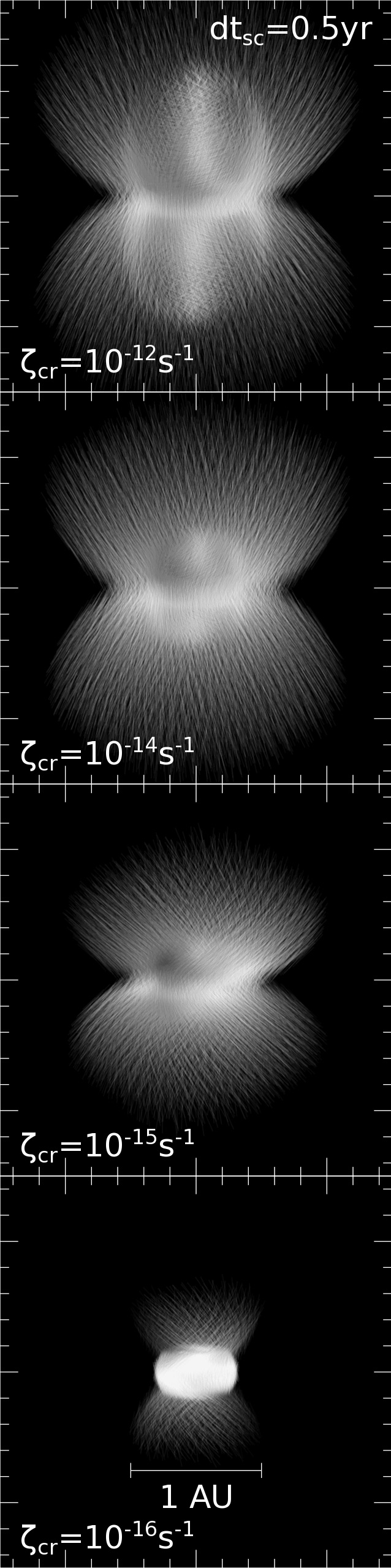}
\includegraphics[width=0.495\columnwidth]{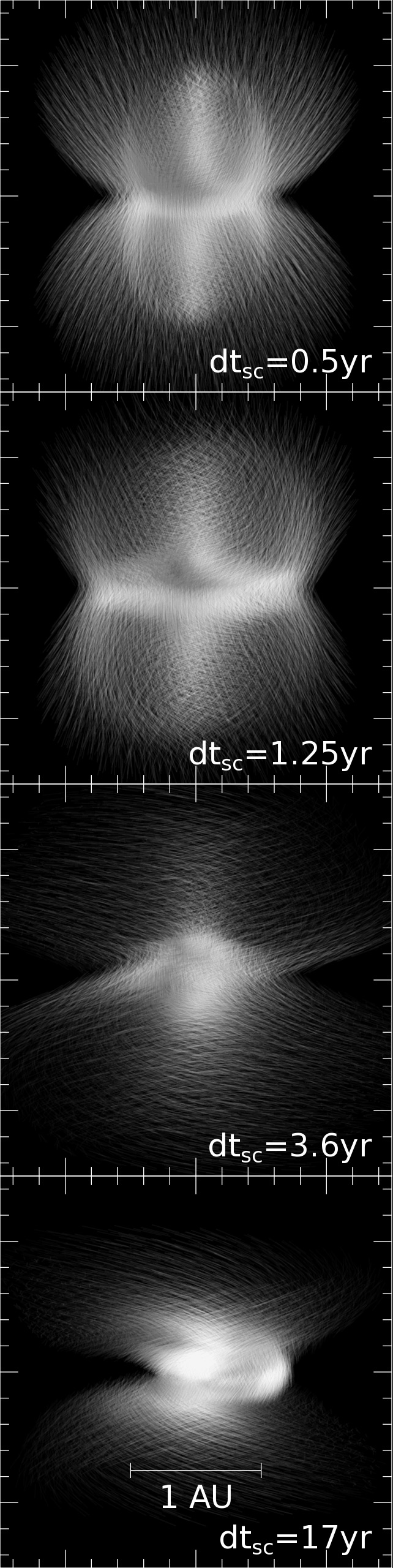}
\caption{Visualisations of the magnetic field geometry in the stellar core outflows, for $12 < |B|/\text{G} < 1.2\times 10^4$.  The images are inclined by 10$^\circ$ out of the page, and the panels have a horizontal dimension of 3 au.  The left-hand column is at \dtscapprox{0.5}, while the right-hand column is at the final time of each simulation, as listed in the bottom-right corner.  At \dtscapprox{0.5}, the magnetic field lines are more tightly wound for models with lower ionisation rates due to the reduced magnetic braking.}
\label{fig:shc:B:iron}
\end{figure} 

Fig.~\ref{fig:shc:B:crosssection} shows the magnetic field strengths ($\left|\bm{B}\right|$, $|B_\text{p}|$, $|B_\phi |$ and $|B_\phi/B_\text{p}|$) and plasma $\beta$ in a cross section through the centre of the core at \dtscapprox{0.5} for the ionised models; visualisations of the magnetic field geometry at \dtscapprox{0.5} and at the end of the simulation are shown in Fig.~\ref{fig:shc:B:iron}. 

As with the first hydrostatic core, the magnetic field is strongest in the outflows, with stronger magnetic fields associated with the stronger outflows and hence with higher initial ionisation rates.  These outflows are also magnetic tower flows. However, unlike the first core outflows, the magnetic field in the stellar core outflows is strongly dominated by the toroidal component, which can be up to \sm$10 - 100$ times stronger than the poloidal component. In the ideal MHD model of \citet{BatTriPri2014}, a combination of the Lorentz force and thermal pressure were found to be responsible for driving the small-scale, fast outflows.  Fast outflows are only formed in our high ionisation rate models, and these are found to have significant thermal pressure.

Model \zetam{16} has not formed an outflow by \dtscapprox{0.5}, and $B_\phi \sim 100 B_\text{p}$ in the gas pressure supported rotating disc that has formed.  Its subsequent evolution is qualitatively similar to \zetam{14}:  As this model evolves, the winding becomes less tight (i.e. $|B_\phi |$ decreases) and by \dtscapprox{17}, $|B_\phi/B_\text{p}| < 1$ in the disc but $\gtrsim 1$ in the outflows.  A strong toroidal component of the magnetic field forms above and below the midplane, which, in the long term, may be crucial for producing a collimated jet.

In the ideal MHD models of \citet{BatTriPri2014}, decreasing the initial mass-to-flux ratio from $\mu_0=20$ to $\mu_0=5$ had minimal effect on the stellar core outflow.  Their stellar core outflows also had stronger toroidal than poloidal components, and the ratio $|B_\phi/B_\text{p}|$ decreased with decreasing mass-to-flux ratio.  In our non-ideal MHD models, the ratio $|B_\phi/B_\text{p}|$ in the outflows tends to decrease with decreasing ionisation rate; decreasing ionisation rates lead to weaker magnetic fields, thus this trend for decreasing ionisation rates at a fixed initial mass-to-flux ratio is opposite that of decreasing mass-to-flux ratios in ideal MHD.  However, we must be cautious since in \cite{BatTriPri2014} all three ideal MHD models have similar stellar core outflows at the comparison time of \dtscapprox{1}, while the outflows from our non-ideal MHD models vary significantly at \dtscapprox{0.5}.

At \dtscapprox{0.5}, the maximum density between \zetam{12} and \zetam{16} differs by a factor of \appx83, and the temperature differs by a factor of \appx6.  Since the thermal ionisation rate is dependent only on density and temperature, the ionisation fractions are highest in  \zetam{12}, with $n_s/(n_\text{i} + n_\text{n}) \approx 0.30$ in the core; for comparison, the fractions in \zetam{16} are \appx0.012.  The ionisation fraction approximately traces the temperature profile --- the highest ionisation fractions are in the hottest part of the outflows (c.f. Fig.~\ref{fig:shc:temperature}, which shows the cross section of the gas temperature). 

Despite the ionisation fractions differing by a factor of \appx25, the non-ideal MHD coefficients remain similar for all models in the core due to their dependence on the magnetic field strength, which varies by a factor of \appx140 between \zetam{12} and \zetam{16}.  However, since $|\eta| < 10^6$ cm$^2$ s$^{-1}$, non-ideal MHD is no longer playing an important role in the evolution of the stellar core.  The stellar core is now only indirectly affected by non-ideal MHD --- the cool accreting gas is less ionised and has weaker magnetic field strengths in \zetam{16} than in \zetam{12}.

In summary, as the stellar core evolves, the maximum magnetic field strength decreases, with the maximum value being in the gas surrounding the core.  Unlike the first core outflows, the stellar core outflows contain strong toroidal magnetic fields and are dominated by gas pressure.

\subsubsection{The structure of the stellar cores}
\cite{BatTriPri2014} used ideal MHD, but varied the initial mass-to-flux ratio.  They found that the stellar core properties were remarkably similar for their models with $\mu_0 = 5$, 10 and 20.  At \dtscapprox{1}, all three models had similar central densities and temperatures; the radial velocity profiles were also similar, although the maximum infall speed was \sm2 km s$^{-1}$ faster for their $\mu_0=5$ model than their $\mu_0=20$ model (see their fig. 13). Moreover, all three models produced collimated stellar core outflows, although the outflow was slightly faster in their $\mu_0=5$ model (see their fig. 11).  They concluded that this similarity was a result of the gas that collapses to form the stellar core having essentially `universal' properties since it must first be hot enough for molecular hydrogen to dissociate; the later evolution, however, would depend on the details of the accretion.

With the inclusion of non-ideal MHD effects, the characteristics of the stellar core are dependent on the initial cosmic ray ionisation rate.  Although all our models have the same density and temperature at the beginning of the stellar core phase, their evolution diverges almost immediately due to very different accretion rates and they have noticeably different masses even by \dtscapprox{0.5}.  Notably, in our models, stellar core outflows can be broadened or suppressed by decreasing the initial cosmic ray ionisation rate.  

Thus, unlike in ideal MHD, we find that the stellar core phase does not have a universal set of properties and that the impact of the cosmic ray ionisation rate must be carefully taken into account.

\subsubsection{Gas temperatures}

\begin{figure*} 
\centering
\includegraphics[width=0.9\textwidth]{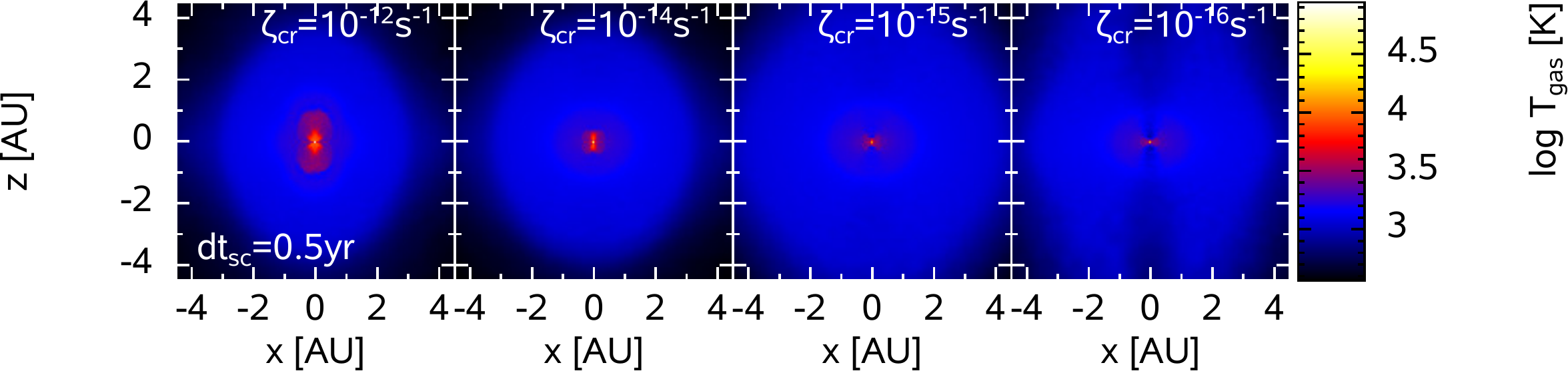}
\caption{Gas temperatures in the stellar cores: Gas temperature cross sections taken through the centre of the stellar core and parallel to the rotation axis at \dtscapprox{0.5} after the formation of the stellar core.  The temperatures are hottest in the stellar core and generally fall off with distance.  However, in the highly ionised models the gas in the fast, collimated, small-scale outflows is also hot (temperatures ranging from $2000-20000$~K).  This plot is qualitatively similar to that of the ionisation fraction, $n_s/(n_\text{i} + n_\text{n})$ where $s\in\{\text{e},\text{i}\}$ since the ionisation fraction is dependent only on temperature and density for $T \gtrsim 1000$~K.}
\label{fig:shc:temperature}
\end{figure*}  

As we have mentioned earlier, the temperatures of the gas associated with different phases of the collapse are almost independent of the ionisation rate (e.g.~Fig.~\ref{fig:VSrho}).  The only substantial differences in temperature structure between the calculations are found following stellar core formation.  Fig.~\ref{fig:shc:temperature} shows the gas temperature in cross sections through the stellar cores at \dtscapprox{0.5}.  At this time, the temperature at the centre of the stellar core is \appx6 times hotter in \zetam{12} than in \zetam{16} due to the greater stellar core mass.  The main difference, however, is of the outflowing gas on au-scales.  The centres of the outflows in \zetam{12} and \zetam{14} are very hot, with $2000 \lesssim T/\text{K} \lesssim 20000$.  By contrast, in \zetam{16}, outside of the small circumstellar disc surrounding the stellar core, the gas temperatures smoothly decrease from a maximum of $\approx 2000$~K as the radius increases.

\subsection{Hall effect and the initial direction of the magnetic field}
\label{res:Halldirection}

The Hall effect depends on the initial orientation of the magnetic field with respect to the axis of rotation \citep{BraWar2012}.  Previous studies have confirmed that, given our initial counter-clockwise rotation, the Hall effect promotes disc formation for $B_{0,z} < 0$ and discourages it for $B_{0,z} > 0$ \citep{TsukamotoEtAl2015a,WurPriBat2016,TsukamotoEtAl2017}. The models we discussed in the previous sections all used initial conditions that promote disc formation.  Here, we briefly present the results of a non-ideal MHD model using \zetaeq{-16} and $B_{0,z} > 0$, and compare it to its counterpart with $B_{0,z} < 0$, which we name \zetams{16}{+} and \zetams{16}{-}, respectively.  Both of these models are calculated using the implicit Ohmic resistivity algorithm to speed up the calculations (see Appendix~\ref{app:implicit}), thus, although very similar, \zetams{16}{-} is not identical to \zetam{16} which has been discussed above.

\begin{figure}
\centering
\includegraphics[width=\columnwidth]{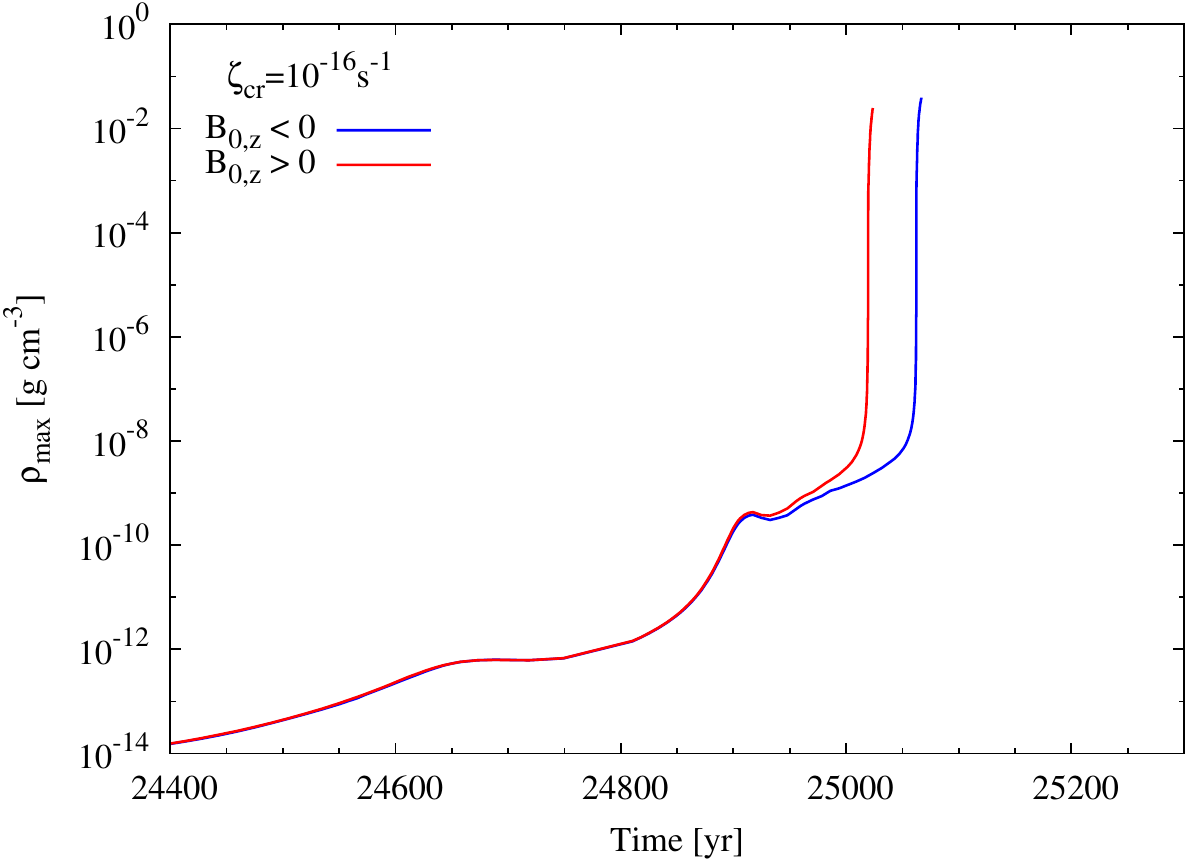}
\caption{Effect of the Hall effect on the time for collapse:  The evolution of the maximum density during the collapse of a molecular cloud core for non-ideal MHD models in which the direction of the magnetic field is reversed.  Both models have \zetaeq{-16}, but in one $B_{0,z} < 0$ (\zetams{16}{-}; blue), while in the other $B_{0,z} > 0$ (\zetams{16}{+}; red).  When $B_{0,z} < 0$ the Hall effect acts against magnetic braking, while with $B_{0,z} > 0$ strong magnetic braking allows the gas to collapse more rapidly.}
\label{fig:hall:rhoVtime}
\end{figure} 

\begin{figure*}
\centering
\includegraphics[width=0.25\textwidth]{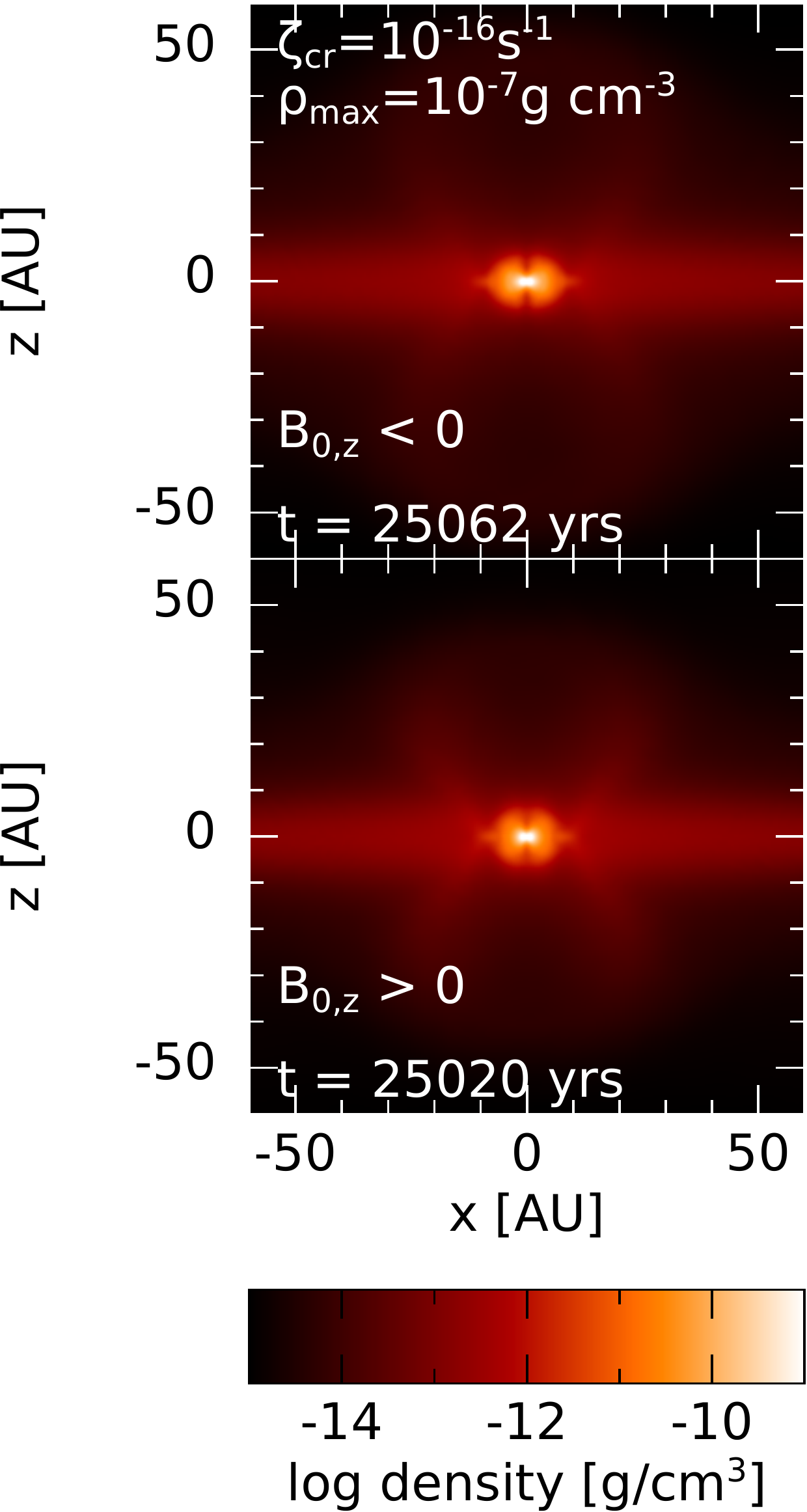}
\includegraphics[width=0.17\textwidth,trim={4cm 0 0 0},clip]{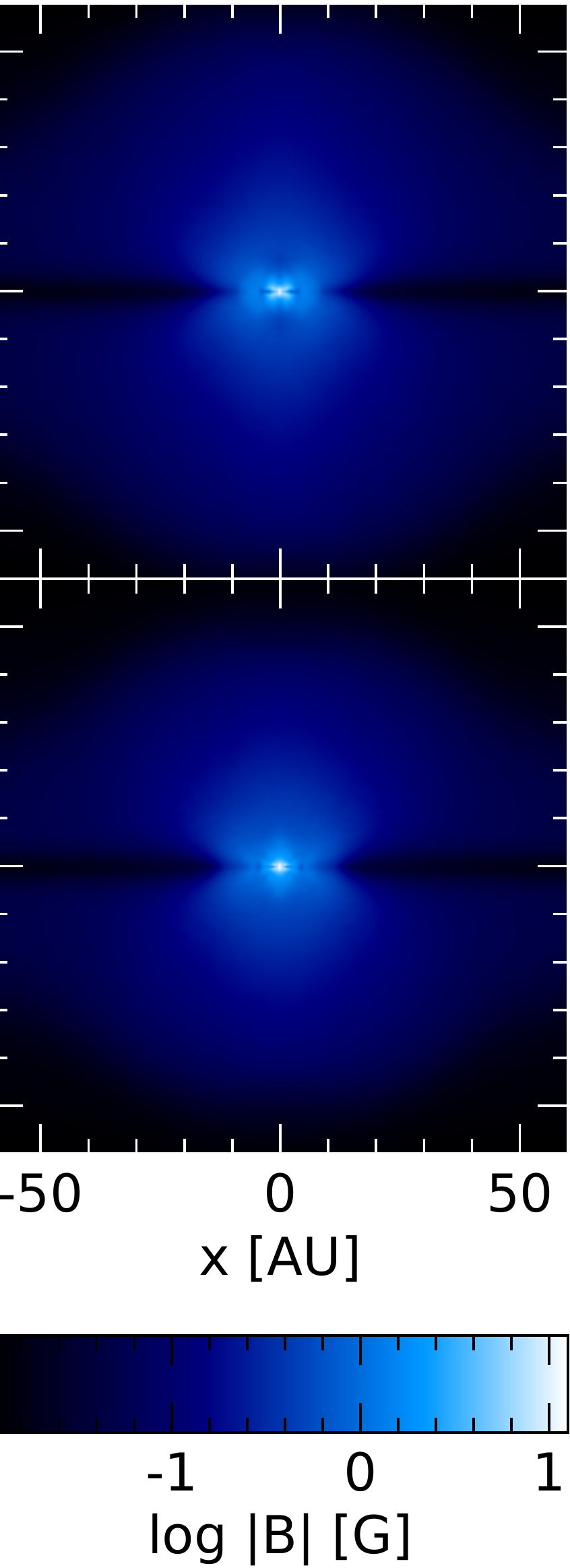}
\includegraphics[width=0.17\textwidth,trim={4cm 0 0 0},clip]{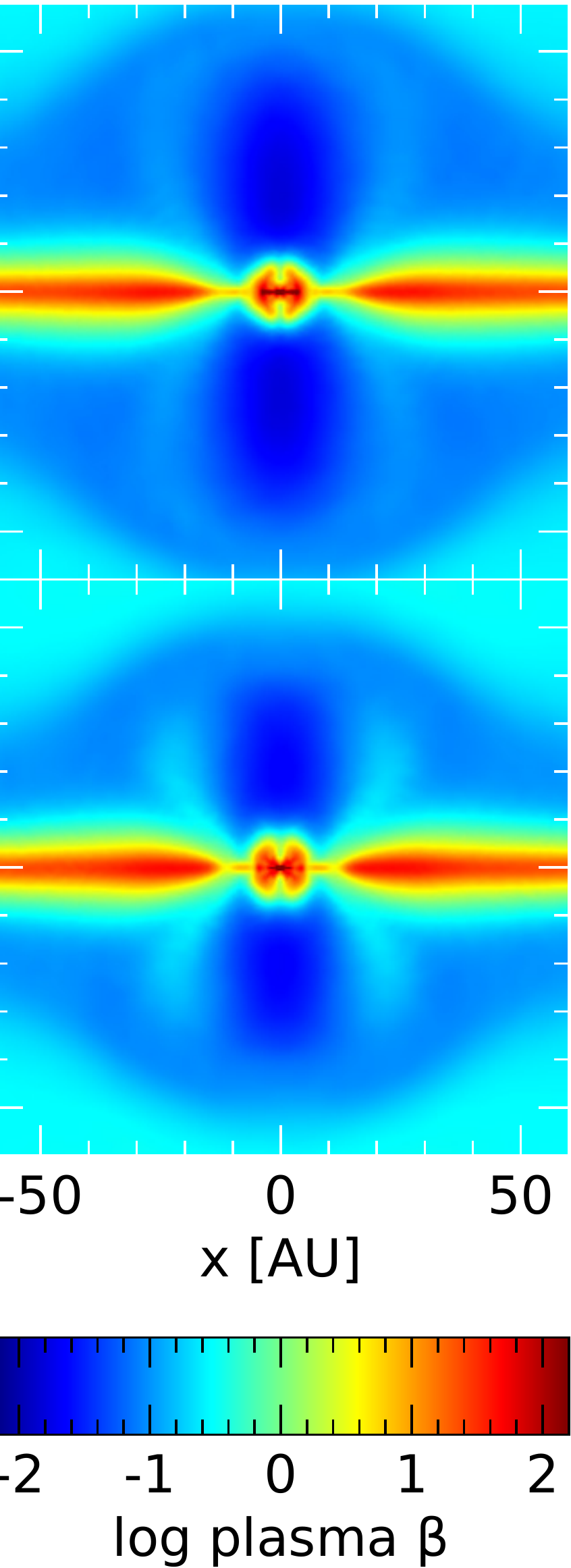}
\includegraphics[width=0.17\textwidth,trim={4cm 0 0 0},clip]{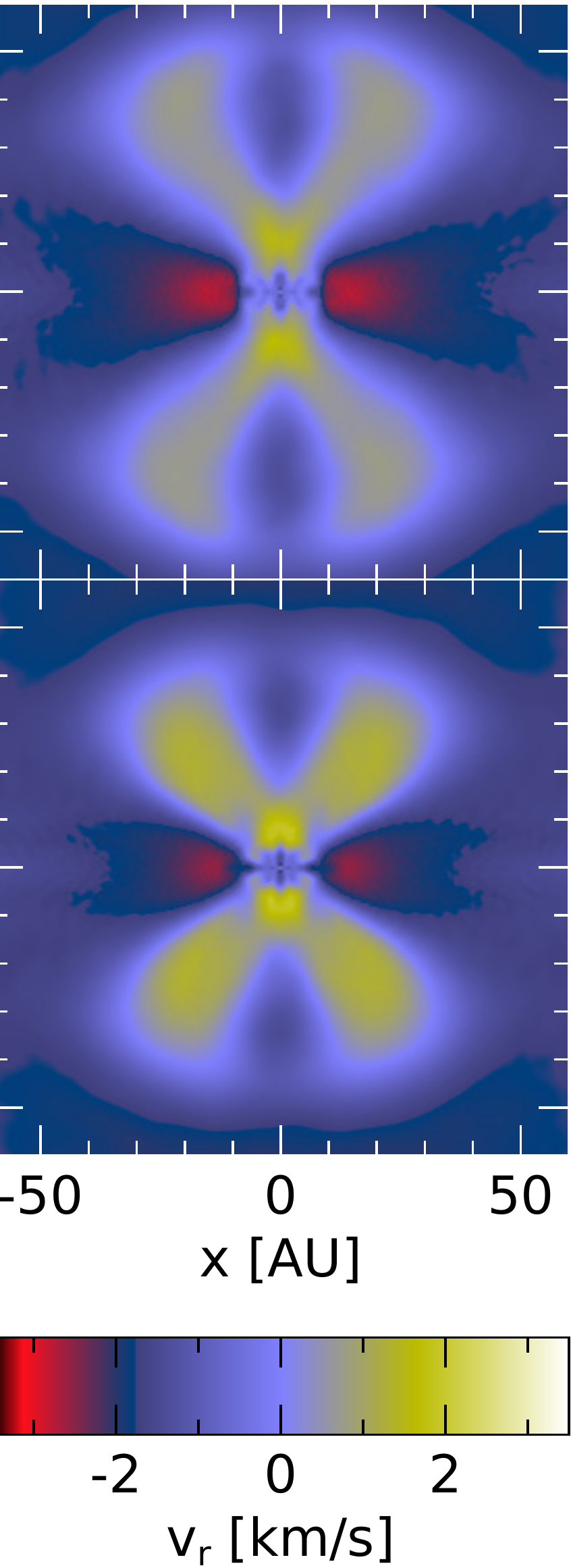}
\includegraphics[width=0.17\textwidth,trim={4cm 0 0 0},clip]{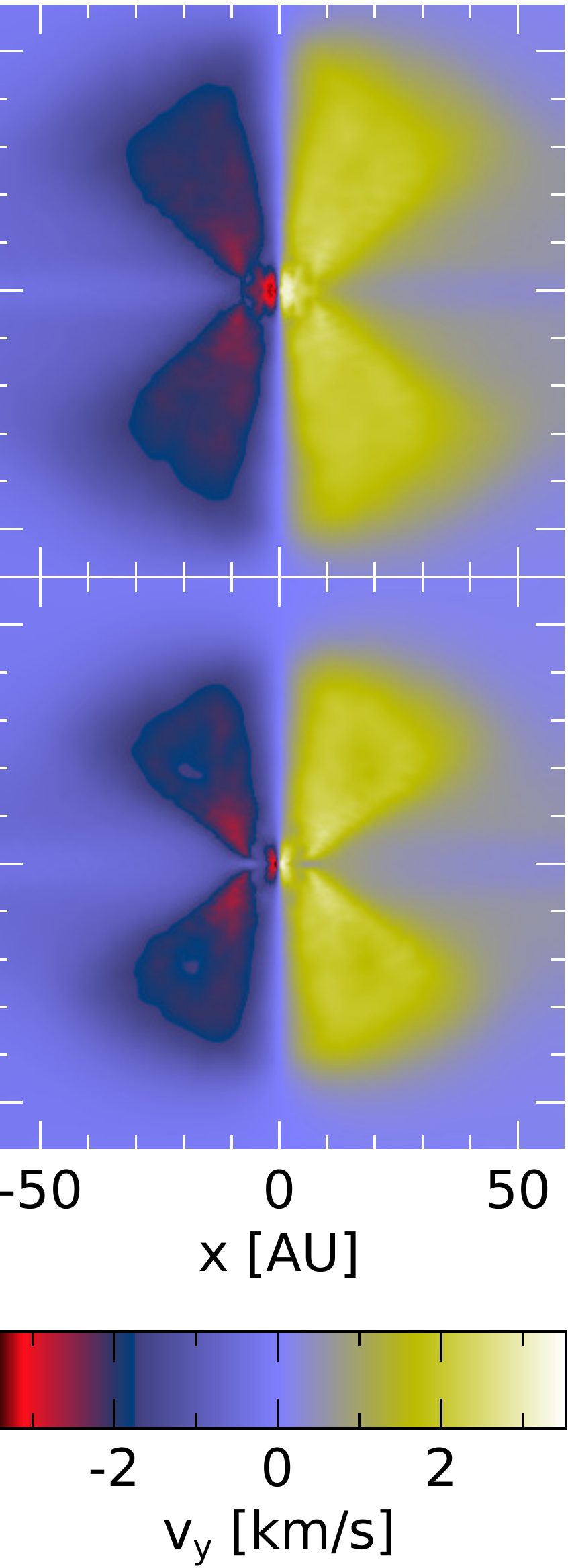}
\caption{Influence of the Hall effect on the outflows from the first core:  From left to right we provide cross sections of the gas density, magnetic field strength, plasma $\beta$, radial velocity and rotational velocity for the models with \zetaeq{16} at \rhoapprox{-7} for \zetams{16}{-} (top) and \zetams{16}{+} (bottom).  Colour scales differ from those in the Section~\ref{sec:fhc} for clarity.  The outflows are very similar except that in model \zetams{16}{+} the outflow has not had as long to propagate as in model \zetams{16}{-} and, thus, it is slightly smaller at \rhoapprox{-7}.}
\label{fig:hall:fhc}
\end{figure*} 

\subsubsection{The first hydrostatic core}

Fig.~\ref{fig:hall:rhoVtime} shows the evolution of the maximum density of the two models, which begins to diverge during the first collapse phase at \rhoapprox{-10}.  Model \zetams{16}{-} remains in the first collapse phase longer, reaching \rhoapprox{-7} 42~yr after \zetams{16}{+}.  Thus, from the point of view of the lifetime of first hydrostatic cores, setting $B_{0,z} > 0$ has a similar effect to increasing the initial ionisation rate.  

Fig.~\ref{fig:hall:fhc} shows the cross sections of the density, magnetic field strength, plasma $\beta$, radial velocity and rotational velocity for \zetams{16}{-} and \zetams{16}{+} at the end of the first hydrostatic core phase and beginning of the second collapse (\rhoapprox{-7}). At this density, both models have similar structures, although \zetams{16}{-} has more angular momentum and has had  additional evolution time so it has developed a more oblate first core and more extended outflows.  The midplane magnetic field strengths are similar for both models, however, the magnetic field strength is weaker in the envelope and stronger in the inner regions of \zetams{16}{+} than \zetams{16}{-}, and the magnetic tower is more magnetically dominated in \zetams{16}{-}. Thus, again, aligning the magnetic field with the rotation axis gives a similar result to increasing the ionisation rate (i.e.\ both result in larger central magnetic field strengths).

\begin{figure}
\centering
\includegraphics[width=\columnwidth]{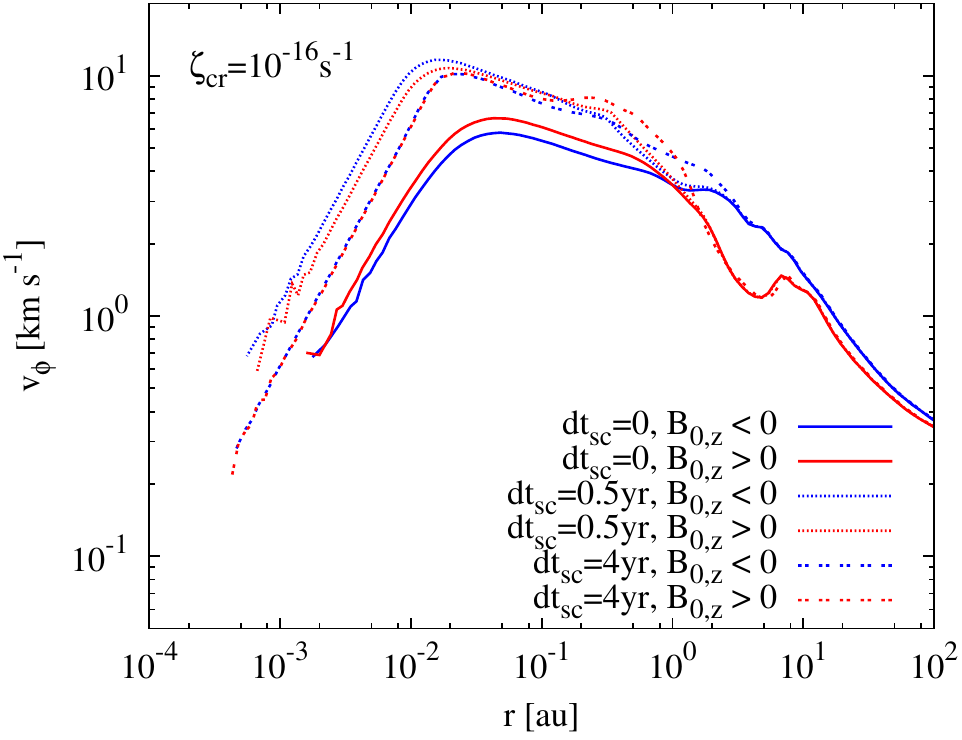}
\caption{The influence of the Hall effect on angular momentum: Azimuthally-averaged azimuthal velocity for the gas within 20$^\circ$ of the midplane at d$t_\text{sc}=0$ (solid), \dtscapprox{0.5} (dotted) and \dtscapprox{4} (short-dashed) for \zetams{16}{-} (blue) and \zetams{16}{+} (red).  The outer regions ($r \gtrsim 1$~au) do not evolve significantly between \rhoapprox{-7} and \dtscapprox{4}.  The decrease in $v_\phi$ for \zetams{16}{+} ($r \approx 1-10$~au) occurs in the transition region where the Hall coefficient switches sign.   Model \zetams{16}{+} has a slightly larger rotational velocity for $r \lesssim 1$~au until shortly after the formation of the stellar core, after which the rotational velocity is larger for \zetams{16}{-}.  As the evolution continues, the differences decreases, with both models having similar rotational profiles for $r \lesssim 0.1$~au by \dtscapprox{4}.}
\label{fig:hall:vphi}
\end{figure} 

\begin{figure*}
\centering
\includegraphics[width=0.25\textwidth]{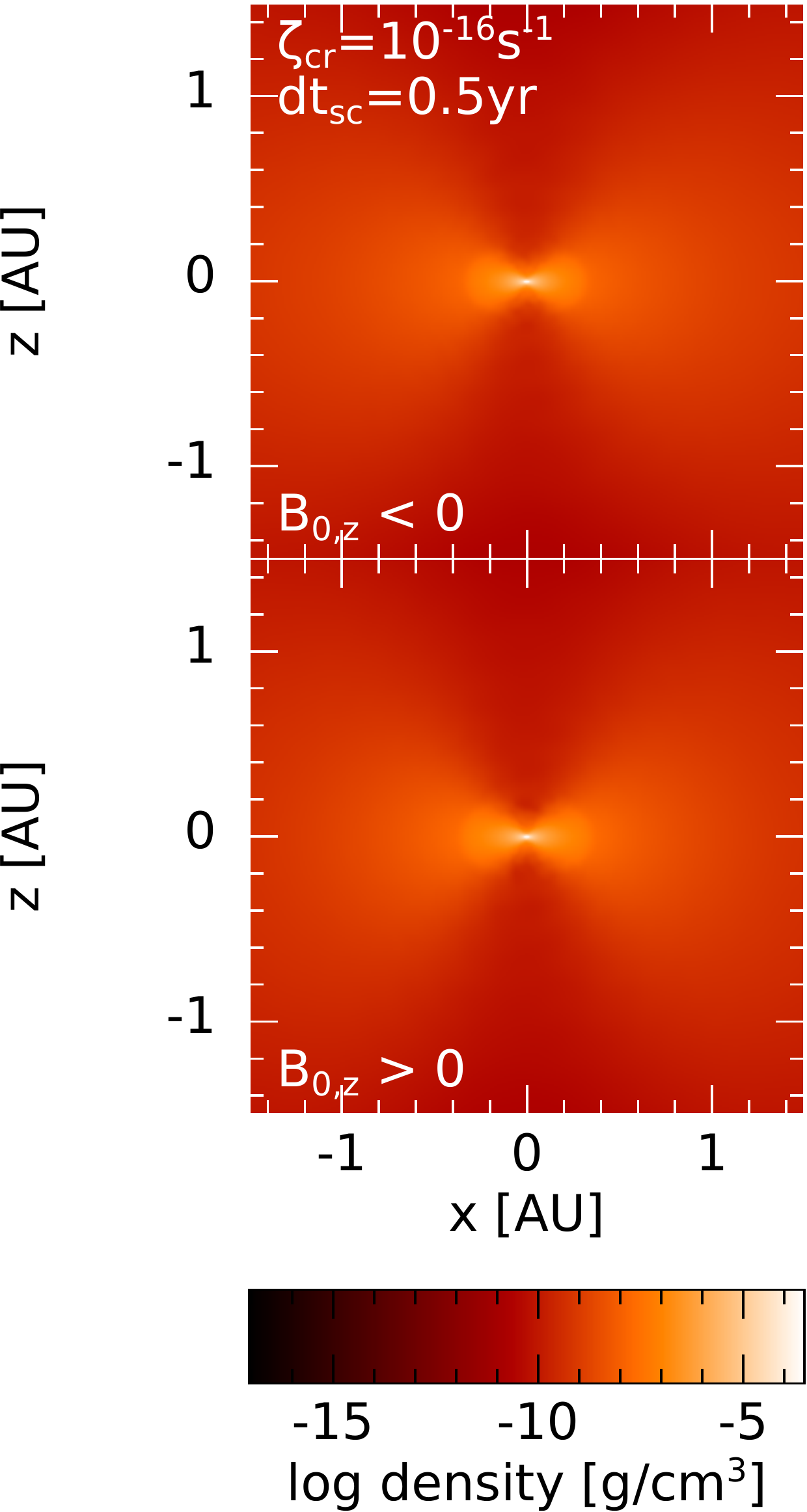}
\includegraphics[width=0.17\textwidth,trim={4cm 0 0 0},clip]{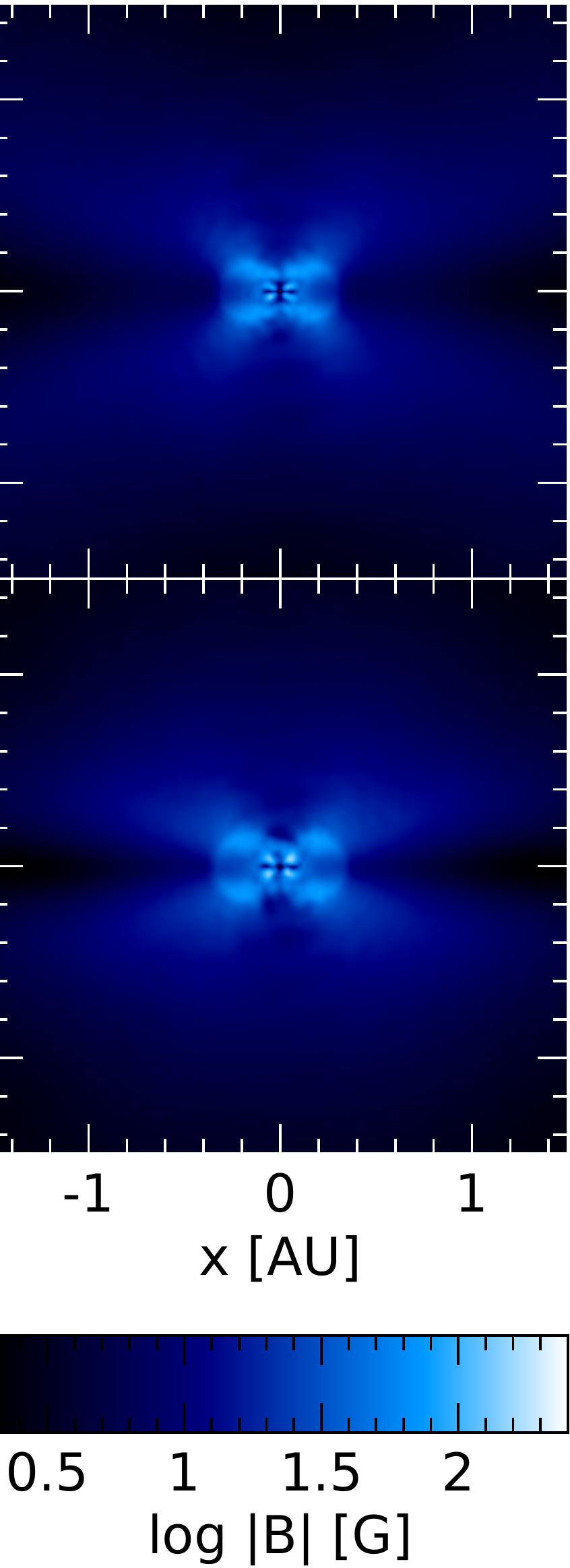}
\includegraphics[width=0.17\textwidth,trim={4cm 0 0 0},clip]{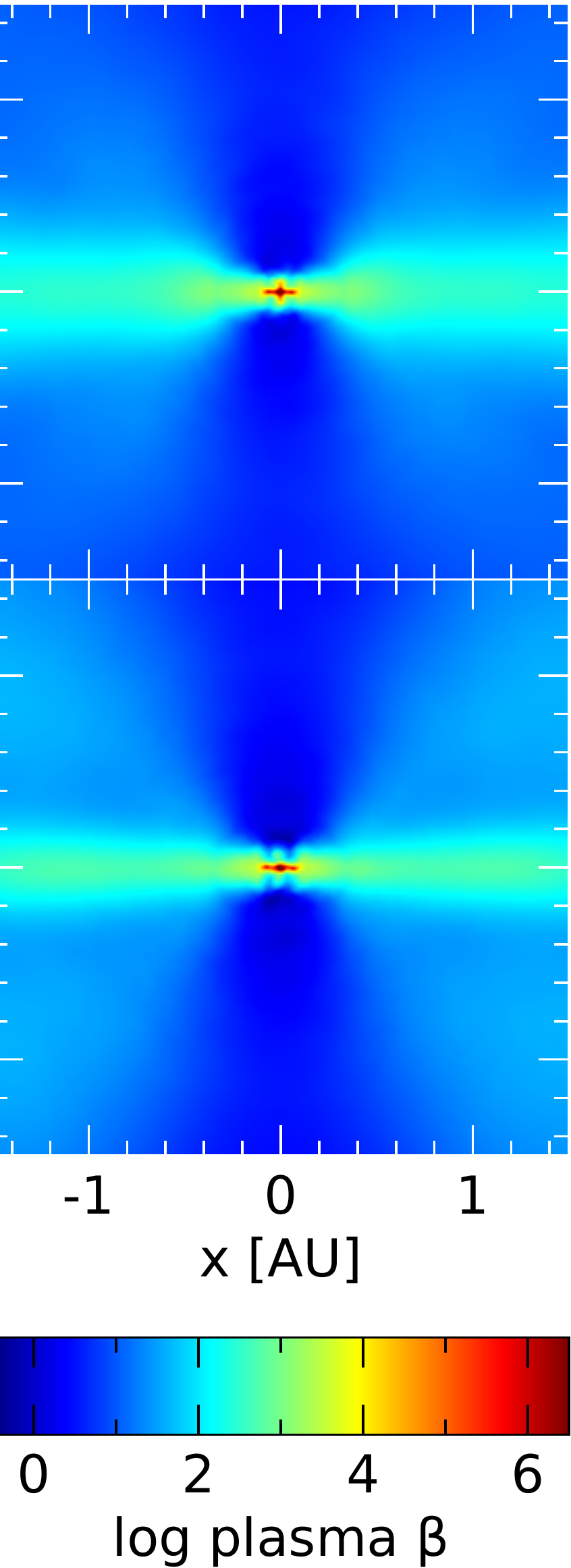}
\includegraphics[width=0.17\textwidth,trim={4cm 0 0 0},clip]{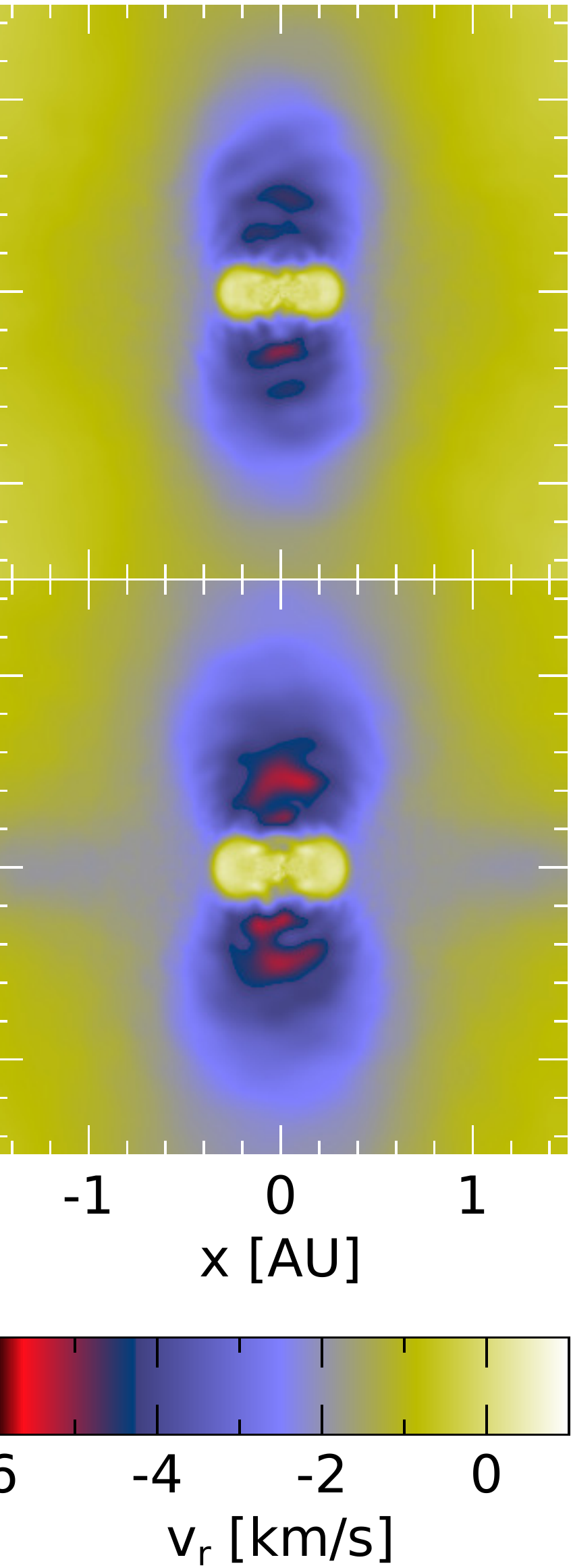}
\includegraphics[width=0.17\textwidth,trim={4cm 0 0 0},clip]{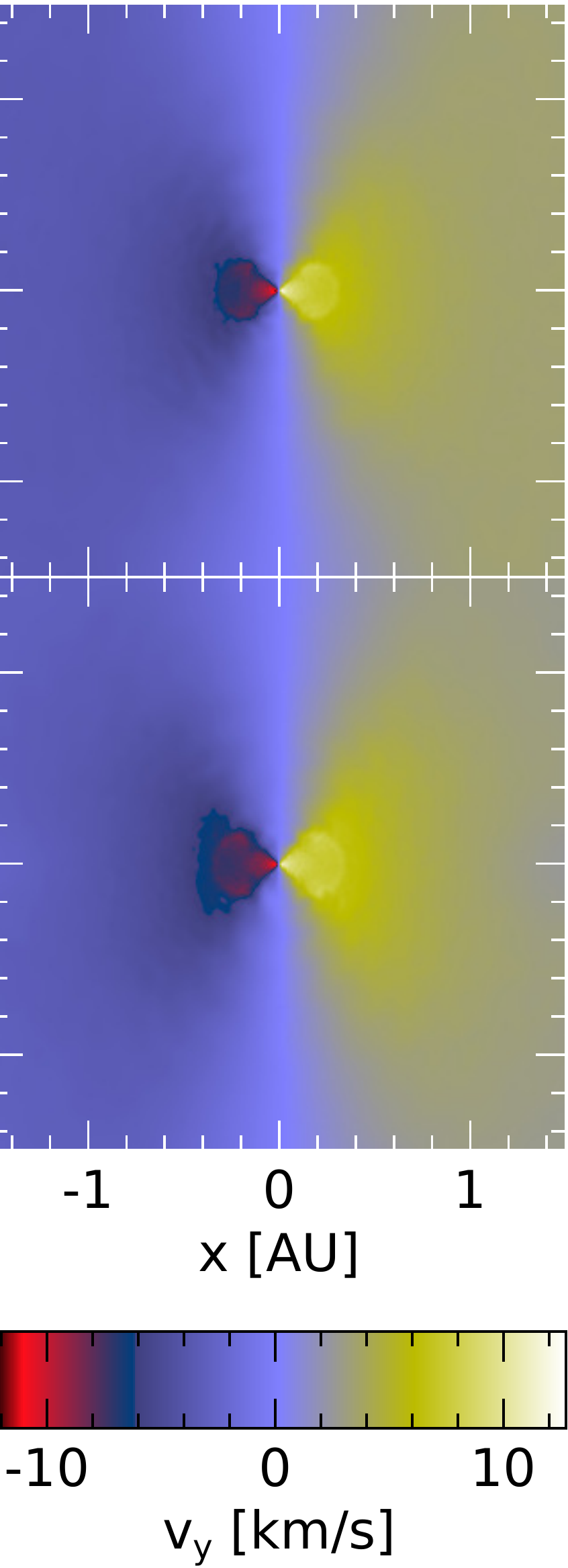}
\caption{Influence of the Hall effect on the structure in the vicinity of the stellar core:  From left to right are cross sections of the gas density, magnetic field strength, plasma $\beta$, radial velocity and rotational velocity for the models with \zetaeq{16} at \dtscapprox{0.5} for \zetams{16}{-} (top) and \zetams{16}{+} (bottom).  Colour scales differ from those in the Section~\ref{sec:shc} for clarity.   Model \zetams{16}{+} has a slightly more vertically extended circumstellar disc, but there are no outflows at this time on these spatial scales in either model (the radial velocities above and below the disc are negative).}
\label{fig:hall:shc}
\end{figure*} 

\subsubsection{The stellar core}

Fig.~\ref{fig:hall:shc} shows the cross sections of the density, magnetic field strength, plasma $\beta$,  radial velocity and azimuthal velocity at \dtscapprox{0.5}.  At this time, the stellar core in \zetams{16}{-} is rotating \sm20 per cent faster, is slightly more dense and has a weaker central magnetic field strength than \zetams{16}{+}.  Both models have similar disc scale heights but the vertical infall velocity is faster for \zetams{16}{+}.

By \dtscapprox{4}, the gas structure and flow around the core differs between the two models, which results in different stellar accretion rates.  Although this may have implications for the evolution of the spin rates of young stellar objects (YSOs; see review by \citealp{Bouvieretal2014}), our models end very early in the Class 0 phase and do not progress far enough for us to predict the long term effect of the Hall effect on the spin of YSOs.

For the duration of our simulations, the non-ideal MHD coefficients are similar for both \zetams{16}{-} and \zetams{16}{+}.  Thus, the Hall effect acts in the opposite sense for the two models.  

\subsubsection{Magnetic braking}
\label{res:hall:magbrake}

The model with $B_{0,z} < 0$ should have less magnetic braking since the Hall effect will induce a rotation in the same direction as the initial rotation of the cloud.  Indeed, the azimuthal speed, $v_\phi$ (see Fig.~\ref{fig:hall:vphi}), in \zetams{16}{+} is significantly lower than in \zetams{16}{-} at radii $1 \lesssim r/\text{au} \lesssim 7$; this decrease in azimuthal velocity is similar to \zetam{15} \ (see Fig.~\ref{fig:radialprofiles}).   This is the transition region where the Hall effect switches from negative to positive.  Since $\eta_\text{HE} > 0$ and  $B_{0,z} < 0$ for \zetam{15}, the Hall effect at $r \gtrsim 7$~au is contributing to the toroidal magnetic field in the same direction for \zetams{16}{+} and \zetam{15}, but the effect is stronger for \zetams{16}{+} due to its lower ionisation rate.  This decrease in rotational velocity directly leads to the faster radial infall and overall rate of evolution.

In the inner regions during the first core collapse, the rotating core of \zetams{16}{-} is slightly more diffuse, thus to conserve angular momentum, the larger core rotates slightly slower than \zetams{16}{+}.

As the stellar cores evolve, both continue to collapse and to spin up.  By \dtscapprox{0.5}, the core of \zetams{16}{-} is more dense and is rotating faster than \zetams{16}{+}.  Magnetic braking occurs in the core after this time to decrease the spin rate.  Since the magnetic field is stronger in the inner core of \zetams{16}{-} ($r < 0.01$~au) due to its higher density, this model undergoes more magnetic braking, thus by \dtscapprox{4}, the rotational profiles of the cores has converged, such that the azimuthal velocity for $r \lesssim 0.1$~au differs by less than three per cent (Fig.~\ref{fig:hall:vphi}).
\section{Conclusions}
\label{sec:conclusion}

We have presented a suite of radiation non-ideal magnetohydrodynamic simulations studying the collapse of a molecular cloud through the first and second core phases to stellar densities.  Our models were initialised as 1 M$_\odot$, spherically symmetric, rotating molecular cloud cores with magnetic field strengths such that the initial mass-to-flux ratio was $\mu_0=5$, corresponding to $B_0 = 1.63\times 10^{-4}$~G.  For most calculations, the magnetic field was initially anti-parallel to the rotation axis to maximise the influence of the Hall effect.

We included all three non-ideal MHD terms (ambipolar diffusion, Ohmic resistivity, and the Hall effect), with coefficients calculated by the \textsc{Nicil} library, and analysed 4 different cosmic ray ionisation rates, $\zeta_\text{cr}$.  At low densities and temperatures, the cosmic ray ionisation rate is primarily responsible for ionising the elements, whereas at high temperatures thermal ionisation dominates.  

We find that non-ideal MHD processes have significant effects during the star formation process.
Our key results are as follows:
\begin{enumerate}

\item Non-ideal MHD models with cosmic ray ionisation rates of \zetage{-12} yield results indistinguishable from ideal MHD. 

\item Non-ideal MHD yields longer-lived first hydrostatic cores and second core phases.

\item During the first hydrostatic core phase, the evolution of temperature with increasing density is similar for all models, but magnetic fields are weaker in models with lower ionisation rates.  

\item Large-scale outflows during the first hydrostatic core phase have similar conical morphologies on scales of tens of au, regardless of the level of ionisation; these outflows are slower and broader with lower ionisation rates.  With low ionisation rates there is also the development of an outflow from the poles of the first core that is not present with ideal MHD or high ionisation rates.  These outflows are magnetic tower flows in which the poloidal and toroidal components of the magnetic field have comparable strengths.

\item In contrast to the mild dependence of the first core outflow on the ionisation rate, the structure of the outflows on au-scales varies strongly with the ionisation rate.  With ideal MHD or high ionisation rates, a fast ($\approx 14$~km~s$^{-1}$) collimated outflow is launched from the vicinity of the stellar core (scales $<0.1$~au) immediately after its formation.  However, with the lowest ionisation rate (\zetaeq{-16}) there is no outflow from the vicinity of the stellar core soon after its formation.  Instead, a small circumstellar disc is formed and a slower ($\approx 3-4$~km~s$^{-1}$) conical outflow develops from au-scales that merges into the larger outflow from the first core.

\item At first core formation, the magnetic field strengths are independent of the ionisation rate, but by the end of the first core phase the field strength is an order of magnitude lower with \zetaeq{-16} compared to the ideal MHD or \zetaeq{-12} models.  The maximum field strengths are attained at the formation of the stellar core vary from $B_\text{max} \approx 10^5$ to $5\times10^3$ G for the models with $\zeta_\text{cr} = 10^{-12}$ and $10^{-16}$ s$^{-1}$, respectively.  

\item Due the Hall effect, changing the direction of the initial magnetic field to be aligned with the axis of rotation decreases the lifetime of the first hydrostatic core phase.  The change from $B_\text{0,z} > 0$ to $B_\text{0,z} < 0$ results in a faster spinning stellar core, however, the overall morphologies are relatively unaffected by the initial direction of the magnetic field.

\end{enumerate}

\section*{Acknowledgements}

JW and MRB acknowledge support from the European Research Council under the European Commission's Seventh Framework Programme (FP7/2007- 2013 grant agreement no. 339248).  DJP and JW were funded by Australian Research Council grants FT130100034 and DP130102078.  The calculations for this paper were performed on the University of Exeter Supercomputer, a DiRAC Facility jointly funded by STFC, the Large Facilities Capital Fund of BIS, and the University of Exeter.  We used {\sc splash} \citep{Price2007} for the column density figures. 

\bibliography{NIcollapse.bib}

\begin{thebibliography}{}
\makeatletter
\relax
\def\mn@urlcharsother{\let\do\@makeother \do\$\do\&\do\#\do\^\do\_\do\%\do\~}
\def\mn@doi{\begingroup\mn@urlcharsother \@ifnextchar [ {\mn@doi@}
  {\mn@doi@[]}}
\def\mn@doi@[#1]#2{\def\@tempa{#1}\ifx\@tempa\@empty \href
  {http://dx.doi.org/#2} {doi:#2}\else \href {http://dx.doi.org/#2} {#1}\fi
  \endgroup}
\def\mn@eprint#1#2{\mn@eprint@#1:#2::\@nil}
\def\mn@eprint@arXiv#1{\href {http://arxiv.org/abs/#1} {{\tt arXiv:#1}}}
\def\mn@eprint@dblp#1{\href {http://dblp.uni-trier.de/rec/bibtex/#1.xml}
  {dblp:#1}}
\def\mn@eprint@#1:#2:#3:#4\@nil{\def\@tempa {#1}\def\@tempb {#2}\def\@tempc
  {#3}\ifx \@tempc \@empty \let \@tempc \@tempb \let \@tempb \@tempa \fi \ifx
  \@tempb \@empty \def\@tempb {arXiv}\fi \@ifundefined
  {mn@eprint@\@tempb}{\@tempb:\@tempc}{\expandafter \expandafter \csname
  mn@eprint@\@tempb\endcsname \expandafter{\@tempc}}}

\bibitem[\protect\citeauthoryear{{Alexander}}{{Alexander}}{1975}]{Alexander1975}
{Alexander} D.~R.,  1975, \apjs, \href
  {http://adsabs.harvard.edu/abs/1975ApJS...29..363A} {29, 363}

\bibitem[\protect\citeauthoryear{{Alexiades}, {Amiez}  \&
  {Gremaud}}{{Alexiades} et~al.}{1996}]{AleAmiGre1996}
{Alexiades} V.,  {Amiez} G.,   {Gremaud} P.-A.,  1996, \mn@doi [Commun. Numer.
  Meth. Eng.] {10.1002/(SICI)1099-0887(199601)12:1<31::AID-CNM950>3.0.CO;2-5},
  12, 31

\bibitem[\protect\citeauthoryear{{Allen}, {Li}  \& {Shu}}{{Allen}
  et~al.}{2003}]{AllLiShu2003}
{Allen} A.,  {Li} Z.-Y.,   {Shu} F.~H.,  2003, \mn@doi [\apj] {10.1086/379243},
  \href {http://adsabs.harvard.edu/abs/2003ApJ...599..363A} {599, 363}

\bibitem[\protect\citeauthoryear{{Asplund}, {Grevesse}, {Sauval}  \&
  {Scott}}{{Asplund} et~al.}{2009}]{AsplundEtAl2009}
{Asplund} M.,  {Grevesse} N.,  {Sauval} A.~J.,   {Scott} P.,  2009, \mn@doi
  [\araa] {10.1146/annurev.astro.46.060407.145222}, \href
  {http://adsabs.harvard.edu/abs/2009ARA%26A..47..481A} {47, 481}

\bibitem[\protect\citeauthoryear{{Banerjee} \& {Pudritz}}{{Banerjee} \&
  {Pudritz}}{2006}]{BanPud2006}
{Banerjee} R.,  {Pudritz} R.~E.,  2006, \mn@doi [\apj] {10.1086/500496}, \href
  {http://adsabs.harvard.edu/abs/2006ApJ...641..949B} {641, 949}

\bibitem[\protect\citeauthoryear{{Basu} \& {Mouschovias}}{{Basu} \&
  {Mouschovias}}{1994}]{BasMou1994}
{Basu} S.,  {Mouschovias} T.~C.,  1994, \mn@doi [\apj] {10.1086/174611}, \href
  {http://adsabs.harvard.edu/abs/1994ApJ...432..720B} {432, 720}

\bibitem[\protect\citeauthoryear{{Bate}}{{Bate}}{1998}]{Bate1998}
{Bate} M.~R.,  1998, \mn@doi [\apjl] {10.1086/311719}, \href
  {http://adsabs.harvard.edu/abs/1998ApJ...508L..95B} {508, L95}

\bibitem[\protect\citeauthoryear{{Bate}}{{Bate}}{2010}]{Bate2010}
{Bate} M.~R.,  2010, \mn@doi [\mnras] {10.1111/j.1745-3933.2010.00839.x}, \href
  {http://cdsads.u-strasbg.fr/abs/2010MNRAS.404L..79B} {404, L79}

\bibitem[\protect\citeauthoryear{{Bate}}{{Bate}}{2011}]{Bate2011}
{Bate} M.~R.,  2011, \mn@doi [\mnras] {10.1111/j.1365-2966.2011.19386.x}, \href
  {http://adsabs.harvard.edu/abs/2011MNRAS.417.2036B} {417, 2036}

\bibitem[\protect\citeauthoryear{{Bate} \& {Burkert}}{{Bate} \&
  {Burkert}}{1997}]{BatBur1997}
{Bate} M.~R.,  {Burkert} A.,  1997, \mnras, \href
  {http://adsabs.harvard.edu/abs/1997MNRAS.288.1060B} {288, 1060}

\bibitem[\protect\citeauthoryear{{Bate}, {Bonnell}  \& {Price}}{{Bate}
  et~al.}{1995}]{BatBonPri1995}
{Bate} M.~R.,  {Bonnell} I.~A.,   {Price} N.~M.,  1995, MNRAS, \href
  {http://adsabs.harvard.edu/abs/1995MNRAS.277..362B} {277, 362}

\bibitem[\protect\citeauthoryear{{Bate}, {Tricco}  \& {Price}}{{Bate}
  et~al.}{2014}]{BatTriPri2014}
{Bate} M.~R.,  {Tricco} T.~S.,   {Price} D.~J.,  2014, \mn@doi [\mnras]
  {10.1093/mnras/stt1865}, \href
  {http://cdsads.u-strasbg.fr/abs/2014MNRAS.437...77B} {437, 77}

\bibitem[\protect\citeauthoryear{{Benz}}{{Benz}}{1990}]{Benz1990}
{Benz} W.,  1990, in {Buchler} J.~R.,  ed., Numerical Modelling of Nonlinear
  Stellar Pulsations Problems and Prospects.. Kluwer, Dordrecht, p.~269

\bibitem[\protect\citeauthoryear{{Boley}, {Hartquist}, {Durisen}  \&
  {Michael}}{{Boley} et~al.}{2007}]{Boleyetal2007}
{Boley} A.~C.,  {Hartquist} T.~W.,  {Durisen} R.~H.,   {Michael} S.,  2007,
  \mn@doi [\apjl] {10.1086/512235}, \href
  {http://adsabs.harvard.edu/abs/2007ApJ...656L..89B} {656, L89}

\bibitem[\protect\citeauthoryear{{B{\o}rve}, {Omang}  \& {Trulsen}}{{B{\o}rve}
  et~al.}{2001}]{BorOmaTru2001}
{B{\o}rve} S.,  {Omang} M.,   {Trulsen} J.,  2001, \mn@doi [\apj]
  {10.1086/323228}, \href {http://adsabs.harvard.edu/abs/2001ApJ...561...82B}
  {561, 82}

\bibitem[\protect\citeauthoryear{{Bouvier}, {Matt}, {Mohanty}, {Scholz},
  {Stassun}  \& {Zanni}}{{Bouvier} et~al.}{2014}]{Bouvieretal2014}
{Bouvier} J.,  {Matt} S.~P.,  {Mohanty} S.,  {Scholz} A.,  {Stassun} K.~G.,
  {Zanni} C.,  2014, \mn@doi [Protostars and Planets VI]
  {10.2458/azu_uapress_9780816531240-ch019}, \href
  {http://adsabs.harvard.edu/abs/2014prpl.conf..433B} {pp 433--450}

\bibitem[\protect\citeauthoryear{{Braiding} \& {Wardle}}{{Braiding} \&
  {Wardle}}{2012}]{BraWar2012}
{Braiding} C.~R.,  {Wardle} M.,  2012, \mn@doi [\mnras]
  {10.1111/j.1365-2966.2012.22001.x}, \href
  {http://adsabs.harvard.edu/abs/2012MNRAS.427.3188B} {427, 3188}

\bibitem[\protect\citeauthoryear{{Brookshaw}}{{Brookshaw}}{1985}]{Brookshaw1985}
{Brookshaw} L.,  1985, PASA, \href
  {http://adsabs.harvard.edu/abs/1985PASAu...6..207B} {6, 207}

\bibitem[\protect\citeauthoryear{{B{\"u}rzle}, {Clark}, {Stasyszyn}, {Dolag}
  \& {Klessen}}{{B{\"u}rzle} et~al.}{2011}]{Burzleetal2011b}
{B{\"u}rzle} F.,  {Clark} P.~C.,  {Stasyszyn} F.,  {Dolag} K.,   {Klessen}
  R.~S.,  2011, \mn@doi [\mnras] {10.1111/j.1745-3933.2011.01120.x}, \href
  {http://adsabs.harvard.edu/abs/2011MNRAS.417L..61B} {417, L61}

\bibitem[\protect\citeauthoryear{{Cleary} \& {Monaghan}}{{Cleary} \&
  {Monaghan}}{1999}]{CleMon1999}
{Cleary} P.~W.,  {Monaghan} J.~J.,  1999, \mn@doi [J. Comp. Phys.]
  {10.1006/jcph.1998.6118}, \href
  {http://adsabs.harvard.edu/abs/1999JCoPh.148..227C} {148, 227}

\bibitem[\protect\citeauthoryear{{Commer{\c c}on}, {Hennebelle}, {Audit},
  {Chabrier}  \& {Teyssier}}{{Commer{\c c}on} et~al.}{2010}]{Commerconetal2010}
{Commer{\c c}on} B.,  {Hennebelle} P.,  {Audit} E.,  {Chabrier} G.,
  {Teyssier} R.,  2010, \mn@doi [\aap] {10.1051/0004-6361/200913597}, \href
  {http://adsabs.harvard.edu/abs/2010A%26A...510L...3C} {510, L3+}

\bibitem[\protect\citeauthoryear{{Commer{\c c}on}, {Audit}, {Chabrier}  \&
  {Chi{\`e}ze}}{{Commer{\c c}on} et~al.}{2011}]{Commerconetal2011b}
{Commer{\c c}on} B.,  {Audit} E.,  {Chabrier} G.,   {Chi{\`e}ze} J.-P.,  2011,
  \mn@doi [\aap] {10.1051/0004-6361/201016213}, \href
  {http://adsabs.harvard.edu/abs/2011A%26A...530A..13C} {530, A13+}

\bibitem[\protect\citeauthoryear{{Commer{\c c}on}, {Launhardt}, {Dullemond}  \&
  {Henning}}{{Commer{\c c}on} et~al.}{2012}]{Commerconetal2012}
{Commer{\c c}on} B.,  {Launhardt} R.,  {Dullemond} C.,   {Henning} T.,  2012,
  \mn@doi [\aap] {10.1051/0004-6361/201118706}, \href
  {http://adsabs.harvard.edu/abs/2012A%26A...545A..98C} {545, A98}

\bibitem[\protect\citeauthoryear{{Dedner}, {Kemm}, {Kr{\"o}ner}, {Munz},
  {Schnitzer}  \& {Wesenberg}}{{Dedner} et~al.}{2002}]{Dedneretal2002}
{Dedner} A.,  {Kemm} F.,  {Kr{\"o}ner} D.,  {Munz} C.-D.,  {Schnitzer} T.,
  {Wesenberg} M.,  2002, \mn@doi [Journal of Computational Physics]
  {10.1006/jcph.2001.6961}, \href
  {http://adsabs.harvard.edu/abs/2002JCoPh.175..645D} {175, 645}

\bibitem[\protect\citeauthoryear{{Fehlberg}}{{Fehlberg}}{1969}]{Fehlberg1969}
{Fehlberg} E.,  1969, NASA Technical Report R-315

\bibitem[\protect\citeauthoryear{{Heiles} \& {Crutcher}}{{Heiles} \&
  {Crutcher}}{2005}]{HeiCru2005}
{Heiles} C.,  {Crutcher} R.,  2005, in {Wielebinski} R.,  {Beck} R.,  eds,
  Lecture Notes in Physics, Berlin Springer Verlag Vol. 664, Cosmic Magnetic
  Fields. p.~137 (\mn@eprint {} {astro-ph/0501550}),
  \mn@doi{10.1007/11369875_7}

\bibitem[\protect\citeauthoryear{{Hennebelle} \& {Fromang}}{{Hennebelle} \&
  {Fromang}}{2008}]{HenFro2008}
{Hennebelle} P.,  {Fromang} S.,  2008, \mn@doi [\aap]
  {10.1051/0004-6361:20078309}, \href
  {http://adsabs.harvard.edu/abs/2008A%26A...477....9H} {477, 9}

\bibitem[\protect\citeauthoryear{{Kato}, {Mineshige}  \& {Shibata}}{{Kato}
  et~al.}{2004}]{KatMinShi2004}
{Kato} Y.,  {Mineshige} S.,   {Shibata} K.,  2004, \mn@doi [\apj]
  {10.1086/381234}, \href {http://adsabs.harvard.edu/abs/2004ApJ...605..307K}
  {605, 307}

\bibitem[\protect\citeauthoryear{{Keith} \& {Wardle}}{{Keith} \&
  {Wardle}}{2014}]{KeiWar2014}
{Keith} S.~L.,  {Wardle} M.,  2014, \mn@doi [\mnras] {10.1093/mnras/stu245},
  \href {http://adsabs.harvard.edu/abs/2014MNRAS.440...89K} {440, 89}

\bibitem[\protect\citeauthoryear{{Krasnopolsky}, {Li}  \&
  {Shang}}{{Krasnopolsky} et~al.}{2011}]{KraLiSha011}
{Krasnopolsky} R.,  {Li} Z.-Y.,   {Shang} H.,  2011, \mn@doi [\apj]
  {10.1088/0004-637X/733/1/54}, \href
  {http://adsabs.harvard.edu/abs/2011ApJ...733...54K} {733, 54}

\bibitem[\protect\citeauthoryear{{Larson}}{{Larson}}{1969}]{Larson1969}
{Larson} R.~B.,  1969, \mnras, \href
  {http://adsabs.harvard.edu/abs/1969MNRAS.145..271L} {145, 271}

\bibitem[\protect\citeauthoryear{{Larson}}{{Larson}}{1972}]{Larson1972}
{Larson} R.~B.,  1972, \mnras, \href
  {http://adsabs.harvard.edu/abs/1972MNRAS.156..437L} {156, 437}

\bibitem[\protect\citeauthoryear{{Li}, {Krasnopolsky}  \& {Shang}}{{Li}
  et~al.}{2011}]{LiKraSha2011}
{Li} Z.-Y.,  {Krasnopolsky} R.,   {Shang} H.,  2011, \mn@doi [\apj]
  {10.1088/0004-637X/738/2/180}, \href
  {http://adsabs.harvard.edu/abs/2011ApJ...738..180L} {738, 180}

\bibitem[\protect\citeauthoryear{{Lynden-Bell}}{{Lynden-Bell}}{2003}]{Lyndenbell2003}
{Lynden-Bell} D.,  2003, \mn@doi [\mnras] {10.1046/j.1365-8711.2003.06506.x},
  \href {http://adsabs.harvard.edu/abs/2003MNRAS.341.1360L} {341, 1360}

\bibitem[\protect\citeauthoryear{{Mac Low} \& {Klessen}}{{Mac Low} \&
  {Klessen}}{2004}]{MacKle2004}
{Mac Low} M.-M.,  {Klessen} R.~S.,  2004, \mn@doi [Reviews of Modern Physics]
  {10.1103/RevModPhys.76.125}, \href
  {http://adsabs.harvard.edu/abs/2004RvMP...76..125M} {76, 125}

\bibitem[\protect\citeauthoryear{{Mac Low}, {Norman}, {Konigl}  \&
  {Wardle}}{{Mac Low} et~al.}{1995}]{MacNorKonWar1995}
{Mac Low} M.-M.,  {Norman} M.~L.,  {Konigl} A.,   {Wardle} M.,  1995, \mn@doi
  [\apj] {10.1086/175477}, \href
  {http://adsabs.harvard.edu/abs/1995ApJ...442..726M} {442, 726}

\bibitem[\protect\citeauthoryear{{Machida}, {Matsumoto}, {Hanawa}  \&
  {Tomisaka}}{{Machida} et~al.}{2005}]{Machidaetal2005}
{Machida} M.~N.,  {Matsumoto} T.,  {Hanawa} T.,   {Tomisaka} K.,  2005, \mn@doi
  [\mnras] {10.1111/j.1365-2966.2005.09327.x}, \href
  {http://adsabs.harvard.edu/abs/2005MNRAS.362..382M} {362, 382}

\bibitem[\protect\citeauthoryear{{Machida}, {Inutsuka}  \&
  {Matsumoto}}{{Machida} et~al.}{2006}]{MacInuMat2006}
{Machida} M.~N.,  {Inutsuka} S.,   {Matsumoto} T.,  2006, \mn@doi [\apjl]
  {10.1086/507179}, \href {http://adsabs.harvard.edu/abs/2006ApJ...647L.151M}
  {647, L151}

\bibitem[\protect\citeauthoryear{{Machida}, {Inutsuka}  \&
  {Matsumoto}}{{Machida} et~al.}{2008}]{MacInuMat2008}
{Machida} M.~N.,  {Inutsuka} S.-i.,   {Matsumoto} T.,  2008, \mn@doi [\apj]
  {10.1086/528364}, \href {http://adsabs.harvard.edu/abs/2008ApJ...676.1088M}
  {676, 1088}

\bibitem[\protect\citeauthoryear{{Machida}, {Inutsuka}  \&
  {Matsumoto}}{{Machida} et~al.}{2010}]{MacInuMat2010}
{Machida} M.~N.,  {Inutsuka} S.-i.,   {Matsumoto} T.,  2010, \mn@doi [\apj]
  {10.1088/0004-637X/724/2/1006}, \href
  {http://adsabs.harvard.edu/abs/2010ApJ...724.1006M} {724, 1006}

\bibitem[\protect\citeauthoryear{{Marchand}, {Masson}, {Chabrier},
  {Hennebelle}, {Commer{\c c}on}  \& {Vaytet}}{{Marchand}
  et~al.}{2016}]{MarchandEtAl2016}
{Marchand} P.,  {Masson} J.,  {Chabrier} G.,  {Hennebelle} P.,  {Commer{\c
  c}on} B.,   {Vaytet} N.,  2016, \mn@doi [\aap] {10.1051/0004-6361/201526780},
  \href {http://adsabs.harvard.edu/abs/2016A%26A...592A..18M} {592, A18}

\bibitem[\protect\citeauthoryear{{Masunaga} \& {Inutsuka}}{{Masunaga} \&
  {Inutsuka}}{1999}]{MasInu1999}
{Masunaga} H.,  {Inutsuka} S.-I.,  1999, \mn@doi [\apj] {10.1086/306608}, \href
  {http://adsabs.harvard.edu/abs/1999ApJ...510..822M} {510, 822}

\bibitem[\protect\citeauthoryear{{Masunaga} \& {Inutsuka}}{{Masunaga} \&
  {Inutsuka}}{2000}]{MasInu2000}
{Masunaga} H.,  {Inutsuka} S.-I.,  2000, \mn@doi [\apj] {10.1086/308439}, \href
  {http://adsabs.harvard.edu/abs/2000ApJ...531..350M} {531, 350}

\bibitem[\protect\citeauthoryear{{Mellon} \& {Li}}{{Mellon} \&
  {Li}}{2008}]{MelLi2008}
{Mellon} R.~R.,  {Li} Z.-Y.,  2008, \mn@doi [\apj] {10.1086/587542}, \href
  {http://adsabs.harvard.edu/abs/2008ApJ...681.1356M} {681, 1356}

\bibitem[\protect\citeauthoryear{{Mestel}}{{Mestel}}{1999}]{Mestel1999}
{Mestel} L.,  1999, {Stellar magnetism}.
Clarendon, Oxford

\bibitem[\protect\citeauthoryear{{Morris} \& {Monaghan}}{{Morris} \&
  {Monaghan}}{1997}]{MorMon1997}
{Morris} J.~P.,  {Monaghan} J.~J.,  1997, J.\ Comp.\ Phys., 136, 41

\bibitem[\protect\citeauthoryear{{Mouschovias} \& {Spitzer}}{{Mouschovias} \&
  {Spitzer}}{1976}]{MouSpi1976}
{Mouschovias} T.~C.,  {Spitzer} Jr. L.,  1976, \mn@doi [\apj] {10.1086/154835},
  \href {http://adsabs.harvard.edu/abs/1976ApJ...210..326M} {210, 326}

\bibitem[\protect\citeauthoryear{{Pollack}, {McKay}  \&
  {Christofferson}}{{Pollack} et~al.}{1985}]{PolMcKChr1985}
{Pollack} J.~B.,  {McKay} C.~P.,   {Christofferson} B.~M.,  1985, \mn@doi
  [Icarus] {10.1016/0019-1035(85)90069-7}, \href
  {http://adsabs.harvard.edu/abs/1985Icar...64..471P} {64, 471}

\bibitem[\protect\citeauthoryear{{Pollack}, {Hollenbach}, {Beckwith},
  {Simonelli}, {Roush}  \& {Fong}}{{Pollack} et~al.}{1994}]{PollackEtAl1994}
{Pollack} J.~B.,  {Hollenbach} D.,  {Beckwith} S.,  {Simonelli} D.~P.,  {Roush}
  T.,   {Fong} W.,  1994, \mn@doi [\apj] {10.1086/173677}, \href
  {http://adsabs.harvard.edu/abs/1994ApJ...421..615P} {421, 615}

\bibitem[\protect\citeauthoryear{{Price}}{{Price}}{2007}]{Price2007}
{Price} D.~J.,  2007, \mn@doi [Publ. Astron. Soc. Australia] {10.1071/AS07022},
  \href {http://adsabs.harvard.edu/abs/2007PASA...24..159P} {24, 159}

\bibitem[\protect\citeauthoryear{{Price}}{{Price}}{2012}]{Price2012}
{Price} D.~J.,  2012, \mn@doi [Journal of Computational Physics]
  {10.1016/j.jcp.2010.12.011}, \href
  {http://adsabs.harvard.edu/abs/2012JCoPh.231..759P} {231, 759}

\bibitem[\protect\citeauthoryear{{Price} \& {Bate}}{{Price} \&
  {Bate}}{2007}]{PriBat2007}
{Price} D.~J.,  {Bate} M.~R.,  2007, \mn@doi [\mnras]
  {10.1111/j.1365-2966.2007.11621.x}, \href
  {http://adsabs.harvard.edu/abs/2007MNRAS.377...77P} {377, 77}

\bibitem[\protect\citeauthoryear{{Price} \& {Monaghan}}{{Price} \&
  {Monaghan}}{2004}]{PriMon2004b}
{Price} D.~J.,  {Monaghan} J.~J.,  2004, \mn@doi [\mnras]
  {10.1111/j.1365-2966.2004.07346.x}, \href
  {http://adsabs.harvard.edu/abs/2004MNRAS.348..139P} {348, 139}

\bibitem[\protect\citeauthoryear{{Price} \& {Monaghan}}{{Price} \&
  {Monaghan}}{2005}]{PriMon2005}
{Price} D.~J.,  {Monaghan} J.~J.,  2005, \mn@doi [\mnras]
  {10.1111/j.1365-2966.2005.09576.x}, \href
  {http://adsabs.harvard.edu/abs/2005MNRAS.364..384P} {364, 384}

\bibitem[\protect\citeauthoryear{{Price} \& {Monaghan}}{{Price} \&
  {Monaghan}}{2007}]{PriMon2007}
{Price} D.~J.,  {Monaghan} J.~J.,  2007, \mn@doi [\mnras]
  {10.1111/j.1365-2966.2006.11241.x}, \href
  {http://adsabs.harvard.edu/abs/2007MNRAS.374.1347P} {374, 1347}

\bibitem[\protect\citeauthoryear{{Price}, {Tricco}  \& {Bate}}{{Price}
  et~al.}{2012}]{PriTriBat2012}
{Price} D.~J.,  {Tricco} T.~S.,   {Bate} M.~R.,  2012, \mn@doi [\mnras]
  {10.1111/j.1745-3933.2012.01254.x}, \href
  {http://adsabs.harvard.edu/abs/2012MNRAS.423L..45P} {423, L45}

\bibitem[\protect\citeauthoryear{{Saigo} \& {Tomisaka}}{{Saigo} \&
  {Tomisaka}}{2006}]{SaiTom2006}
{Saigo} K.,  {Tomisaka} K.,  2006, \mn@doi [\apj] {10.1086/504028}, \href
  {http://adsabs.harvard.edu/abs/2006ApJ...645..381S} {645, 381}

\bibitem[\protect\citeauthoryear{{Saigo}, {Tomisaka}  \& {Matsumoto}}{{Saigo}
  et~al.}{2008}]{SaiTomMat2008}
{Saigo} K.,  {Tomisaka} K.,   {Matsumoto} T.,  2008, \mn@doi [\apj]
  {10.1086/523888}, \href {http://adsabs.harvard.edu/abs/2008ApJ...674..997S}
  {674, 997}

\bibitem[\protect\citeauthoryear{{Sch{\"o}nke} \& {Tscharnuter}}{{Sch{\"o}nke}
  \& {Tscharnuter}}{2011}]{SchTsc2011}
{Sch{\"o}nke} J.,  {Tscharnuter} W.~M.,  2011, \mn@doi [\aap]
  {10.1051/0004-6361/201015734}, \href
  {http://adsabs.harvard.edu/abs/2011A%26A...526A.139S} {526, A139+}

\bibitem[\protect\citeauthoryear{{Spitzer} \& {Tomasko}}{{Spitzer} \&
  {Tomasko}}{1968}]{SpiTom1968}
{Spitzer} Jr. L.,  {Tomasko} M.~G.,  1968, \mn@doi [\apj] {10.1086/149610},
  \href {http://adsabs.harvard.edu/abs/1968ApJ...152..971S} {152, 971}

\bibitem[\protect\citeauthoryear{{Stamatellos}, {Whitworth}, {Bisbas}  \&
  {Goodwin}}{{Stamatellos} et~al.}{2007}]{Stamatellosetal2007}
{Stamatellos} D.,  {Whitworth} A.~P.,  {Bisbas} T.,   {Goodwin} S.,  2007,
  \mn@doi [\aap] {10.1051/0004-6361:20077373}, \href
  {http://adsabs.harvard.edu/abs/2007A%26A...475...37S} {475, 37}

\bibitem[\protect\citeauthoryear{{Tomida}, {Tomisaka}, {Matsumoto}, {Ohsuga},
  {Machida}  \& {Saigo}}{{Tomida} et~al.}{2010a}]{Tomidaetal2010a}
{Tomida} K.,  {Tomisaka} K.,  {Matsumoto} T.,  {Ohsuga} K.,  {Machida} M.~N.,
  {Saigo} K.,  2010a, \mn@doi [\apjl] {10.1088/2041-8205/714/1/L58}, \href
  {http://adsabs.harvard.edu/abs/2010ApJ...714L..58T} {714, L58}

\bibitem[\protect\citeauthoryear{{Tomida}, {Machida}, {Saigo}, {Tomisaka}  \&
  {Matsumoto}}{{Tomida} et~al.}{2010b}]{Tomidaetal2010b}
{Tomida} K.,  {Machida} M.~N.,  {Saigo} K.,  {Tomisaka} K.,   {Matsumoto} T.,
  2010b, \mn@doi [\apjl] {10.1088/2041-8205/725/2/L239}, \href
  {http://adsabs.harvard.edu/abs/2010ApJ...725L.239T} {725, L239}

\bibitem[\protect\citeauthoryear{{Tomida}, {Tomisaka}, {Matsumoto}, {Hori},
  {Okuzumi}, {Machida}  \& {Saigo}}{{Tomida} et~al.}{2013}]{Tomidaetal2013}
{Tomida} K.,  {Tomisaka} K.,  {Matsumoto} T.,  {Hori} Y.,  {Okuzumi} S.,
  {Machida} M.~N.,   {Saigo} K.,  2013, \mn@doi [\apj]
  {10.1088/0004-637X/763/1/6}, \href
  {http://adsabs.harvard.edu/abs/2013ApJ...763....6T} {763, 6}

\bibitem[\protect\citeauthoryear{{Tomisaka}}{{Tomisaka}}{1998}]{Tomisaka1998}
{Tomisaka} K.,  1998, \mn@doi [\apjl] {10.1086/311504}, \href
  {http://adsabs.harvard.edu/abs/1998ApJ...502L.163T} {502, L163}

\bibitem[\protect\citeauthoryear{{Tomisaka}}{{Tomisaka}}{2002}]{Tomisaka2002}
{Tomisaka} K.,  2002, \mn@doi [\apj] {10.1086/341133}, \href
  {http://adsabs.harvard.edu/abs/2002ApJ...575..306T} {575, 306}

\bibitem[\protect\citeauthoryear{{Tricco} \& {Price}}{{Tricco} \&
  {Price}}{2012}]{TriPri2012}
{Tricco} T.~S.,  {Price} D.~J.,  2012, \mn@doi [Journal of Computational
  Physics] {10.1016/j.jcp.2012.06.039}, \href
  {http://adsabs.harvard.edu/abs/2012JCoPh.231.7214T} {231, 7214}

\bibitem[\protect\citeauthoryear{{Tricco} \& {Price}}{{Tricco} \&
  {Price}}{2013}]{TriPri2013}
{Tricco} T.~S.,  {Price} D.~J.,  2013, \mn@doi [\mnras]
  {10.1093/mnras/stt1776}, \href
  {http://adsabs.harvard.edu/abs/2013MNRAS.436.2810T} {436, 2810}

\bibitem[\protect\citeauthoryear{{Tricco}, {Price}  \& {Bate}}{{Tricco}
  et~al.}{2016}]{TriPriBat2016}
{Tricco} T.~S.,  {Price} D.~J.,   {Bate} M.~R.,  2016, \mn@doi [Journal of
  Computational Physics] {10.1016/j.jcp.2016.06.053}, \href
  {http://adsabs.harvard.edu/abs/2016JCoPh.322..326T} {322, 326}

\bibitem[\protect\citeauthoryear{{Tscharnuter}}{{Tscharnuter}}{1987}]{Tscharnuter1987}
{Tscharnuter} W.~M.,  1987, \aap, \href
  {http://adsabs.harvard.edu/abs/1987A%26A...188...55T} {188, 55}

\bibitem[\protect\citeauthoryear{{Tscharnuter}, {Sch{\"o}nke}, {Gail},
  {Trieloff}  \& {L{\"u}ttjohann}}{{Tscharnuter}
  et~al.}{2009}]{Tscharnuteretal2009}
{Tscharnuter} W.~M.,  {Sch{\"o}nke} J.,  {Gail} H.,  {Trieloff} M.,
  {L{\"u}ttjohann} E.,  2009, \mn@doi [\aap] {10.1051/0004-6361/200912120},
  \href {http://adsabs.harvard.edu/abs/2009A%26A...504..109T} {504, 109}

\bibitem[\protect\citeauthoryear{{Tsukamoto}, {Iwasaki}, {Okuzumi}, {Machida}
  \& {Inutsuka}}{{Tsukamoto} et~al.}{2015a}]{TsukamotoEtAl2015b}
{Tsukamoto} Y.,  {Iwasaki} K.,  {Okuzumi} S.,  {Machida} M.~N.,   {Inutsuka}
  S.,  2015a, \mn@doi [\mnras] {10.1093/mnras/stv1290}, \href
  {http://adsabs.harvard.edu/abs/2015MNRAS.452..278T} {452, 278}

\bibitem[\protect\citeauthoryear{{Tsukamoto}, {Iwasaki}, {Okuzumi}, {Machida}
  \& {Inutsuka}}{{Tsukamoto} et~al.}{2015b}]{TsukamotoEtAl2015a}
{Tsukamoto} Y.,  {Iwasaki} K.,  {Okuzumi} S.,  {Machida} M.~N.,   {Inutsuka}
  S.,  2015b, \mn@doi [\apjl] {10.1088/2041-8205/810/2/L26}, \href
  {http://adsabs.harvard.edu/abs/2015ApJ...810L..26T} {810, L26}

\bibitem[\protect\citeauthoryear{{Tsukamoto}, {Okuzumi}, {Iwasaki}, {Machida}
  \& {Inutsuka}}{{Tsukamoto} et~al.}{2017}]{TsukamotoEtAl2017}
{Tsukamoto} Y.,  {Okuzumi} S.,  {Iwasaki} K.,  {Machida} M.~N.,   {Inutsuka}
  S.-i.,  2017, preprint, \href
  {http://adsabs.harvard.edu/abs/2017arXiv170604363T} {} (\mn@eprint {arXiv}
  {1706.04363})

\bibitem[\protect\citeauthoryear{{Umebayashi} \& {Nakano}}{{Umebayashi} \&
  {Nakano}}{1981}]{UmeNak1981}
{Umebayashi} T.,  {Nakano} T.,  1981, \pasj, \href
  {http://adsabs.harvard.edu/abs/1981PASJ...33..617U} {33, 617}

\bibitem[\protect\citeauthoryear{{Vaytet}, {Audit}, {Chabrier}, {Commer{\c
  c}on}  \& {Masson}}{{Vaytet} et~al.}{2012}]{Vaytetetal2012}
{Vaytet} N.,  {Audit} E.,  {Chabrier} G.,  {Commer{\c c}on} B.,   {Masson} J.,
  2012, \mn@doi [\aap] {10.1051/0004-6361/201219427}, \href
  {http://adsabs.harvard.edu/abs/2012A%26A...543A..60V} {543, A60}

\bibitem[\protect\citeauthoryear{{Vaytet}, {Chabrier}, {Audit}, {Commer{\c
  c}on}, {Masson}, {Ferguson}  \& {Delahaye}}{{Vaytet}
  et~al.}{2013}]{Vaytetetal2013}
{Vaytet} N.,  {Chabrier} G.,  {Audit} E.,  {Commer{\c c}on} B.,  {Masson} J.,
  {Ferguson} J.,   {Delahaye} F.,  2013, \mn@doi [\aap]
  {10.1051/0004-6361/201321423}, \href
  {http://adsabs.harvard.edu/abs/2013A%26A...557A..90V} {557, A90}

\bibitem[\protect\citeauthoryear{{Wardle} \& {Ng}}{{Wardle} \&
  {Ng}}{1999}]{WarNg1999}
{Wardle} M.,  {Ng} C.,  1999, \mn@doi [\mnras]
  {10.1046/j.1365-8711.1999.02211.x}, \href
  {http://adsabs.harvard.edu/abs/1999MNRAS.303..239W} {303, 239}

\bibitem[\protect\citeauthoryear{{Whitehouse} \& {Bate}}{{Whitehouse} \&
  {Bate}}{2006}]{WhiBat2006}
{Whitehouse} S.~C.,  {Bate} M.~R.,  2006, \mn@doi [\mnras]
  {10.1111/j.1365-2966.2005.09950.x}, \href
  {http://cdsads.u-strasbg.fr/abs/2006MNRAS.367...32W} {367, 32}

\bibitem[\protect\citeauthoryear{{Whitehouse}, {Bate}  \&
  {Monaghan}}{{Whitehouse} et~al.}{2005}]{WhiBatMon2005}
{Whitehouse} S.~C.,  {Bate} M.~R.,   {Monaghan} J.~J.,  2005, \mn@doi [\mnras]
  {10.1111/j.1365-2966.2005.09683.x}, \href
  {http://adsabs.harvard.edu/abs/2005MNRAS.364.1367W} {364, 1367}

\bibitem[\protect\citeauthoryear{{Wurster}}{{Wurster}}{2016}]{Wurster2016}
{Wurster} J.,  2016, \mn@doi [\pasa] {10.1017/pasa.2016.34}, \href
  {http://adsabs.harvard.edu/abs/2016PASA...33...41W} {33, e041}

\bibitem[\protect\citeauthoryear{{Wurster}, {Price}  \& {Ayliffe}}{{Wurster}
  et~al.}{2014}]{WurPriAyl2014}
{Wurster} J.,  {Price} D.,   {Ayliffe} B.,  2014, \mn@doi [\mnras]
  {10.1093/mnras/stu1524}, \href
  {http://adsabs.harvard.edu/abs/2014MNRAS.444.1104W} {444, 1104}

\bibitem[\protect\citeauthoryear{{Wurster}, {Price}  \& {Bate}}{{Wurster}
  et~al.}{2016}]{WurPriBat2016}
{Wurster} J.,  {Price} D.~J.,   {Bate} M.~R.,  2016, \mn@doi [\mnras]
  {10.1093/mnras/stw013}, \href
  {http://adsabs.harvard.edu/abs/2016MNRAS.457.1037W} {457, 1037}

\bibitem[\protect\citeauthoryear{{Wurster}, {Price}  \& {Bate}}{{Wurster}
  et~al.}{2017}]{WurPriBat2017}
{Wurster} J.,  {Price} D.~J.,   {Bate} M.~R.,  2017, \mn@doi [\mnras]
  {10.1093/mnras/stw3181}, \href
  {http://adsabs.harvard.edu/abs/2017MNRAS.466.1788W} {466, 1788}

\makeatother
\end{thebibliography}
\appendix
\section{Implicit Resistivity}
\label{app:implicit}

\subsection{Algorithm}
For the implicit solver, we solve only the resistive part of the magnetic field evolution, namely
\begin{equation}
\left(\frac{{\rm d}\bm{B}}{{\rm d}t}\right)_\text{resist} = \nabla \cdot (\eta \nabla \bm{B}_a).
\end{equation}
 For the discretisation in time, either backwards Euler or Crank-Nicolson methods can be chosen, both of which are unconditionally stable for this problem.  Crank-Nicolson is more accurate than backwards Euler, but its convergence properties are not as robust for very large time steps.  We have used the Crank-Nicolson method for this paper and have had not any problems to date; if problems are found in the future either backward Euler or a hybrid method could be used.

Introducing the quantity $\mathcal{F}$, where $\mathcal{F}=1/2$ corresponds to Crank-Nicolson, and $\mathcal{F}=1$ corresponds to backward Euler, and discretising in space using the standard expression for the Laplacian in SPH  \citep[e.g.][]{Brookshaw1985,CleMon1999,Price2012}, our discrete equation is given by
\begin{eqnarray}
\frac{\bm{B}^{n+1}_a - \bm{B}^n_a}{\Delta t} & = & - \sum_b \frac{m_b}{\rho_b}\eta_{ab}  \left[ \mathcal{F} (\bm{B}^{n+1}_a - \bm{B}^{n+1}_b) - \right. \notag \\
 &~& \left. (1- \mathcal{F}) (\bm{B}^{n}_a - \bm{B}^{n}_b) \right] \overline{F}_{ab}, \label{eq:dBdt}
 \end{eqnarray}
where $\eta_{ab} \equiv (\eta_a + \eta_b)$, the superscript $n$ denotes the time step number, and we symmetrise the kernel using
\begin{equation}
\overline{F}_{ab} \equiv \frac12 \left[ \frac{\vert \nabla_a W_{ab} (h_a) \vert}{\vert r_{ab} \vert \Omega_a} + \frac{\vert \nabla_a W_{ab} (h_b) \vert}{\vert r_{ab} \vert \Omega_b} \right].
\end{equation}
In the above the $\Omega$ terms are the usual variable smoothing length correction terms \citep[e.g.][]{PriMon2007} and $\bm{r}_{ab} \equiv \bm{r}_a - \bm{r}_b$. Rearranging (\ref{eq:dBdt}), we find
\begin{equation}
\bm{B}^{n+1}_a = \frac{ \bm{B}^n_a + \Delta t_a \rho_a \bm{C}}{ 1 + \Delta t_a \rho_a  D},
\label{eq:Bupdate}
\end{equation}
where
\begin{align}
\bm{C} & \equiv \sum_b \frac{m_b}{\rho_a \rho_b} {\eta}_{ab} \left[ \mathcal{F}  (\bm{B}^{n+1}_b) + (1- \mathcal{F}) (\bm{B}^{n}_a - \bm{B}^{n}_b) \right] \overline{F}_{ab} , \\
D & \equiv \sum_b \frac{m_b}{\rho_a \rho_b} {\eta}_{ab} \overline{F}_{ab}.
\end{align}
We then solve (\ref{eq:Bupdate}) by fixed point iteration until $\vert \bm{B}_a^n - \bm{B}_a^{n-1}\vert < \epsilon$. We also check that Equation~\ref{eq:dBdt} is solved to the same tolerance. By default we use $\epsilon = 10^{-6}$. Each iteration requires recomputing $\bm{C}$ using the value of the magnetic field on neighbouring particles obtained from the previous iteration.

\begin{figure}
\centering
\includegraphics[width=\columnwidth]{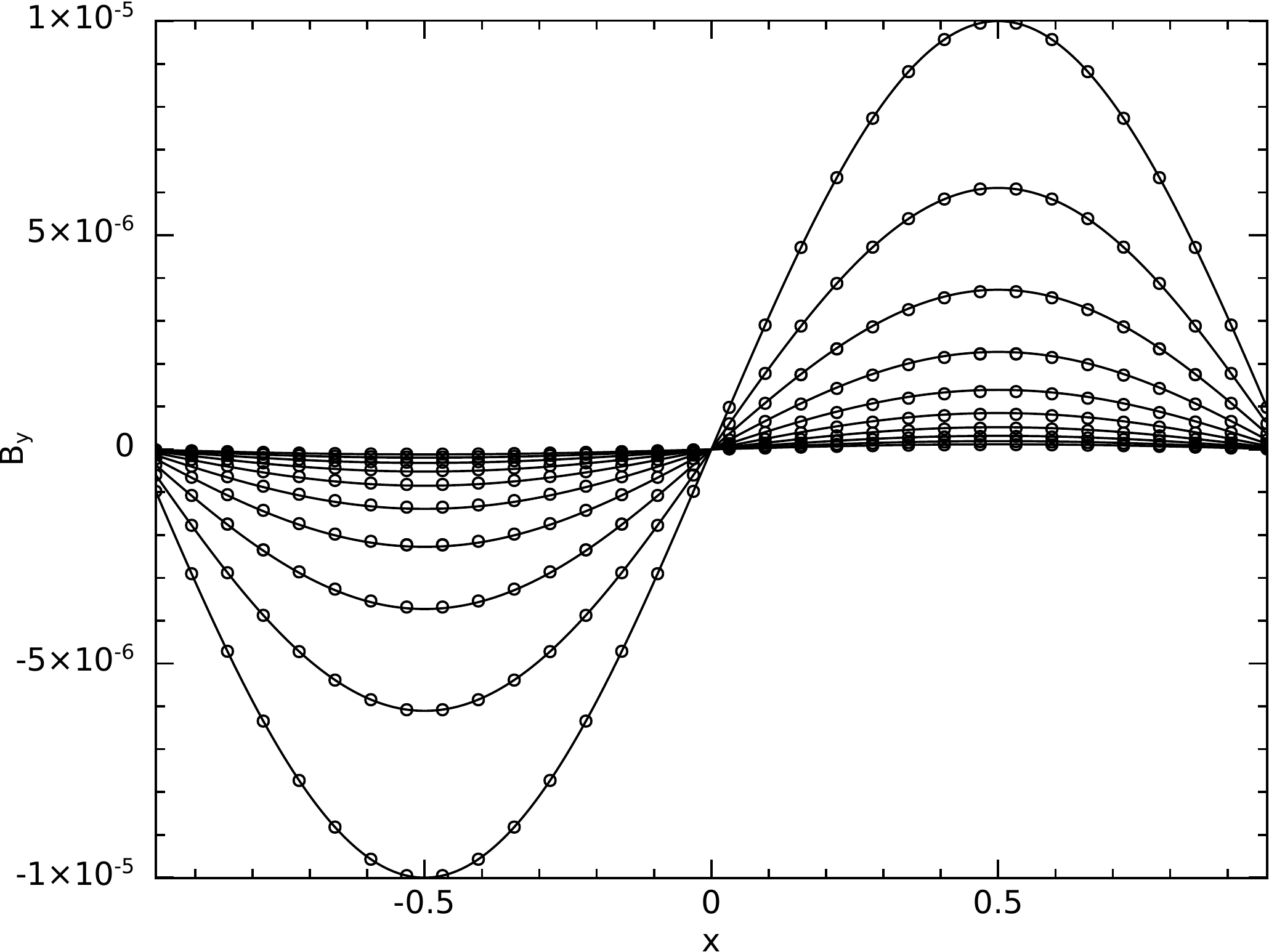}
\caption{The decay of a sinusoidal magnetic field modelled using the implicit resistivity method.  The points give the values of $B_{\rm y}$ on the SPH particles at 10 different times, in increments of ${\rm d}t=0.05$ in code units.  The solid lines give the analytical solution.}
\label{fig:implicit_decay}
\end{figure}

Taking iterations can be slow in general because it involves recomputing the neighbours for each particle. To mitigate this we follow the implementation of the implicit flux-limited diffusion method by \citet{WhiBatMon2005}. That is, we store, for all particle pairs, all of the required terms which do not change value during the iterations. Our use of individual particle timesteps further complicates matters. In this case we update $\bm{B}$ only on \emph{active} particles. For efficiency we only store neighbours within $2 h_a$ for active particles and compute the `gather' contribution from the $h_b$ term by `giving back' a contribution to neighbours during the update step. Inactive particles within either $2h_a$ or $2h_b$ are always counted as neighbours but do not receive an update. To implement this requires splitting $\overline{F}_{ab}$ into separate terms. That is, we store the following quantities for all pairs
\begin{align}
v_1^{ab} & = \frac12 \frac{m_b}{\rho_a\rho_b} {\eta}_{ab} \frac{\vert \nabla_a W_{ab} (h_a) \vert}{\vert r_{ab} \vert \Omega_a}, \\
v_2^{ab} & = \frac12 \frac{m_a}{\rho_a\rho_b} {\eta}_{ab} \frac{\vert \nabla_a W_{ab} (h_a) \vert}{\vert r_{ab} \vert \Omega_a}, \\
v_3^{ab} & = \frac12 \frac{m_b}{\rho_a\rho_b} {\eta}_{ab} \frac{\vert \nabla_a W_{ab} (h_b) \vert}{\vert r_{ab} \vert \Omega_b}.
\end{align}
Computing $\bm{C}$ for a given pair of particles $a$ and $b$ then proceeds as follows
\begin{align}
\bm{C}_a & = \bm{C}_a + 
\begin{cases}
v_1^{ab}  \bm{E}_{ab}  &  \textrm{if $b$ is active}; \\
\left(v_1^{ab} + v_3^{ab} \right) \bm{E}_{ab} & \textrm{if $b$ is inactive};
\end{cases} \\
\bm{C}_b & = \bm{C}_b + v_2^{ab} \bm{E}_{ba} \hspace{1.8cm} \textrm{if $b$ is active},
\end{align}
where $\bm{E}_{ab} = \left[ \mathcal{F}  \bm{B}^*_b + (1- \mathcal{F}) (\bm{B}^{n}_a - \bm{B}^{n}_b) \right]$ and $\bm{B}^*$ represents the updated magnetic field from the previous iteration. Typically the update converges in less than 10 iterations.

\subsection{Tests of implicit resistivity}

To test the implementation of the implicit resistivity, we modelled the decay of a magnetic field with a sinusoidal amplitude in a periodic box.  The three-dimensional cubic box had dimensions ${x,y,z}=[-1,1]$ and contained 32768 particles on a cubic lattice (32 particles per dimension) with a uniform density of $4.6\times 10^{-4}$ in code units.  The initial magnetic field was $\bm{B} = 10^{-5} \sin(\pi x) ~ \hat{\bm{y}}$.  The analytic solution is $B_{\rm y}=10^{-5} \sin(\pi x) \exp(-\pi^2 \eta t)$.  The numerical and analytical solutions at ten different times for $\eta=1$ are plotted in Fig.~\ref{fig:implicit_decay}.  Various different values of $\eta$ were tested.

\subsection{Comparison of implicit and explicit resistivity}
\begin{figure}
\centering
\includegraphics[width=\columnwidth]{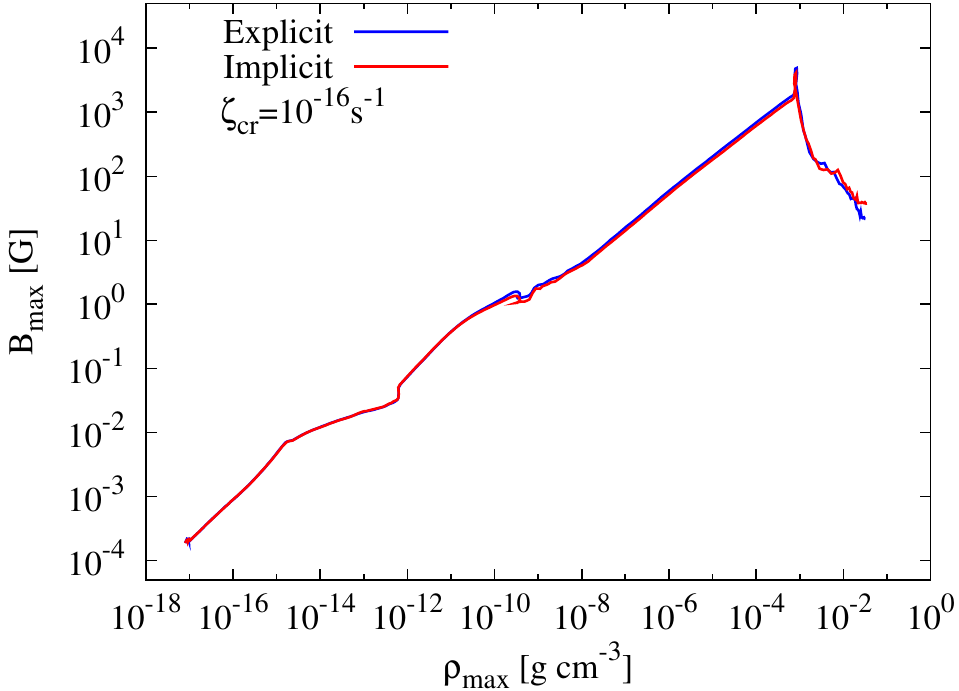}
\caption{Explicit vs implicit resistivity: Evolution of the maximum magnetic field strength versus maximum density for the collapsing molecular cloud cores  The maximum magnetic field strength agrees within 20 per cent at all densities, with the largest discrepancies during the stellar core evolution.}
\label{fig:app:rhoB}
\end{figure}

\begin{figure}
\centering
\includegraphics[width=0.5\columnwidth]{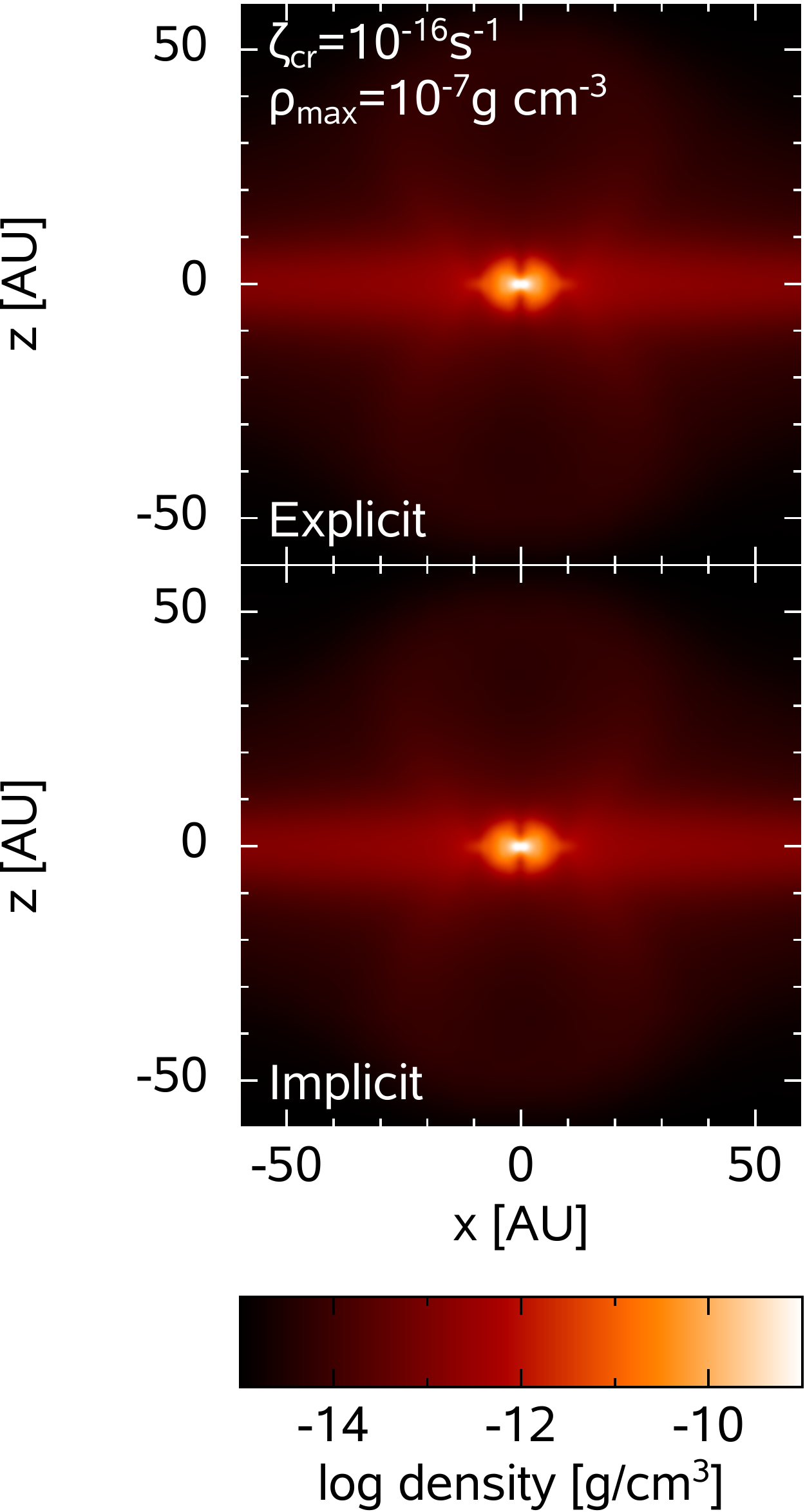}
\includegraphics[width=0.363\columnwidth,trim={3.5cm 0 0 0},clip]{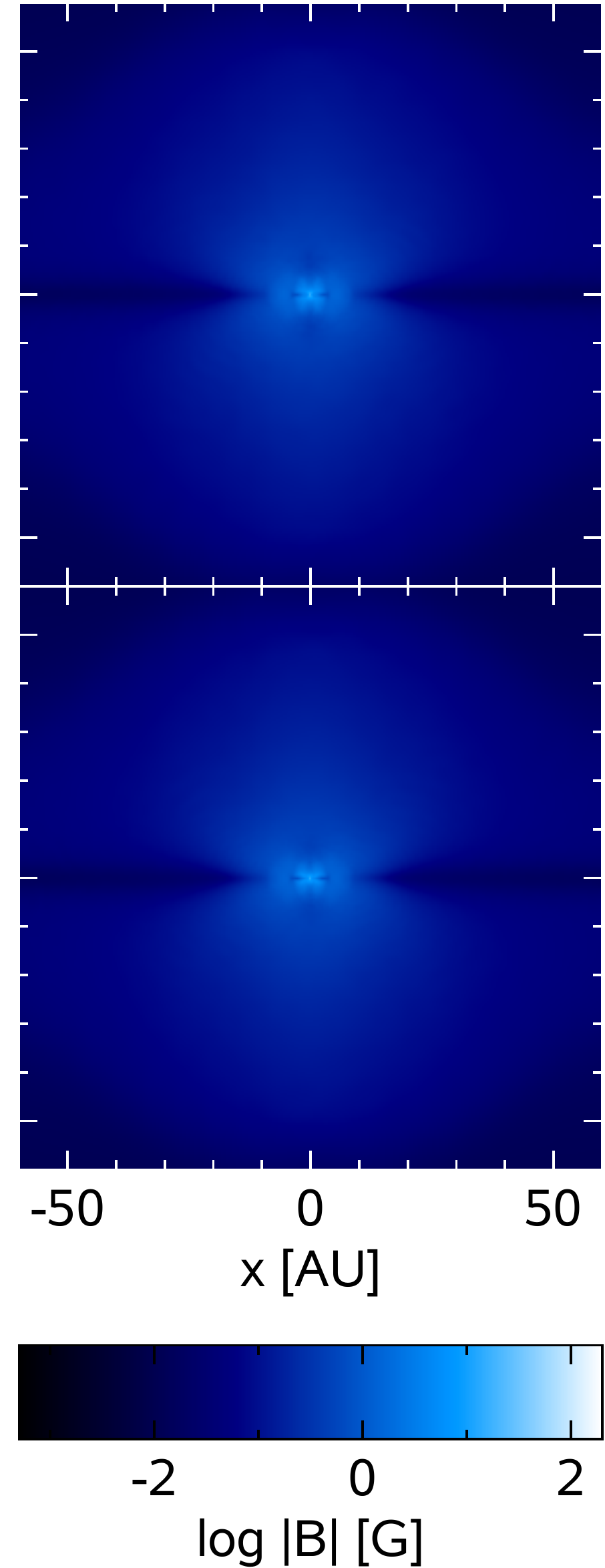}
\caption{Explicit vs implicit resistivity: Gas density (left) and magnetic field strength (right) cross sections taken through the centre of the first core and parallel to the rotation axis when \rhoapprox{-7} at the end of the first collapse phase.  The results are indistinguishable from one other.}
\label{fig:app:fhc:rhoB}
\end{figure}

\begin{figure}
\centering
\includegraphics[width=0.5\columnwidth]{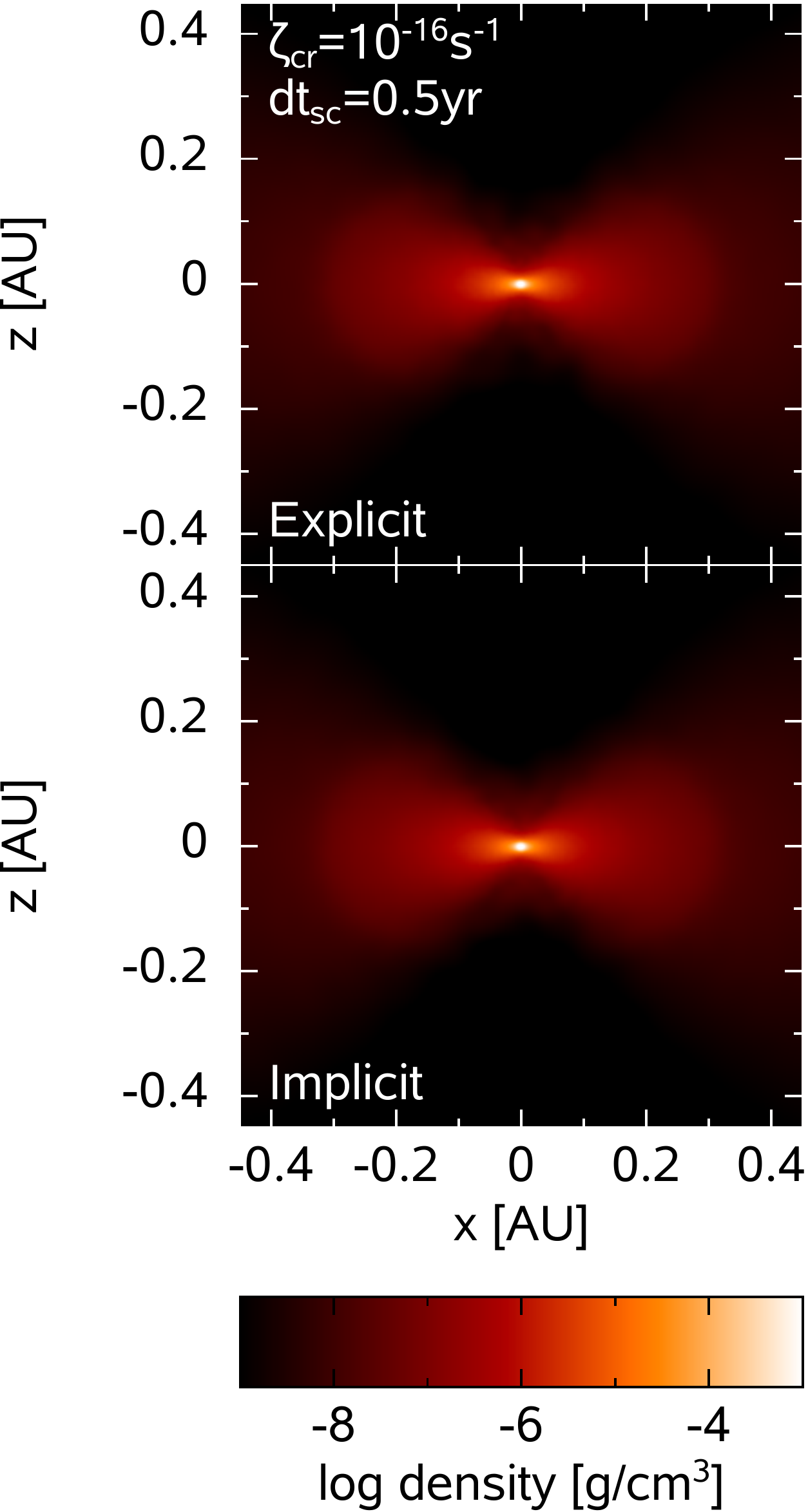}
\includegraphics[width=0.363\columnwidth,trim={3.5cm 0 0 0},clip]{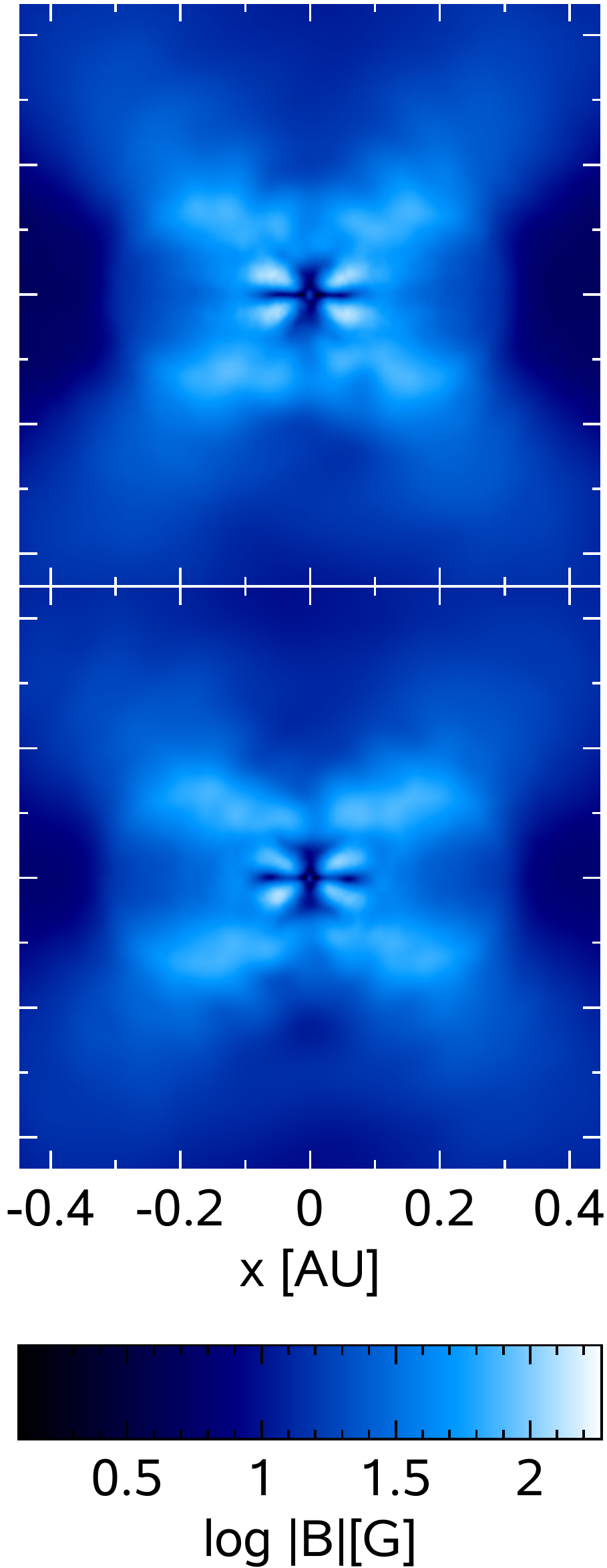}
\caption{Explicit vs implicit resistivity: Gas density (left) and magnetic field strength (right) cross sections taken through the centre of the stellar core and parallel to the rotation axis at \dtscapprox{0.5} after the formation of the stellar core.  Frame sizes are chosen to show the detail of the circumstellar disc surrounding the stellar core.  As with earlier times, the structures are the same, and the results are almost indistinguishable from one another.}
\label{fig:app:shc:rhoB}
\end{figure}

We present two versions of \zetam{16} (i.e. our model with \zetaeq{-16}) using both the implicit Ohmic resistivity (used in Section~\ref{res:Halldirection}) and explicit Ohmic resistivity (used in the rest of our paper).  Fig.~\ref{fig:app:rhoB} plots the maximum magnetic fields strength as a function of the maximum density for \dtsclt{4}.  Fig.~\ref{fig:app:fhc:rhoB} shows cross sections of density and magnetic field strength through the centre of the first core and parallel to the rotation axis at the end of the first core phase at \rhoapprox{-7}.  Fig.~\ref{fig:app:shc:rhoB} shows the cross sections at \dtscapprox{0.5}.  The scales on the plots have been altered to emphasise the comparison, and do not necessarily match the scales used in previous sections of this paper.

Ohmic resistivity becomes important during the first collapse.  The maximum field strengths in the two calculations are slightly different with the implicit calculation producing sightly lower maximum values (Fig.~\ref{fig:app:rhoB}).  By the end of the first core phase, the mean and maximum magnetic field strengths are \appx1 per cent lower when using implicit Ohmic resistivity.  After the formation of the stellar core, the maximum and mean magnetic field strengths of the two calculations agree to within \appx30 per cent. 

Despite the small differences in the maximum field strength, the morphologies of the first and stellar cores and the outflows are almost indistinguishable from one another (Fig.~\ref{fig:app:fhc:rhoB} and \ref{fig:app:shc:rhoB}).  Thus, we are confident that the implicit Ohmic resistivity can be used to speed up the simulations without adversely affecting the solution.


\label{lastpage}
\end{document}